\documentclass[12pt]{article}
\usepackage[T1]{fontenc}
\usepackage{amsmath}
\usepackage{amsfonts}
\usepackage{graphicx}
\usepackage[mathscr]{euscript}
\usepackage{rotating}
\usepackage{pdflscape}
\usepackage[hmargin=2.5cm,vmargin=2.5cm]{geometry}
\usepackage{booktabs,caption}
\usepackage{natbib}
\usepackage{setspace}
\usepackage{bbm}
\usepackage{amsthm}
\usepackage{enumerate}
\usepackage{threeparttable}
\usepackage{verbatim}
\usepackage{caption}
\usepackage{url}
\usepackage{amssymb}
\usepackage{subcaption,array}
\usepackage{adjustbox}
\usepackage[titletoc,title]{appendix}

\usepackage[hidelinks,breaklinks]{hyperref}
\hypersetup{colorlinks=true,allcolors=blue}

\usepackage{framed}
\usepackage[dvipsnames]{xcolor}

\theoremstyle{definition}

\graphicspath{ {./figures/} }

\begin{document}

\title{\textbf{Decentralized Decision-Making in Retail Chains:\\Evidence from Inventory Management}\thanks{We are grateful for the valuable comments received from Heski Bar-Isaac, Loren Brandt, Avi Goldfarb, Jiaying Gu, George Hall, Jordi Mondria, Bob Miller, John Rust, Steven Stern, Junichi Suzuki, and Kosuke Uetake, as well as from seminar participants at Cambridge, Stony Brook, Toronto, UBC-Sauder, EARIE conference, IIOC conference, CEA conference, and the Georgetown conference honoring John Rust. The data used in this paper was obtained through a request made under the Canadian \textit{Access to Information Act}. We sincerely appreciate the generous assistance provided by LCBO personnel.}}

\author{
    Victor Aguirregabiria\footnote{Department of Economics, University of Toronto. 150 St. George Street, Toronto, ON, M5S 3G7, Canada,  \href{mailto: victor.aguirregabiria@utoronto.ca}{victor.aguirregabiria@utoronto.ca}.} 
    \\ \emph{University of Toronto, CEPR} 
    \and 
    Francis Guiton \footnote{Ph.D. Candidate, Department of Economics, University of Toronto. 150 St. George Street, Toronto, ON, M5S 3G7, Canada, \href{mailto: francis.guiton@mail.utoronto.ca} { francis.guiton@mail.utoronto.ca}.} \\ \emph{University of Toronto}}

\maketitle

\thispagestyle{empty}

\begin{abstract}
    \noindent 
    This paper investigates the impact of decentralizing inventory decision-making in multi-establishment firms using data from a large retail chain. Analyzing two years of daily data, we find significant heterogeneity among the inventory decisions made by 634 store managers. By estimating a dynamic structural model, we reveal substantial heterogeneity in managers’ perceived costs. Moreover, we observe a correlation between the variance of these perceptions and managers’ education and experience. Counterfactual experiments show that centralized inventory management reduces costs by eliminating the impact of managers’ skill heterogeneity. However, these benefits are offset by the negative impact of delayed demand information. 

    \vspace{0.4cm}
    
    \vspace{0.4cm}
\end{abstract}

\newpage

\setcounter{page}{1}

\begin{onehalfspacing}

\section{Introduction}

Multi-establishment firms can adopt various decision-making structures ranging from centralized decisions at the headquarters to a more decentralized approach where decision authority is delegated to individual establishments. Determining the optimal decision-making process for a firm involves weighing different trade-offs. A decentralized decision-making structure empowers local managers to leverage valuable information specific to their respective stores. This information, which may be difficult or time-consuming to communicate to headquarters, can be utilized effectively at the local level. However, decentralization also entails granting autonomy to heterogeneous managers who possess varied skills. This heterogeneity can lead to suboptimal outcomes for the overall firm. When determining the degree of decentralization in their decision-making structure, multi-establishment firms must evaluate this trade-off. They need to consider the benefits of local knowledge and timely decision-making against the potential challenges posed by managerial heterogeneity. 

In this study, we investigate the inventory management decisions made by store managers at the \textit{Liquor Control Board of Ontario} (LCBO) and examine the impact on the firm's performance of delegating these decisions to the store level. The LCBO is a provincial government enterprise responsible for alcohol sales throughout Ontario. As a decentralized retail chain, each store has a degree of autonomy in its decision-making process. Specifically, store managers have discretion in two key areas of inventory management: assortment decisions (i.e., determining which products to offer) and replenishment decisions (i.e., when and how much to order for each product). Replenishment decisions involve forming expectations about future demand to determine the optimal order quantity and timing. To conduct our analysis, we utilize a comprehensive dataset obtained from the LCBO, encompassing daily information on inventories, orders, sales, stockouts, and prices for every store and product (SKU) from October 2011 to October 2013 (677 working days).\footnote{These data were obtained under the Access to Information Act, with assistance from the LCBO personnel.} Additionally, we supplement our main dataset with information from LCBO reports and gather data on store managers' education and experience from professional networking platforms such as \textit{LinkedIn}.

The LCBO data and framework provide a unique setting to study inventory management due to the simple pricing mechanism employed, where prices are set as a fixed markup over wholesale cost. This feature of the market allows us to focus specifically on the inventory setting problem without the need to incorporate a model where equilibrium prices are explicitly determined by inventory decisions. By abstracting from the complex relationship between prices and inventory decisions, we can concentrate our analysis on understanding the factors influencing inventory management within the LCBO retail chain.

By employing descriptive evidence and the estimation of reduced-form models of inventory decision rules, we first show substantial heterogeneity across store managers’ replenishment decisions. Observable store characteristics -- such as demand level, size, category, and geographic location -- explain less than half of the differences across store managers in their inventory decision rules.

To gain a deeper understanding of the factors contributing to this heterogeneity, we propose and estimate a dynamic structural model of inventory management. The model allows for differences across stores in demand, storage costs, stockout costs, and ordering costs. Leveraging the high-frequency nature of the daily data, we obtain precise estimates of holding cost, stockout cost, and ordering costs at both the individual product (SKU) and store levels. Our findings reveal significant heterogeneity across stores in all the revealed-preference cost parameters. Remarkably, observable store characteristics only account for less than $50\%$ of this heterogeneity. Furthermore, we uncover a correlation between the remaining heterogeneity and managers' education and experience, suggesting that manager characteristics contribute to this residual variation. We interpret this unexplained heterogeneity as the result of local managers' idiosyncratic perceptions regarding store-level costs.

Using the estimated structural model, we quantify the impact of store manager heterogeneity on inventory outcomes. Overall, eliminating the idiosyncratic heterogeneity in cost parameters produces significant effects on inventory management. Specifically, we observe a $6$-day increase in the waiting time between orders, a decrease in the average order amount equivalent to $1.5$ days of average sales, and a $21\%$ reduction in the inventory-to-sales ratio. However, the frequency of stockouts remains largely unaffected. These findings indicate that if the idiosyncratic component of costs represents a biased perception by store managers, it has a substantial negative impact on the firm's profitability. This is due to increased storage and ordering costs while having little effect on stockouts and revenue generation.

Finally, we conduct an evaluation of the effects associated with centralizing the decision-making of inventory management at the LCBO headquarters. To simulate this counterfactual experiment, we take into account information provided by company reports, which indicate that store-level sales information is processed by the headquarters with a one-week delay. The main trade-off examined in this experiment revolves around the fact that a centralized inventory management system eliminates the influence of store managers' heterogeneous skills and biased perceptions of costs. However, it also relinquishes the advantages derived from store managers' just-in-time information about demand and inventories. Our findings reveal that implementing a centralized inventory management system would result in a substantial reduction in ordering and storage costs, with an average cost decrease of $23\%$ and a $3.7\%$ reduction for the median store. Despite the significant cost reduction, this benefit is nearly completely offset by the negative impact on profits due to the delayed information about demand. Consequently, the net effect on profits is modest, with a mere $2\%$ increase in annual profit for LCBO, equivalent to $\$34$ million. We further explore the implications of this trade-off for designing a more efficient inventory system that incorporates elements of both centralized and decentralized approaches.

This paper contributes to the growing empirical literature exploring the trade-offs between centralization and decentralization of decision-making in multi-division firms. Notably, \citeauthor{dellavigna_gentzkow_2019} (\citeyear{dellavigna_gentzkow_2019}) observe that most retail chains in the US employ uniform pricing across their stores, despite substantial differences in demand elasticities and potential gains from third-degree price discrimination. The authors discuss possible explanations for this phenomenon. In a study of a major international airline company, \citeauthor{hortacsu_natan_2021} (\citeyear{hortacsu_natan_2021}) analyze a pricing system that combines decision rights across different organizational teams. They find that despite employing advanced techniques, the pricing system fails to internalize consumer substitution effects, exhibits persistent biases in demand forecasting, and does not adapt to changes in opportunity costs. These inefficiencies are primarily attributed to limited coordination between teams. Examining decentralization decisions from a diverse set of firms across 11 OECD countries, \citeauthor{aghion_bloom_2021} (\citeyear{aghion_bloom_2021}) show that firms that delegate decision power from central headquarters to plant managers exhibited better performance during the Great Recession compared to similar firms with more centralized structures. The empirical evidence presented in their study supports the interpretation that the value of local information increases during turbulent economic times. Our paper contributes to this literature by being, to the best of our knowledge, the first empirical study to examine the trade-offs related to the (de)centralization of inventory management within retail chains, and specifically the role of store managers' heterogeneous skills.

This paper also contributes to the empirical literature on dynamic structural models of inventory behavior. Previous contributions in this area include works by \citeauthor{aguirregabiria_1999} (\citeyear{aguirregabiria_1999}), \citeauthor{hall_rust_2000} (\citeyear{hall_rust_2000}), \citeauthor{kryvtsov_midrigan_2013} (\citeyear{kryvtsov_midrigan_2013}), and \citeauthor{bray_yao_2019} (\citeyear{bray_yao_2019}) in the context of firm inventories, as well as \citeauthor{eberly_1994} (\citeyear{eberly_1994}), \citeauthor{attanasio_2000} (\citeyear{attanasio_2000}), and \citeauthor{adda_cooper_2000} (\citeyear{adda_cooper_2000}) in the domain of household purchases of durable products. We contribute to this literature by using high-frequency (daily) data at the granular store and product level to estimate cost parameters using a dynamic structural model.

Finally, our paper contributes to the literature on structural models with boundedly rational firms. Most of this literature studies firms' entry/exit decisions (\citeauthor{goldfarb_xiao_2011}, \citeyear{goldfarb_xiao_2011}; and 
\citeauthor{aguirregabiria_magesan_2020}, \citeyear{aguirregabiria_magesan_2020}), 
pricing decisions (\citeauthor{huang_ellickson_2022}, \citeyear{huang_ellickson_2022}; \citeauthor{ellison_snyder_2018}, \citeyear{ellison_snyder_2018}), and bidding behavior in auctions (\citeauthor{hortacsu_puller_2008}, \citeyear{hortacsu_puller_2008};
\citeauthor{doraszelski_lewis_2018}, \citeyear{doraszelski_lewis_2018}; \citeauthor{hortacsu_luco_2019}, \citeyear{hortacsu_luco_2019}). To the best of our knowledge, our paper represents the first investigation into bounded rationality in firms' inventory decisions. This research also contributes to the existing literature by combining revealed-preference estimates of managers' perceived costs with a decomposition of these costs into the objective component explained by store characteristics and the subjective component associated with managers' education and experience. 

The rest of the paper is organized as follows. Section \ref{sec:firm_data} describes the institutional background of the LCBO and presents the dataset and descriptive evidence. Section \ref{sec:s_s_rules} presents evidence of managers following $(S,s)$ decision rules and illustrates the heterogeneity in these $(S,s)$ thresholds across store managers. Section \ref{sec:structural}
presents the structural model and its estimation. The counterfactual experiments to evaluate the effects of decentralization are described in section \ref{sec:counterfactuals}. We summarize and conclude in Section \ref{sec:conclusion}.

\section{Firm and data \label{sec:firm_data}}

\subsection{LCBO retail chain\label{sec:lcbo}}

\textbf{History.} LCBO was founded in 1927 as part of the passage of the Ontario Liquor Licence Act.\footnote{The information in this section originates from various archived documents from the LCBO. General information about the company and its organization is based on the company's annual reports \citeauthor{lcbo_2011_2012} (\citeyear{lcbo_2011_2012}) and \citeauthor{lcbo_2012_2013} (\citeyear{lcbo_2012_2013}), and the collective agreement between the LCBO and OPSEU. Information regarding headquarters' order recommendations (\textit{Suggested Order Quantities}, SOQs) originates from the report \citeauthor{lcbo_2016} (\citeyear{lcbo_2016}). Additional information regarding the role of store managers originates from an interview we conducted with an LCBO store manager from a downtown Toronto store.} This act established that LCBO was a crown corporation of the provincial government of Ontario. Today, the wine retail industry in Ontario is a triopoly - consisting of 634 LCBO stores, 164 Wine Rack stores, and 100 Wine Shop stores. Despite its government ownership, LCBO is a profit maximizing company. As described in its governing act, part of its mandate is "generating maximum profits to fund government programs and priorities".\footnote{See
https://www.lcbo.com/content/lcbo/en/corporate-pages/about/aboutourbusiness.html.}

\medskip

\noindent \textbf{Store managers.} According to the LCBO, store managers are responsible for managing their "store, sales and employees to reflect [their] customers’ needs and business goals", with a particular focus on the inventory management of their store. Managers must oversee their store's overall inventory level to ensure that daily demand is met. For incentive purposes, part of the store managers' remuneration depends on the overall sales performance of their store. The managers' pay is therefore closely tied to their stores' profits. In order to satisfy daily demand, managers periodically restock their shelves by ordering products from the nearest distribution center. The order is then delivered by trucks according to a pre-determined route and schedule. In addition to the ordering decisions, managers are also responsible for their store's product assortment, as they must decide which products to offer at their store in order to cater to local demand. Inventory management at LCBO, therefore, entails a dual responsibility for store managers: providing products that are in high demand and keeping these products stocked on the shelves. 

\medskip

\noindent \textbf{Classification of stores.} 
The LCBO classifies its stores into six categories, AAA, AA, A, B, C, and D, ranging from the highest to the lowest. These classifications primarily reflect variations in store size and product assortment. However, there are also differences in the consumer shopping experience across these classifications, with the AAA and AA stores being considered flagship stores.

\medskip

\noindent \textbf{Headquarters.} Headquarters are in charge of the assignment of store managers across the different stores. Assignments are occasionally shuffled due to managers being promoted (demoted) to higher- (lower-) classified stores, with seniority being a main factor in the promotion decision. Another responsibility of headquarters is to assist store managers in their inventory decisions. Headquarters use forecasting techniques to provide recommendations to managers regarding how much to order for each product at their store. In the company's internal jargon, these recommendations are referred to as \textit{Suggested Order Quantities} (SOQs). For each store and product, headquarters generate order recommendations based on the previous week's sales and inventory information.\footnote{More specifically, order recommendations are a function of the \textit{Average Rate of Sale} (ARS) of the product from the previous week, and of seasonal brand factors.} Importantly, this entails that headquarters process store-level information with a weekly delay. Headquarters do not use \textit{just-in-time} daily information that store managers may be using in their replenishment decisions. This informational friction may play a role in the optimal allocation of decision rights.

\medskip

\noindent \textbf{Pricing.} LCBO and its competitors are subject to substantial pricing restrictions. Prices must be the same across all stores in all markets for a given store-keeping unit (SKU). There is no price variation across the LCBO and its competitors. Retail prices are determined on a fixed markup over the wholesale price set by wine distributors. Furthermore, the percentage markup applies to all the SKUs within broadly defined categories.\footnote{See \citeauthor{aguirregabiria_ershov_2016} (\citeyear{aguirregabiria_ershov_2016}) for further details about markups at LCBO.}

\medskip

\noindent \textbf{No franchising system.} The LCBO operates its stores without adopting a franchising system. Instead, all store managers are employees of the LCBO. As a result, store managers are not required to pay any franchise fees, fees per order, or any other types of fees to the firm. The absence of a franchising system ensures that the store managers operate within the organizational structure of the LCBO as employees, without the additional financial obligations associated with a franchising arrangement.

\medskip

\noindent \textbf{Union.} Most employees at the LCBO are unionized under the Ontario Public Service Employees Union (OPSEU). As of 2022, the OPSEU "represents more than 8,000 workers at the Liquor Control Board of Ontario", with their main goal being to "establish and continue harmonious relations between the [LCBO] and the employees". Members of the union include store managers, retail workers, warehouse workers, and corporate workers.

\subsection{Data from LCBO \label{sec:data}}

Our analysis combines three data sources: the main dataset provided by the LCBO; data on store managers' experience and education collected from the social media platform \textit{LinkedIn}; and consumers' socioeconomic characteristics from the 2011 Census of Population.

We use a comprehensive and rich dataset obtained from the LCBO, encompassing daily information on inventories, sales, deliveries, and prices of every product sold at each LCBO store. The dataset covers a period of two years, specifically from October 2011 to October 2013, spanning a total of 677 days. With a total of 634 stores operating across Ontario and an extensive product range consisting of over 20,000 different items, our dataset comprises approximately 720 million observations. Moreover, the dataset includes additional valuable information, such as product characteristics, store characteristics (including location, size, and store category), and the store manager's name.

Table \ref{tab:summary} presents summary statistics. The average store has an assortment of  $2,029$ items. Weekly sales per store amount to $12,909$ units, translating to an average of $6.36$ units sold per item. The average weekly revenue per store is $\$162,250$, resulting in $\$80$ in revenue per item per week and $\$12.6$ in revenue per unit sold. Regarding deliveries, stores receive shipments at approximately $5.37$ days per week, with total weekly deliveries containing an average of $12,258$ units. Stockout events occur, on average, $415$ times per week across all stores. 

Notably, these figures exhibit significant variation across different store types. Larger stores, as expected, generate higher weekly revenues. For instance, \textit{AAA} and \textit{AA} stores generate average weekly revenues of $\$874,367$ and $\$518,067$, respectively, while \textit{C} and \textit{D} stores generate average weekly revenues of $\$59,328$ and $\$23,913$, respectively. Stockout events appear to be more prevalent in larger stores compared to smaller ones. The average \textit{AAA} store experiences $1,203$ stockout events per week, whereas the average \textit{D} store encounters $204$ stockout events per week. Additionally, larger stores tend to place orders more frequently and in larger quantities. On average, \textit{AAA} and \textit{AA} stores receive orders $6.30$ and $6.27$ days per week, totaling $51,340$ and $35,711$ units, respectively. In contrast, \textit{C} and \textit{D} stores receive orders $5.09$ and $3.83$ days per week, amounting to $4,395$ and $1,660$ units, respectively.

The bottom panel in Table \ref{tab:summary} presents inventory-to-(daily)sales ratios and ordering frequencies. These statistics are closely related to the $(S,s)$ decision rules that we analyze in Section \ref{sec:s_s_rules}. 
At the store-product level, the \textit{inventory-to-sales ratio} before and after an order corresponds to the thresholds \textit{s} and \textit{S}, respectively. On average, stores maintain enough inventory to meet product demand for approximately $23$ days, initiate an order when there is sufficient inventory for about $9$ days, and the ordered quantity covers sales for around $18$ days. In terms of ordering frequency, the average store and product place an order approximately once every two weeks, equivalent to a frequency of $0.07 \simeq 1/14$.

Average inventory-to-sales ratios tend to decrease with store size/type, although this difference is influenced by the composition effect arising from varying product assortments across store types. For the remainder of the paper, to account for this composition effect and reduce the computational burden of estimating our model across numerous products, we focus on a working sample consisting of a few products carried by all stores. This approach allows us to manage the complexity associated with different product assortments while maintaining the robustness of our analysis.

\begin{table}[ht]
\caption{Summary Statistics \label{tab:summary}}
\centering
\resizebox{\textwidth}{!}{\begin{tabular}{r|ccccccc}
\hline \hline 
& \multicolumn{6}{c}{\textit{Type of store}}
\\
& \textit{All} & \textit{AAA} 
& \textit{AA} & \textit{A} 
& \textit{B} & \textit{C} 
& \textit{D}
\\ \hline 
& \textit{Mean} & \textit{Mean} 
& \textit{Mean} & \textit{Mean} 
& \textit{Mean} & \textit{Mean} 
& \textit{Mean}
\\
& \textit{(st.dev)} & \textit{(st.dev)} 
& \textit{(st.dev)} & \textit{(st.dev)} 
& \textit{(st.dev)} & \textit{(st.dev)} 
& \textit{(st.dev)}
\\ \hline
\multicolumn{1}{l|}
{\textbf{Number of Observations}}
&  &
&  & 
&  & 
&
\\ 
\textit{Number of Stores} 
& 634 & 5
& 25 & 148
& 157 & 164
& 135
\\
\textit{Number of Unique Products} 
& 22,327 & 18,200
& 18,879 & 20,860
& 17,769 & 13,527
& 9,198
\\
\textit{Number of Days} 
& 677 & 676
& 676 & 676
& 677 & 676
& 675
\\
\hline
\multicolumn{1}{l|}{\textbf{Sales \& Stockouts Per Store}}
&  & 
&  & 
&  &
&
\\
\textit{Revenue per week (\$)} 
& 162,249 & 874,367
& 518,067 & 320,263 & 162,588 & 59,327 & 23,912
\\
& (158,268) & (255,172)
& (111,241) & (75,387) & (47,385) & (21,159) & (10,503)
\\
\textit{Units sold per week} 
& 12,908 & 57,160
& 38,369 & 25,665 & 14,068 & 4,522 & 1,730
\\
& (11,885) & (8,992)
& (5,399) & (5,405) & (4,554) & (1,820) & (819)
\\
\textit{Stockout events per week} 
& 414  & 1,202
& 912 & 699 & 371 & 273 & 204
\\
& (326) & (429)
& (279) & (306) & (152) & (252) & (200)
\\ \hline
\multicolumn{1}{l|}{\textbf{Inventories Per Store}}
&  & 
&  & 
&  &
&
\\
\textit{Number of products} 
& 2,028 & 6,477
& 4,852 & 3,555 & 2,102 & 1,098 & 742
\\
& (1,416) & (1,018)
& (745) & (937) & (706) & (291) & (190)
\\
\textit{Delivery days per week} 
& 5.37  & 6.30
& 6.27 & 6.23 & 6.06 & 5.09 & 3.83
\\
& (1.08)& (0.01)
& (0.05) & (0.11) & (0.31) & (0.66) & (0.88)
\\
\textit{Delivery units per week} 
& 12,257 & 51,339
& 35,710 & 24,350 & 13,476 & 4,395 & 1,660
\\
& (11,108) & (6,872)
& (4,848) & (5,063) & (4,478) & (1,792) & (814)
\\ \hline
\multicolumn{1}{l|}{\textbf{Inventory Ratios Per Store}}
&  & 
&  & 
&  &
&
\\
\textit{Inventory to (daily) sales ratio} 
& 23.31 & 15.67 & 15.74 & 16.89 & 18.97 & 25.21 & 34.23
\\
& (9.86) & (2.16) & (2.74) & (3.32) & (5.24) & (7.20) & (11.85)
\\
\textit{Inventory to sales ratio after order} 
& 18.48 & 8.78 & 10.34 & 12.18 & 15.13 & 21.65 & 26.87
\\
& (9.16) & (0.62) & (1.55) & (2.76) & (4.40) & (7.38) & (11.72)
\\
\textit{Inventory to sales ratio before order} 
& 8.62 & 5.31 & 6.83 & 7.82 & 8.88 & 9.23 & 8.92
\\
& (3.59) & (0.60) & (1.40) & (2.39) & (2.97) & (3.66) & (4.99)
\\
\textit{Ordering frequency} 
& 0.07 & 0.10
& 0.10 & 0.09
& 0.08 & 0.06
& 0.04
\\
& (0.02) & (0.01) & (0.01) & (0.01) & (0.02) & (0.01) & (0.02)
\\
\hline \hline
\end{tabular}}
\end{table}

\subsection{Working sample}

In our econometric models, estimated in Sections \ref{sec:s_s_rules} and \ref{sec:structural}, the parameters are unrestricted at the store-product level. Considering that our dataset comprises nearly 2 million store-product pairs, estimating these models for every store-product combination would be exceedingly time-consuming. To save time while maintaining the integrity of our analysis, we have employed a different approach. Specifically, we estimate the econometric models for every store in our dataset but limited the analysis to a selected subset of 5 products. 

We employ two criteria to determine the product basket for our analysis. Firstly, we select products that exhibit high sales across all LCBO stores, ensuring that their inventory decisions significantly impact the firm's overall profitability. Secondly, we include products from each broad category to account for product-level heterogeneity. These categories encompass white wine, red wine, vodka, whisky, and rum. Table \ref{tab:sku_summary} provides an overview of the five selected products that satisfy these criteria. By focusing on this subset, we capture a diverse range of products that are representative of the different categories while also being impactful in terms of their sales performance.

\medskip

\begin{table}[ht]
\caption{SKUs - Working Sample \label{tab:sku_summary}}
\centering
\resizebox{\textwidth}{!}{\begin{tabular}{r|ccccc}
\hline \hline 
& \textit{SKU} & \textit{SKU} 
& \textit{SKU} & \textit{SKU} 
& \textit{SKU}
\\
& \textit{\#67} & \textit{\#117} 
& \textit{\#340380} & \textit{\#550715} 
& \textit{\#624544}
\\ \hline 
\multicolumn{1}{l|}
{\textbf{Product Information}}
&  &
&  & 
&  
\\ 
\textit{Name} 
& \textit{Smirnoff} & \textit{Bacardi Superior} & \textit{Two Oceans}  & \textit{Forty Creek} 
& \textit{Yellow Tail}  
\\
 & \textit{Vodka} & \textit{White Rum} & \textit{Sauvignon Blanc} & \textit{Barrel Select Whisky} & \textit{Shiraz}
 \\
 &  &
&  & 
&  
\\ 
\textit{Category} 
& Vodka & Rum
& White Whine & Whisky
& Red Wine 
\\
&  &
&  & 
&  
\\ 
\textit{Average Retail Price (\$)} 
& 25.28 & 24.93
& 9.98 & 25.86
& 11.83
\\
&  &
&  & 
&  
\\ 
\hline \hline
\end{tabular}}
\end{table}

Table \ref{tab_working_summary} provides summary statistics for our working sample. On average, the stores in our working sample sell $91$ units per week, equivalent to $18$ units per SKU. The average weekly revenue per store is $\$1,687$, resulting in $\$347$ in revenue per SKU per week and $\$18$ in revenue per unit sold. Regarding deliveries, stores in our working sample receive shipments approximately 2 days per week, with each delivery containing an average of $88$ units. The average number of stockout events per week per store is $0.37$. 

Similar to Table \ref{tab:summary}, these numbers exhibit significant variation across different store types. Larger stores generate higher average weekly revenues compared to smaller ones. For instance, the average \textit{AAA} and \textit{AA} stores generate weekly revenues of $\$5,197$ and $\$4,511$, respectively, while the smaller \textit{C} and \textit{D} stores generate average weekly revenues of $\$821$ and $\$302$, respectively. Contrary to the full sample, stockout events appear to occur more frequently in smaller stores than in larger ones within our working sample. The average \textit{D} store experiences $0.56$ stockout events per week, while the average \textit{AAA} store encounters $0.14$ stockout events per week. Delivery patterns in our working sample follow a similar trend to Table \ref{tab:summary}, with larger stores placing larger and more frequent orders. The average \textit{AAA} store receives deliveries 4 days per week, totaling 272 units per week, whereas the average \textit{D} store only receives deliveries 0.6 days per week, amounting to 15 units per week.

\medskip

\begin{table}[ht]
\caption{Summary Statistics - Working Sample \label{tab_working_summary}}
\centering
\resizebox{\textwidth}{!}{\begin{tabular}{r|ccccccc}
\hline \hline 
& \multicolumn{7}{c}{\textit{Type of store}}
\\
& \textit{All} & \textit{AAA} 
& \textit{AA} & \textit{A} 
& \textit{B} & \textit{C} 
& \textit{D}
\\ \hline 
& \textit{Mean} & \textit{Mean} 
& \textit{Mean} & \textit{Mean} 
& \textit{Mean} & \textit{Mean} 
& \textit{Mean}
\\
& \textit{(st.dev)} & \textit{(st.dev)} 
& \textit{(st.dev)} & \textit{(st.dev)} 
& \textit{(st.dev)} & \textit{(st.dev)} 
& \textit{(st.dev)}
\\ \hline
\multicolumn{1}{l|}
{\textbf{Number of Observations}}
&  &
&  & 
&  & 
&
\\ 
\textit{Number of Stores} 
& 634 & 5
& 25 & 148
& 157 & 164
& 135
\\
\textit{Number of Unique Products} 
& 5 & 5 & 5 & 5 & 5 & 5 & 5
\\
\textit{Number of Days} 
& 677 & 676
& 676 & 676
& 677 & 676
& 675
\\
\hline
\multicolumn{1}{l|}{\textbf{Sales \& Stockouts Per Store}}
&  & 
&  & 
&  &
&
\\
\textit{Revenue per week (\$)} 
& 1,687 & 5,197
& 4,511 & 3,192
& 1,801 & 821
& 302
\\
& (1,402) & (1,487)
& (1,415) & (835)
& (660) & (443)
& (200)
\\
\textit{Units sold per week} 
& 91 & 274
& 252 & 173
& 98 & 43
& 15
\\
& (75) & (51)
& (65) & (39)
& (33) & (22)
& (10)
\\
\textit{Stockouts events per week} 
& 0.37  & 0.14
& 0.13 & 0.21
& 0.28 & 0.49
& 0.56
\\
& (0.39) & (0.07)
& (0.11) & (0.15)
& (0.23) & (0.35)
& (0.60)
\\ \hline
\multicolumn{1}{l|}{\textbf{Inventories Per Store}}
&  & 
&  & 
&  &
&
\\
\textit{Number of products Offered} 
& 4.85 & 5
& 5 & 5
& 5 & 4.90
& 4.42
\\
& (0.51) & (0)
& (0) & (0)
& (0) & (0.34)
& (0.92)
\\
\textit{Delivery days per week} 
& 1.94  & 4.04
& 3.79 & 3.20
& 2.28 & 1.20
& 0.62
\\
& (1.15) & (0.29)
& (0.24) & (0.49)
& (0.56) & (0.51)
& (0.35)
\\
\textit{Delivery units per week} 
& 88 & 272
& 243 & 167
& 95 & 42
& 15
\\
& (73) & (55)
& (68) & (38)
& (33) & (21)
& (9)
\\ \hline
\multicolumn{1}{l|}{\textbf{Inventory Ratios Per Store}}
&  & 
&  & 
&  &
&
\\
\textit{Inventory to sales ratio} 
& 20.97 & 28.27 & 25.18 & 24.93 & 24.07 & 18.80 & 14.59
\\
& (7.03) & (8.06) & (7.18) & (6.44) & (6.91) & (4.84) & (3.46)
\\
\textit{Inventory to sales ratio after order} 
& 20.80 & 21.02 & 18.85 & 20.09 & 22.48 & 21.46 & 19.17
\\
& (4.94) & (10.69) & (4.11) & (4.46) & (6.18) & (4.07) & (3.81)
\\
\textit{Inventory to sales ratio before order} 
& 13.06 & 17.48 & 15.77 & 16.61 & 16.25 & 10.90 & 7.43
\\
& (5.33) & (7.10) & (3.42) & (3.83) & (4.72) & (3.44) & (2.95)
\\
\textit{Ordering frequency} 
& 0.15 & 0.31 & 0.30 & 0.25 & 0.17 & 0.09 & 0.05
\\
& (0.09) & (0.06) & (0.06) & (0.06) & (0.05) & (0.04) & (0.03)
\\
\hline \hline
\end{tabular}}
\end{table}

\medskip

\begin{figure}[ht]
\begin{center}
\caption{Empirical Distribution (Inverse CDF) of Inventory Outcomes Across Stores \label{fig:inv_het}}
\begin{subfigure}{0.4\textwidth}
    \caption{Stockout Frequency}
    \centering
    \includegraphics[width=5.7cm]{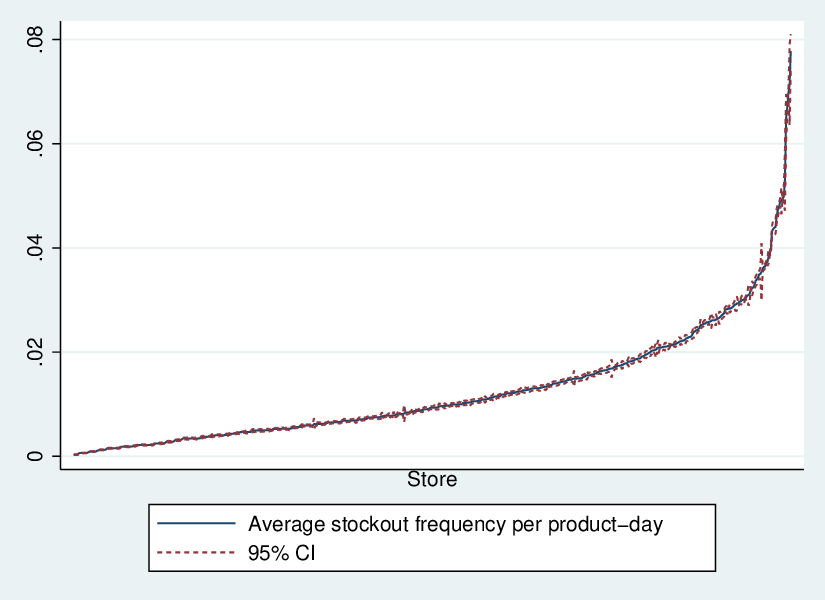}
\end{subfigure}
\vspace{5mm}
\begin{subfigure}{0.4\textwidth}
    \caption{Stockout Cost}
    \centering
    \includegraphics[width=5.7cm]{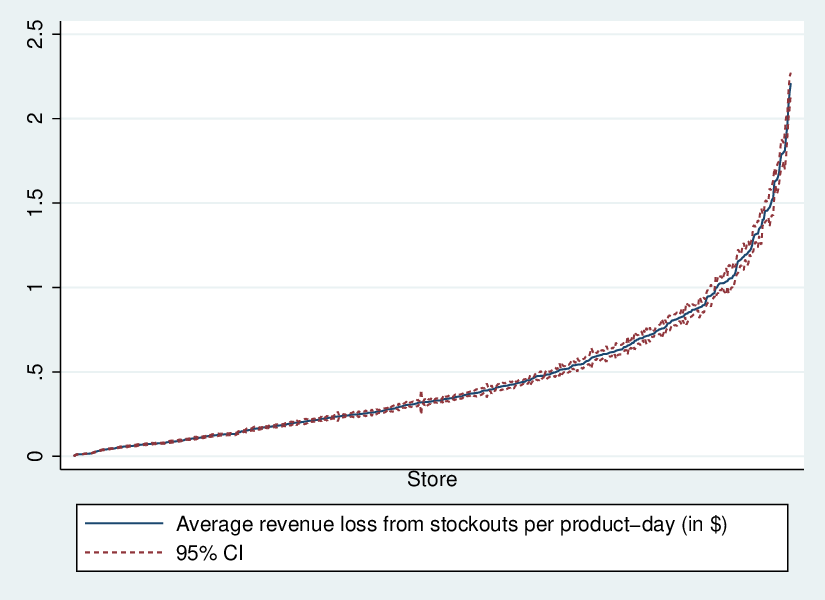}
    \end{subfigure}
\vspace{5mm}
\begin{subfigure}{0.4\textwidth}
    \caption{Ordering Frequency}
    \centering
    \includegraphics[width=5.7cm]{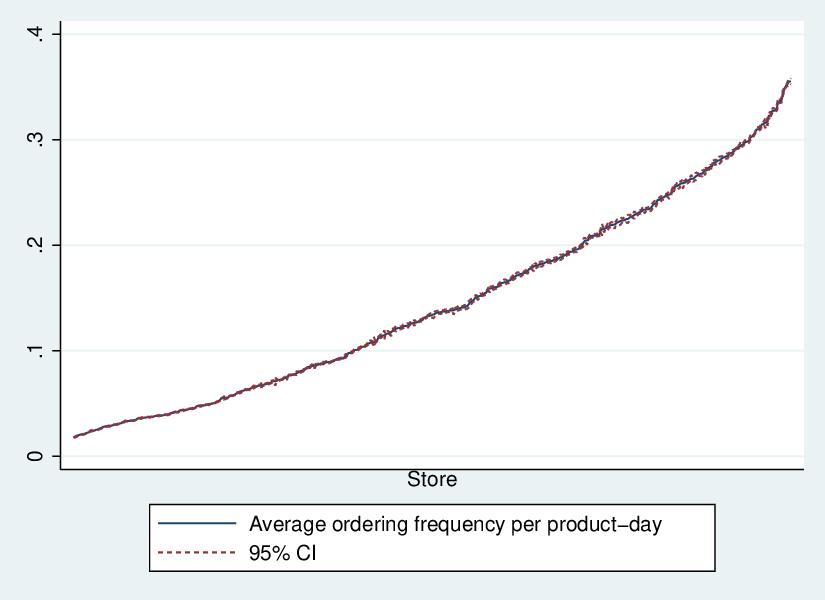}
\end{subfigure}
\vspace{5mm}
\begin{subfigure}{0.4\textwidth}
    \caption{Inventory to Sales Ratio}
    \centering
    \includegraphics[width=5.7cm]{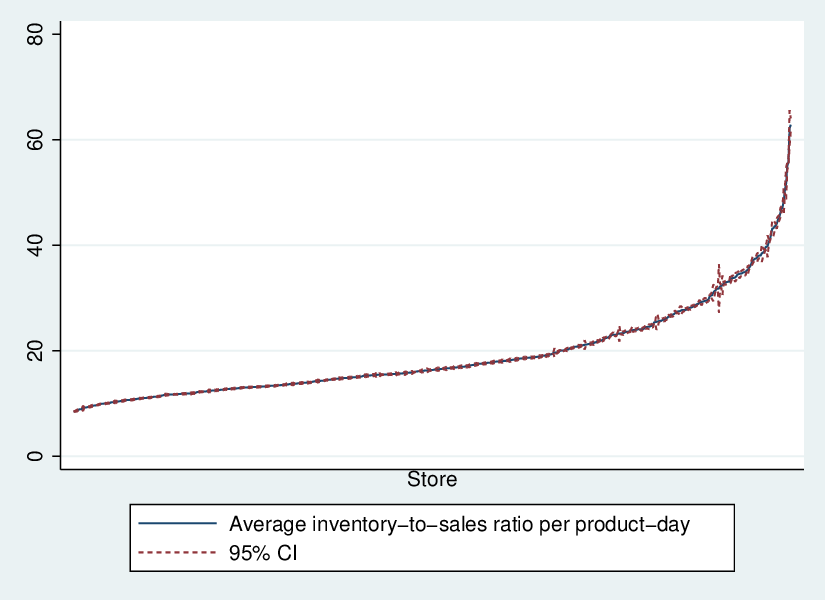}
    \end{subfigure}
\vspace{5mm}
\begin{subfigure}{0.4\textwidth}
    \captionsetup{justification=centering}
    \caption{Inventory to Sales Ratio, \\ After Order}
    \centering
    \includegraphics[width=5.7cm]{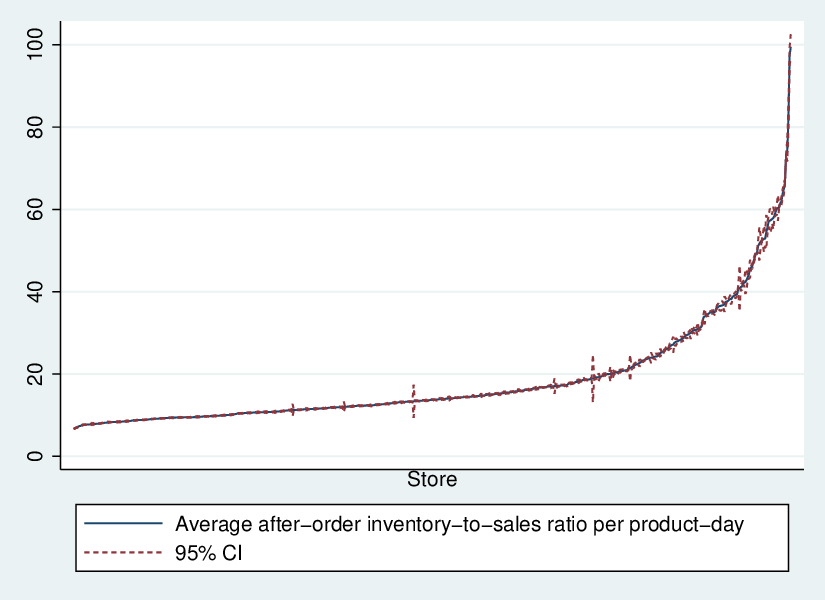}
 \end{subfigure}
\vspace{5mm}
\begin{subfigure}{0.4\textwidth}
    \captionsetup{justification=centering}
    \caption{Inventory to Sales Ratio, \\ Before Order}
    \centering
    \includegraphics[width=5.7cm]{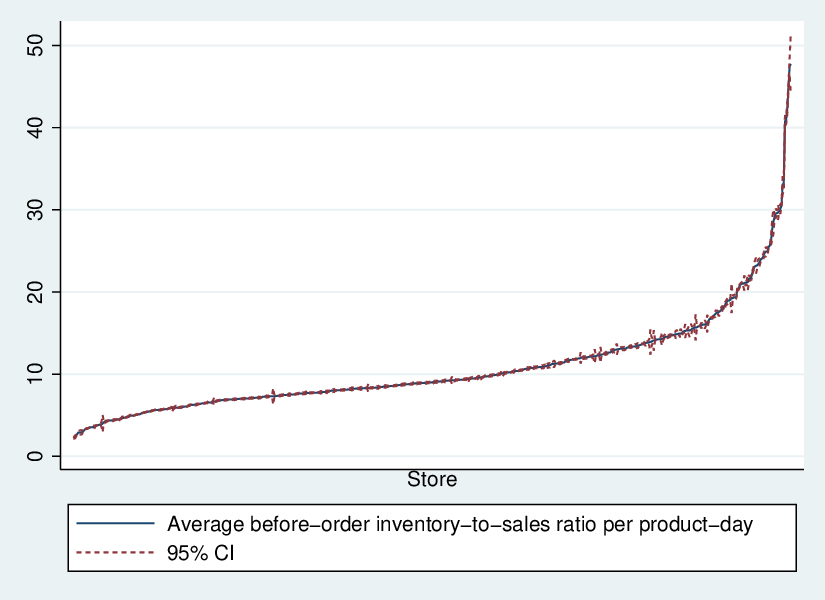}
    \end{subfigure}
\vspace{5mm}
\end{center}
\end{figure}

Figure \ref{fig:inv_het} shows strong heterogeneity across stores in several measures related to inventory management of the five products in our working sample. The figures in panels (a) to (f) are inverse cumulative distributions over stores, together with their $95\%$ confidence bands.\footnote{For every store, the $95\%$ confidence interval is based on the construction of store-product-specific rates. The $95\%$ confidence interval is determined by percentiles $2.5\%$ and $97.5\%$ in this distribution.} 

Panel (a) presents the distribution of the \textit{stockout rate}. For store $i$, we have:
\begin{equation}
    Stockout \text{ } rate_{i} = 
    \frac{\# \text{ } (product,day) \text{ } observations 
    \text{ } with \text{ } 
    stockout \text{ } in \text{ } store \text{ } i}
    {\# \text{ } (product,day) \text{ } observations \text{ } for \text{ } store \text{ } i}
\end{equation}
The figure shows a large spread of stockout rates: the $10^{\text{th}}$ and $90^{\text{th}}$ percentiles are $0.20\%$ and $2.82\%$, respectively. Panel (b) shows substantial heterogeneity across stores in the \textit{revenue-loss per-product-day} generated by stockouts. Indexing products by $j$, and using $\mathcal{J}=5$ to represent the number of products, the \textit{revenue-loss} for store $i$ is:
\begin{equation}
    Revenue \text{ } loss_{i} = 
    \frac{1}{J}
    \sum_{j=1}^{J}
    Stockout \text{ } rate_{i,j} \text{ } \times \text{ } 
    Average \text{ } daily \text{ } revenue \text{ } without \text{ } stockouts_{i,j}
\end{equation}
The $10^{\text{th}}$ and $90^{\text{th}}$ percentiles are $\$0.06$ and $\$1.03$ per product-day, respectively. Aggregated at the annual level and over all the products offered in a store, they imply an average annual revenue-loss of approximately $\$44,000$ at the $10^{\text{th}}$ percentile and $\$760,000$ at the $90^{\text{th}}$ percentile. Panel (c) presents the ordering frequency of stores in our sample calculated as:
\begin{equation}
    Ordering \text{ } frequency_{i} = 
    \frac{\# \text{ } (product,day) \text{ } observations 
    \text{ } with \text{ } an \text{ } order \text{ } in \text{ } store \text{ } i}
    {\# \text{ } (product,day) \text{ }
    observations \text{ } for \text{ } store \text{ } i}
\end{equation}
This ordering rate varies significantly across stores, with the $10^{\text{th}}$ percentile being $3.59\%$ and the $90^{\text{th}}$ being $28.41\%$. Panels (d) to (f) present the empirical distributions for the \textit{inventory-to-sales ratio}, for this ratio just before an order (a measure of the lower threshold \textit{s}), and for this ratio just after an order (a measure of the upper threshold \textit{S}). Indexing days by $t$:
\begin{equation}
    Inventory-to-(daily)sales-ratio_{i} = 
    \frac
    {\sum_{j=1}^{J} \sum_{t=1}^{T}
    Inventory_{i,j,t}}
    {\sum_{j=1}^{J} \sum_{t=1}^{T} Units 
    \text{ } sold_{i,j,t}}
\end{equation}
The distribution of the \textit{inventory-to-sales ratio} shows that stores at the $10^{\text{th}}$ and $90^{\text{th}}$ percentiles hold inventory for $11$ days and $33$ days of average sales, respectively. For the upper threshold \textit{S} (in Panel (e)), the values of these percentiles are $9$ and $37$ days, and for the lower threshold \textit{s} (in Panel (f)) they are $5$ and $19$ days.

Given these substantial differences in inventory outcomes across stores, it is interesting to explore how they vary together. We present these correlations in Appendix \ref{appendix_correlations} (Figure \ref{fig:inv_het_scatter}). The strongest correlation appears for the positive relationship between our measures of the thresholds \textit{S} and \textit{s}. This correlation can be explained by store heterogeneity in storage costs: a higher storage cost implies lower values of both $s$ and $S$. We confirm this conjecture in the estimation of the structural model in Section \ref{sec:structural}.

\clearpage

To investigate the possibility of stockouts at the warehouse level and their potential impact on store-level stockouts, we also analyze aggregate daily deliveries from the warehouse to all 634 stores. Given that the products in our working sample are popular items, we interpret a zero value in aggregate daily deliveries as a stockout event at the warehouse. The results of our analysis, presented in Section \ref{appendix:warehouse_stockouts} in the Appendix, reveal that warehouse stockouts are negligible for our working sample. For each product within the working sample, warehouse stockout events occur on no more than 3 out of the 677 days in the sample, which accounts for less than $0.5\%$ of the observed period. These findings suggest that stockouts observed at the store level are primarily driven by factors other than warehouse-level stockouts.

\subsection{Data on store managers' education and experience}

In addition to the main dataset, we enhance our analysis by incorporating information on store managers' human capital. Leveraging the professional social networking platform \textit{LinkedIn}, we gather data on the education and experience of store managers from their public profiles. Out of the 634 store managers in our dataset, 600 are identifiable by name in the LCBO's records.\footnote{During our sample period, some stores are overseen by interim managers who are not identified by name in our main dataset.} Within this subset, we were able to locate public \textit{LinkedIn} profiles for 143 managers, allowing us to retrieve valuable information about their educational background and work experience.

Table \ref{tab_educ_exper} presents summary statistics for the variables related to store managers' educational background and work experience. Notably, we observe a pattern in which managers with greater experience and higher educational attainment tend to be assigned to higher-classified stores. This finding suggests a positive correlation between manager characteristics and store classification, indicating that the LCBO may allocate more experienced and highly educated managers to stores of higher importance or larger scale. 
In Appendix \ref{sec_man_sto_correlations}, we provide more detailed information on the positive relationship between manager characteristics and the classification of stores within the LCBO retail chain.

\clearpage

\begin{table}[ht]
\begin{center}
\caption{Experience and Education of Store Managers by Store Type \label{tab_educ_exper}}
\resizebox{\textwidth}{!}{\begin{tabular}{r|c|cccccc}
\hline \hline
&  \multicolumn{7}{c}{\textit{Type of store}} \\
\multicolumn{1}{r|}{\textit{Statistic}} &
\multicolumn{1}{c|}{\textit{All}} &
\multicolumn{1}{c}{\textit{AAA}} &
\multicolumn{1}{c}{\textit{AA}} &
\multicolumn{1}{c}{\textit{A}} &
\multicolumn{1}{c}{\textit{B}} &
\multicolumn{1}{c}{\textit{C}} &
\multicolumn{1}{c}{\textit{D}} 
\\
\hline
&  &  &  &  &  &  & \\
\textit{\# of stores}
& 634 & 5 & 25 & 148 & 157 & 163 & 125 \\
\textit{(\%)} 
& (100.0) & (0.8) & (4.0) & (23.3) & (24.7) & (25.9) & (21.3) \\
\textit{\# managers with observed education}
& 86 & 1 & 8 & 30 & 32 & 5 & 10 \\
\textit{(\%)}
& (100.0) & (1.2) & (9.3) & (34.9) & (37.2) & (5.8) & (11.6) \\
\textit{\# managers with observed LCBO exper.}
& 136 & 3 & 9 & 48 & 44 & 20 & 12 \\
\textit{(\%)}
& (100.0) & (2.2) & (6.6) & (35.3) & (32.4) & (14.7) & (8.8) \\
\textit{\# managers with observed industry exper.}
& 72 & 1 & 7 & 30 & 26 & 5 & 3 \\
\textit{(\%)}
& (100.0) & (1.4) & (9.7) & (41.7) & (36.1) & (6.9) & (4.2) \\
&  &  &  &  &  &  & \\
\textit{Dummy highest degree = high school (Mean)}
& 0.093 & 0.000 & 0.000 & 0.067 & 0.094 & 0.200 & 0.200 \\ 
\textit{Dummy highest degree = college (Mean)}
& 0.395 & 0.000 & 0.625 & 0.333 & 0.375 & 0.600 & 0.400 \\ 
\textit{Dummy highest degree = university (Mean)}
& 0.512 & 1.000 & 0.375 & 0.600 & 0.531 & 0.200 & 0.400 \\ 
&  &  &  &  &  &  & \\
\textit{Years LCBO exper. (Mean)}
& 13.7 & 10.0 & 12.3 & 12.2 & 9.7 & 24.7 & 17.8 \\ 
\textit{Years industry (non-LCBO) exper. (Mean)}
& 9.8 & 15.0 & 8.0 & 11.2 & 9.8 & 5.8 & 5.3 \\ 
&  &  &  &  &  &  & \\
\hline \hline
\end{tabular}}
\end{center}
\end{table}

\medskip

\section{(\textit{S,s}) decision rules \label{sec:s_s_rules}}

\subsection{Model \label{sec:ss_model}}

In this section, we study store managers' inventory behavior through the eyes of $(S,s)$ decision rules. In its simplest form, this decision rule involves time-invariant threshold values. When assuming lump-sum (fixed) ordering costs, quasi-K-concavity of the profit function with respect to orders, and time-invariant expected demand, the profit-maximizing inventory decision rule follows a $(S,s)$ structure (\citeauthor{arrow_harris_1951}, \citeyear{arrow_harris_1951}; \citeauthor{scarf_1959}, \citeyear{scarf_1959}; \citeauthor{denardo_1981}, \citeyear{denardo_1981}). The $(S,s)$ rule is characterized by two threshold values: a lower threshold denoted as $s$, which represents the stock level that triggers a new order (known as the "safety stock level"), and an upper threshold denoted as $S$, which indicates the stock level to be achieved when an order is placed. Thus, if $k_t$ represents the stock level at the beginning of day $t$, and $y_t$ represents the orders placed on day $t$, the $(S,s)$ rule can be defined as follows:
\begin{equation}
    y_{t} = 
    \begin{cases} 
        \text{ } S - k_{t} 
        & if \text{ } k_{t} \leq s 
        \\ 
        \text{ } 0 
        & otherwise
    \end{cases}
    \label{eq:ss_model}
\end{equation}
\citeauthor{hadley_whitin_1963} (\citeyear{hadley_whitin_1963}) and \citeauthor{blinder_1981} (\citeyear{blinder_1981}) provide comparative statics for the thresholds $(S,s)$ as functions of the structural parameters in the firm's profit function. They provide the following results:
\begin{equation}
    \begin{array}[c]{lll}
    S = f_{S} 
    \left( 
        d^{e}, \text{ }
        \gamma^{h}, \text{ }
        \displaystyle{\frac{\gamma^{f}}{\gamma^{c}}}, \text{ }
        \displaystyle{\frac{\gamma^{z}}{\gamma^{c}}}
    \right),
    &
    s = f_{s} 
    \left( 
        d^{e}, \text{ }
        \gamma^{h}, \text{ }
        \displaystyle{\frac{\gamma^{f}}{\gamma^{c}}}, \text{ }
        \displaystyle{\frac{\gamma^{z}}{\gamma^{c}}}
    \right),
    &
    S - s = f_{S-s} 
    \left( 
        d^{e}, \text{ }
        \gamma^{h}, \text{ }
        \displaystyle{\frac{\gamma^{f}}{\gamma^{c}}}, \text{ }
        \displaystyle{\frac{\gamma^{z}}{\gamma^{c}}}
    \right)
    \\
    \qquad \qquad 
        + \text{ }
        - \text{ } \text{ } \text{ }
        ? \text{ } \text{ } \text{ } \text{ }
        +
    &
    \qquad \qquad 
        + \text{ }
        - \text{ } \text{ } 
        - \text{ } \text{ } \text{ }
        +
    &
    \qquad \qquad \qquad 
    \text{ } \text{ }
        + \text{ }
        - \text{ } \text{ } 
        + \text{ } \text{ } \text{ }
        ?
    \end{array}
\label{eq:compstatics_ss}
\end{equation}
where $d^{e}$ represents expected demand, $\gamma^{h}$ is the inventory holding cost per period and per unit, 
$\gamma^{f}$ is the fixed (lump-sum) ordering cost, $\gamma^{c}$ is the unit ordering cost, and $\gamma^{z}$ is the stockout cost per period. These $\gamma$'s are the parameters in the structural model that we estimate in Section \ref{sec:structural}. We investigate the predictions of equation (\ref{eq:compstatics_ss}) in Section \ref{sec:counterfactuals}. 

The optimality of the $(S,s)$ decision rule extends to models with state variables that evolve over time according to exogenous Markov processes. Let $\mathbf{z}_{t}$ denote the vector of these exogenous state variables, which can include factors influencing demand, unit ordering costs, wholesale prices, and the product's retail price (when taken as given by the store manager), as is the case in our problem. The optimal decision rule in this context follows a $(S_{t},s_{t})$ structure, where the thresholds $S_{t}$ and $s_{t}$ are time-invariant functions of these state variables: $S_{t} = S(\mathbf{z}_{t})$ and $s_{t} = s(\mathbf{z}_{t})$.

In this section, our empirical approach is inspired by the work of \citeauthor{eberly_1994} (\citeyear{eberly_1994}), \citeauthor{attanasio_2000} (\citeyear{attanasio_2000}), and \citeauthor{adda_cooper_2000} (\citeyear{adda_cooper_2000}). These studies utilize household-level data on durable product purchases, specifically automobiles, to estimate $(S_{t},s_{t})$ decision rules. In these decision rules, the thresholds are functions of household characteristics, prices, and aggregate economic conditions. This approach can be seen as a "semi-structural approach," where the use of $(S_{t},s_{t})$ rules is motivated by a dynamic programming model of optimal behavior. However, the specification of the thresholds as functions of state variables does not explicitly incorporate the structural parameters of the model.

In Section \ref{sec:structural}, we present a full structural approach that explicitly incorporates the structural parameters of the model. Furthermore, in Section \ref{sec:counterfactuals}, we utilize the estimated structural model to conduct counterfactual policy experiments, which address the questions that motivated this paper. However, before delving into the full structural analysis, we find it valuable to explore the data using a more flexible empirical framework that remains consistent with the underlying structural model. We investigate heterogeneity between store managers' inventory decisions by estimating $(S_{t},s_{t})$ rules at the store-product level. These decision rules are consistent with our structural model but they are more flexible. This allows us to gain insights and assess the suitability of the $(S_{t},s_{t})$ decision rules in capturing the inventory behavior of store managers within the LCBO retail chain.

Given that our dataset contains 677 daily observations for every store and product, and that the ordering frequency in the data is high enough to include many orders per store-product, we can estimate the parameters in the $(S_{t},s_{t})$ decision rules at the store-product level. In this section, we omit store and product sub-indexes in variables and parameters, but it should be understood that these sub-indexes are implicitly present.

We consider the following specification for the $(S_{t},s_{t})$ thresholds:\footnote{We attempted to incorporate demand volatility, represented by $\ln \sigma^{2}_{t}$, as an explanatory variable in the decision rule. However, we encountered high collinearity between the time series of $\ln d^{e}_{t}$ (expected demand) and $\ln \sigma^{2}_{t}$, making it challenging to estimate their separate effects on the thresholds. It is worth noting that according to the Negative Binomial distribution, $\ln \sigma^{2}_{t} = \ln d^{e}_{t} + \ln(1+\alpha d^{e}_{t})$. Consequently, we can interpret the effect of $\ln d^{e}_{t}$ on the $(S_{t},s_{t})$ thresholds as the combined impact of both expected demand and volatility.}
\begin{equation}
    \begin{cases} 
        \text{ } 
        S_{t} 
        \text{ } = \text{ }
        \exp\{  
        \beta_{0}^{S} 
        \text{ } + \text{ }
        \beta_{d}^{S} \text{ } 
        \ln d_{t}^{e} 
        \text{ } + \text{ }
        \beta_{p}^{S} \text{ } 
        \ln p_{t}
        \text{ } + \text{ }
        u_{t}^{S}
        \}
        \\ 
        \\
        \text{ } 
        s_{t} 
        \text{ } = \text{ }
        \exp\{  
        \beta_{0}^{s} 
        \text{ } + \text{ }
        \beta_{d}^{s} \text{ } 
        \ln d_{t}^{e} 
        \text{ } + \text{ }
        \beta_{p}^{s} \text{ } 
        \ln p_{t}         
        \text{ } + \text{ }
        u_{t}^{s}
        \}
    \end{cases}
    \label{eq:specification_ss}
\end{equation}
where $p_{t}$ is the product's retail price, $d_{t}^{e}$ is the expected demand, and $u_{t}^{s}$ and $u_{t}^{S}$ represent state variables which are known to the store manager but are unobservable to us as researchers.\footnote{For instance, $u_{t}^{s}$ and $u_{t}^{S}$ may include shocks in fixed and variable ordering costs, or measurement error in our estimate of expected demand.} The vector of exogenous state variables is $\mathbf{z}_{t} = (d_{t}^{e}, p_{t}, u_{t}^{s}, u_{t}^{S})$. The $\beta$'s are reduced form parameters which are constant over time but vary freely across stores and products and are functions of the structural parameters that we present in our structural model in Section \ref{sec:structural}.

Our measure of expected demand is based on an LCBO report regarding the information that headquarters use to construct order recommendations for each store (\citeauthor{lcbo_2016}, \citeyear{lcbo_2016}). Relying on this report, we assume that store managers obtain predictions of demand for each product at their store using information on the product's retail price ($p_{t}$), the average daily sales of the product over the last seven days, (that we represent as $Q^{[-7,-1]}_{t}$), and seasonal dummies. 

Since the observed quantity sold $q_{t}$ has discrete support $\{0, 1, 2, ...\}$, we consider that demand has a Negative Binomial distribution where the logarithm of expected demand at period $t$ has the following form:
\begin{equation}
    \ln d^{e}_{t} = 
    \ln \mathbb{E} 
    \left( 
        q_{t} \text{ } \vert \text{ } 
        p_{t}, Q^{[-7,-1]}_{t} 
    \right)
     = 
    \boldsymbol{\alpha}' \text{ } 
    h\left( 
        \ln p_{t}, \ln Q^{[-7,-1]}_{t} 
    \right)
\label{eq:poisson_expected_sales}
\end{equation}
where $q_{t}$ is the quantity sold of the store-product at day $t$, $\boldsymbol{\alpha}$ is a vector of parameters that are constant over time but vary freely across store-products, and $h\left( \ln p_{t}, \ln Q^{[-7,-1]}_{t} \right)$ is a vector of monomial basis in variables $\ln p_{t}$ and $\ln Q^{[-7,-1]}_{t}$. 

We denote equation (\ref{eq:poisson_expected_sales}) as the \textit{sales forecasting function}. It deserves some explanation. First, it is important to note that this is not a demand function. For this inventory decision problem, managers do not need to know the demand function but only the best possible predictor of future sales given the information they have. Second, this specification ignores substitution effects between products within the same category or across categories. Ignoring substitution effects in demand is fully consistent with LCBO's report and with the firm's price setting, which completely ignores these substitution effects (see \citeauthor{aguirregabiria_ershov_2016}, \citeyear{aguirregabiria_ershov_2016}).\footnote{Recent papers show that the pricing decisions of important multi-product firms do not internalize substitution or cannibalization effects between the firm's own products. See, for instance, \citeauthor{hortacsu_natan_2021} (\citeyear{hortacsu_natan_2021})'s study of the pricing system of a large international airline company,  \citeauthor{dellavigna_gentzkow_2019} (\citeyear{dellavigna_gentzkow_2019}) and \citeauthor{ hitsch_hortacsu_2021} (\citeyear{hitsch_hortacsu_2021}) on uniform pricing at US retail chains, \citeauthor{cho_rust_2010} (\citeyear{cho_rust_2010}) on pricing of car rentals, or \citeauthor{miravete_seim_2020} (\citeyear{miravete_seim_2020}) for liquor stores in Pennsylvania.} In the Appendix (Section \ref{sec:appendix_sales_equation}), we present a summary of the estimation results of the \textit{sales forecasting function} for every store and every product in our working sample.

The $(S_{t},s_{t})$ model in equations (\ref{eq:ss_model}) and (\ref{eq:specification_ss}) implies that the decision of placing an order ($y_{t} > 0$) or not ($y_{t} = 0$) has the structure of a linear-in-parameters binary choice model.
\begin{equation}
    \mathbbm{1}
    \{ y_{t} > 0 \}
    \text{ } = \text{ }
    \mathbbm{1}
    \{
    b_{0}^{s} + 
    b_{k}^{s} \text{ } \ln k_{t} +
    b_{d}^{s} \text{ } \ln d_{t}^{e} +
    b_{p}^{s} \text{ } \ln p_{t} +
    \widetilde{u}_{t}^{s}
    \geq 0
    \},
\label{eq:bc_lowe_s}
\end{equation}
where $\mathbbm{1}\{.\}$ is the indicator function; $\widetilde{u}_{t}^{s} \equiv u_{t}^{s}/\sigma_{u^{s}}$ is the standardized version of $u_{t}^{s}$, as $\sigma_{u^{s}}$ is the standard deviation of $u_{t}^{s}$; and there is the following relationship between $\beta^{s}$ and $b^{s}$ parameters: $b_{k}^{s} = -1/\sigma_{u^{s}}$; $b_{0}^{s} = \beta_{0}^{s}/\sigma_{u^{s}}$; $b_{d}^{s} = \beta_{d}^{s}/\sigma_{u^{s}}$; and $b_{p}^{s} = \beta_{p}^{s}/\sigma_{u^{s}}$. These expressions show that, given the parameters $b^{s}$, we can identify the parameters $\beta^{s}$ and $\sigma_{u^{s}}$. We assume that $\widetilde{u}_{t}^{s}$ has a Normal distribution, such that equation (\ref{eq:bc_lowe_s}) is a Probit model, and we estimate the parameters $b^{s}$ by maximum likelihood. 

Our $(S_{t},s_{t})$ model also implies that in days with positive orders ($y_{t} > 0$)
the logarithm of the total quantity offered, $\ln(k_{t} + y_{t})$, is equal to the logarithm of the upper-threshold, $\ln(S_{t})$, and this implies the following linear-in-parameters (censored) regression model:
\begin{equation}
    \ln(k_{t} + y_{t})
    \text{ } = \text{ }
    \beta_{0}^{S} 
    \text{ } + \text{ }
    \beta_{d}^{S} \text{ } 
    \ln d_{t}^{e}
    \text{ } + \text{ }
    \beta_{p}^{S} \text{ } 
    \ln p_{t}     
    \text{ } + \text{ }
    u_{t}^{S}
    \text{ } \text{ } \text{ }
    \text{if} \text{ }
    y_{t} > 0.
\label{eq:lrm_upper_s}
\end{equation}
Equation (\ref{eq:lrm_upper_s}) includes the selection condition $y_{t} > 0$. That is, the upper-threshold $S_{t}$ is observed only when an order is placed. This selection issue implies that OLS estimation of equation (\ref{eq:lrm_upper_s}) yields inconsistent estimates of the parameters and the threshold itself. However, the $(S_{t},s_{t})$ model implies an exclusion restriction that provides identification of the parameters in equation (\ref{eq:lrm_upper_s}). The inventory level $k_{t}$ affects the binary decision of placing an order or not (as shown in equation (\ref{eq:bc_lowe_s})), but conditional on placing an order, it does not affect the value of the upper-threshold in the right-hand-side of regression equation (\ref{eq:lrm_upper_s}). Therefore, using (\ref{eq:bc_lowe_s}) as the selection equation, we can identify the parameters $\beta^{S}$ in (\ref{eq:lrm_upper_s}) using a Heckman two-step approach.\footnote{This exclusion restriction in $(S,s)$ models have been pointed out by \citeauthor{bertola_guiso_2005} (\citeyear{bertola_guiso_2005}).}

Part of the variation in parameter estimates across stores and products is attributable to estimation error rather than genuine heterogeneity. For any given parameter $b_{i,j}$, where $i$ and $j$ represent store and product indices respectively, let $\widehat{b}_{i,j}$ denote its consistent and asymptotically normal estimate, with an asymptotic variance of $\sigma^{2}_{i,j}$. Using this asymptotic distribution, we can establish a relationship between the variances of $\widehat{b}_{i,j}$ and $b_{i,j}$ across stores and products: $Var(\widehat{b}_{i,j}) = Var(b_{i,j}) + \mathbb{E}(\sigma^{2}_{i,j})$, where $\mathbb{E}(\sigma^{2}_{i,j})$ represents the mean of asymptotic variances $\sigma^{2}_{i,j}$ across stores and products. Since $\mathbb{E}(\sigma^{2}_{i,j})>0$, this equation demonstrates that $Var(\widehat{b}_{i,j})$ overestimates the true dispersion $Var(b_{i,j})$. To mitigate this excess dispersion or spurious heterogeneity arising from estimation error, we employ a \textit{shrinkage estimator}. The details of this estimator are described in section \ref{appendix_shrinkage_estimator} in the Appendix.

\subsection{Estimation of (\textit{S,s}) thresholds \label{sec:ss_estimation}}

\begin{figure}
\begin{center}
\caption{Empirical Distribution (Inverse CDF) of Estimates $\boldsymbol{b^{s}}$ for Lower $s$ Threshold \label{fig:lowerparam_cdf}}
\begin{subfigure}{0.4\textwidth}
    \caption{$b^{s}_{0}$}
    \centering
    \includegraphics[width=5.7cm]{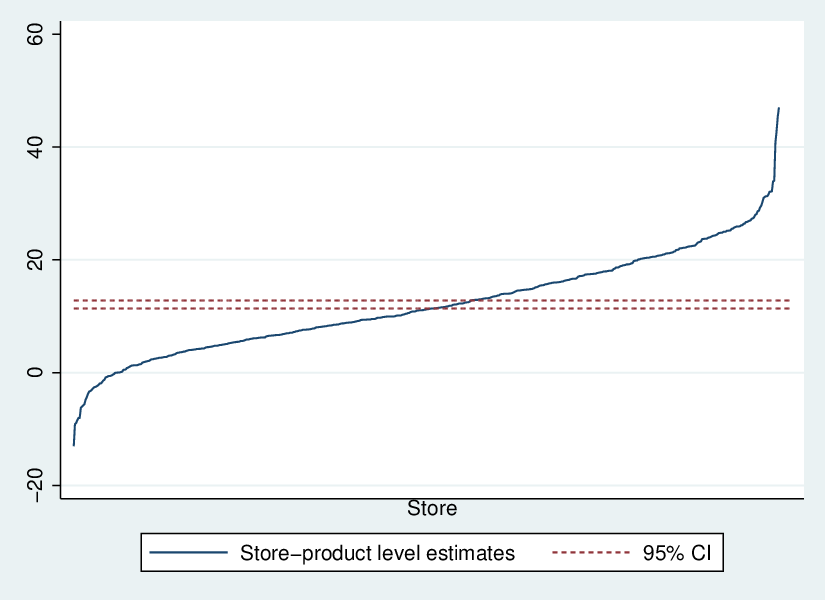}
    \vspace*{1mm}
\end{subfigure}
\begin{subfigure}{0.4\textwidth}
    \caption{$b^{s}_{k}$}
    \centering
    \includegraphics[width=5.7cm]{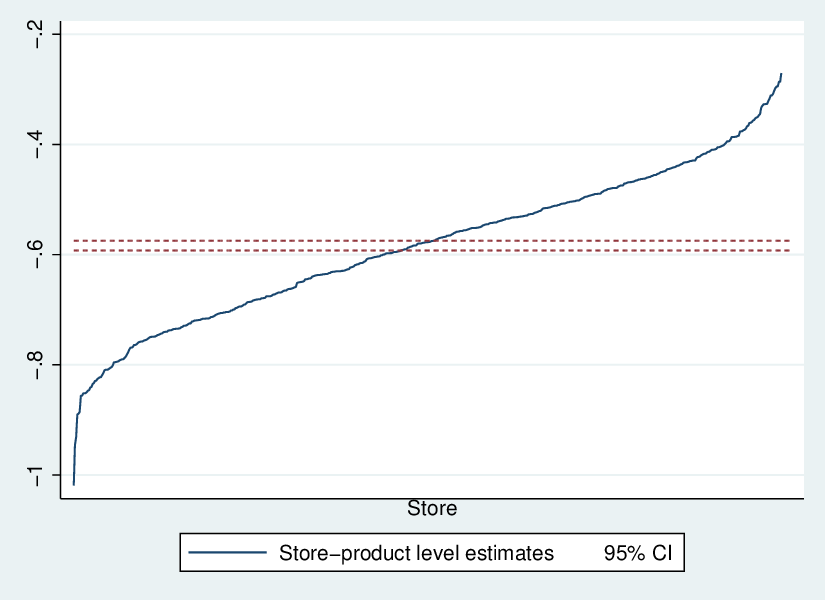}
    \vspace*{1mm}
\end{subfigure}
\begin{subfigure}{0.4\textwidth}
    \caption{$b^{s}_{d}$}
    \centering
    \includegraphics[width=5.7cm]{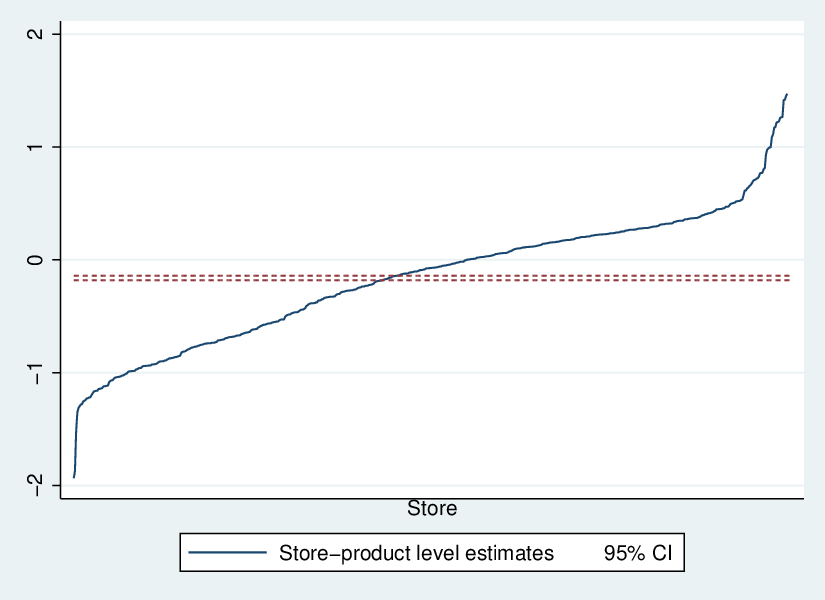}
    \vspace*{1mm}
\end{subfigure}
\begin{subfigure}{0.4\textwidth}
    \caption{$b^{s}_{p}$}
    \centering
    \includegraphics[width=5.7cm]{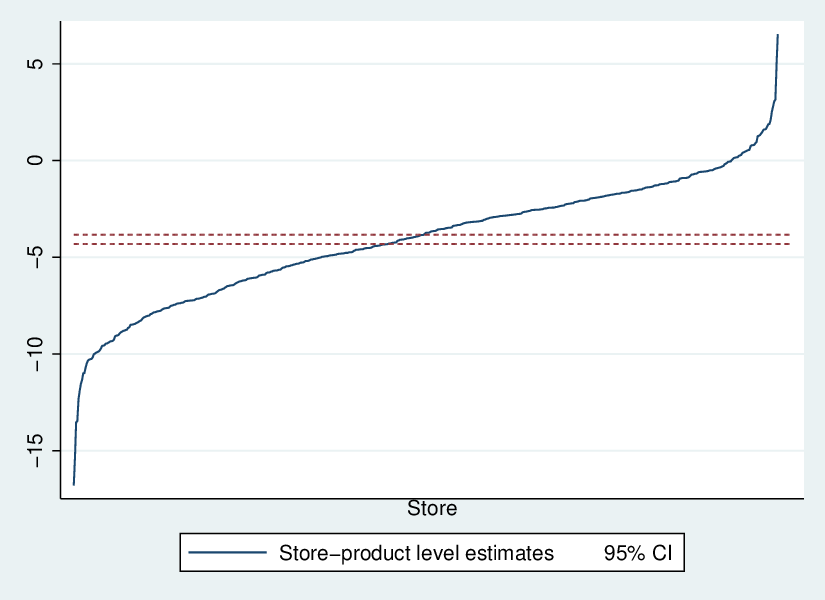}
\end{subfigure}
\end{center}
\end{figure}

Figure \ref{fig:lowerparam_cdf} presents the average estimates of parameters $b^{s}_{0}$, $b^{s}_{k}$, $b^{s}_{d}$, and $b^{s}_{p}$ in the lower threshold for each store, where the average is obtained over the five products in the working sample. We sort stores from the lowest to the largest average estimate such that this curve is the inverse CDF of the average estimate. The red-dashed band around the median of this distribution is the $95\%$ confidence band under the null hypothesis of homogeneity across stores.\footnote{The reported $95\%$ confidence interval incorporates the Bonferroni correction for multiple testing. In this context, the implicit null hypothesis is that every store does not differ significantly from the average store. By applying the Bonferroni correction, we account for the increased probability of observing a significant difference by chance when conducting multiple tests.} The signs of the parameter estimates are for the most part consistent with the predictions of the model. These distributions show that the parameter estimates vary significantly across stores. For $b^{s}_{0}, b^{s}_{k},$, $b^{s}_{d}$, and $b^{s}_{p}$, we have that 95\%, 95\%, 98\%, and 95\% of stores, respectively, lie outside of the Bonferroni confidence interval.

\begin{figure}[ht]
\begin{center}
\caption{Empirical Distribution (Inverse CDF) of Estimates $\boldsymbol{\beta^{S}}$ for Upper $S$ Threshold \label{fig:upperparam_cdf}}
\begin{subfigure}{0.4\textwidth}
    \caption{$\beta^{S}_{0}$}
    \includegraphics[width=5.7cm]{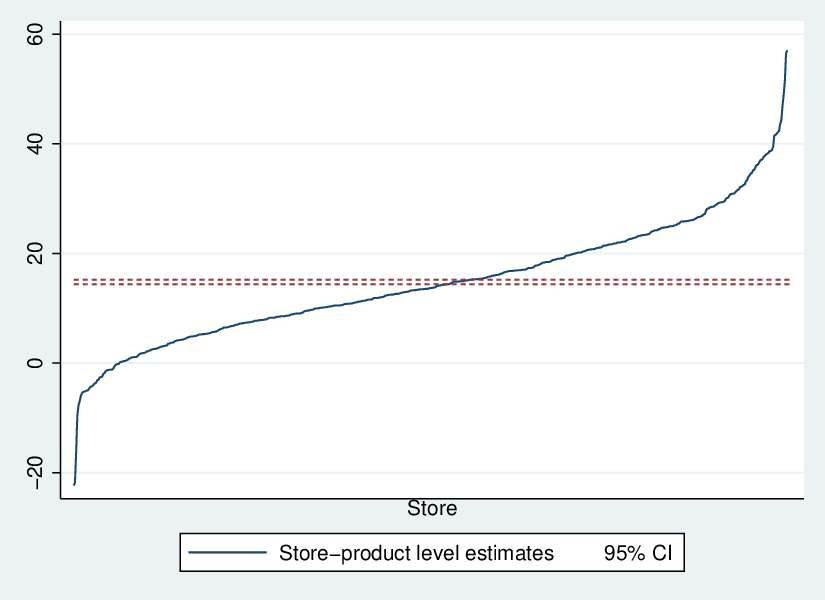}
\end{subfigure}
\begin{subfigure}{0.4\textwidth}
    \caption{$\beta^{S}_{d}$}
    \includegraphics[width=5.7cm]{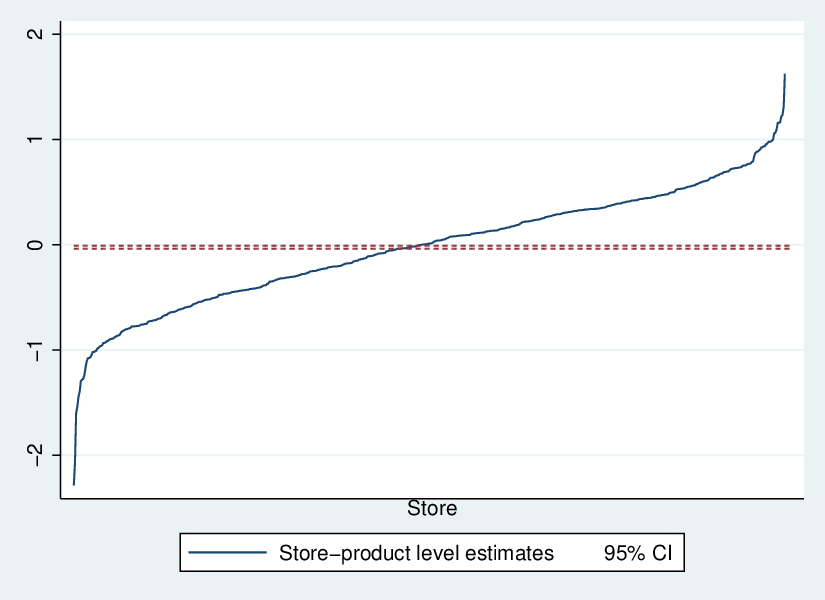}
    \vspace*{2mm}
\end{subfigure}
\begin{subfigure}{0.4\textwidth}
    \caption{$\beta^{S}_{p}$}
    \includegraphics[width=5.7cm]{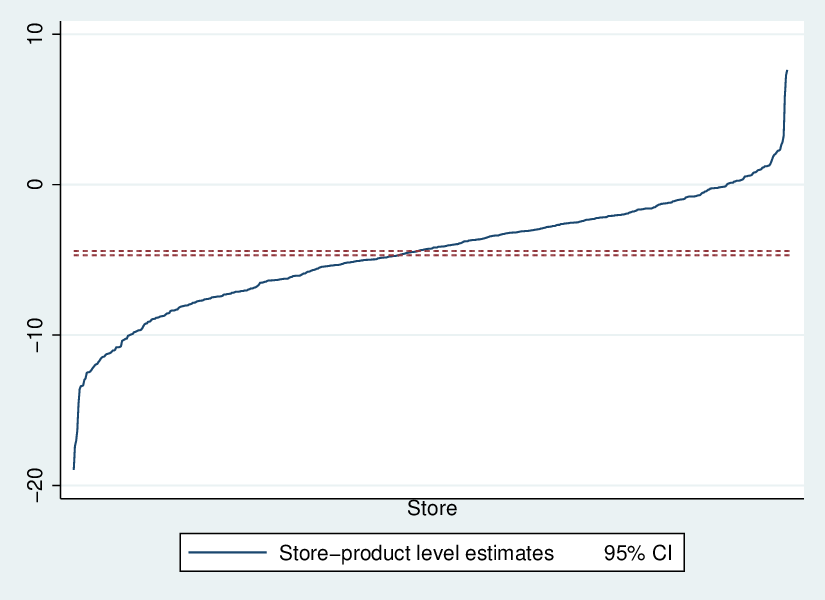}
\end{subfigure}
\end{center}
\end{figure}

Figure \ref{fig:upperparam_cdf} presents the inverse CDF of the store-specific average of the parameters $\beta^{S}_{0}$, $\beta^{S}_{d}$, and $\beta^{S}_{p}$ in the upper threshold, as well as the Bonferroni $95\%$ confidence interval under the null hypothesis of homogeneity. As expected, we have strong evidence of heterogeneity in our estimates. For $\beta^{S}_{0}$, $\beta^{S}_{d}$, and $\beta^{S}_{p}$, approximately 96\%, 97\%, and 97\% of stores lie outside of the confidence bands, respectively.  

\clearpage

\clearpage

Given that we have estimates at the store-product level, we can also explore the heterogeneity in these estimates within stores. Table \ref{tab: param_decomp} presents a decomposition of the variance of parameter estimates into within-store (between products) and between-store variance. The parameters associated with expected demand and the lower threshold inventory parameter show a between-store variance that is at least as large as the within-store variance. For the constant parameters and the price parameters, the variance is larger across products.

\medskip

\begin{table}[ht]
\begin{center}
\caption{Variance Decomposition of Parameter Estimates \label{tab: param_decomp}}

\begin{tabular}{c c c c c c c c}
\hline \hline 
& & 
\\
\multicolumn{1}{c}{\textbf{Variance }} 
& \multicolumn{1}{c}{$\boldsymbol{b^{s}_{0}}$} 
& \multicolumn{1}{c}{$\boldsymbol{b^{s}_{k}}$}
& \multicolumn{1}{c}{$\boldsymbol{b^{s}_{d}}$}
& \multicolumn{1}{c}{$\boldsymbol{b^{s}_{p}}$}
& \multicolumn{1}{c}{$\boldsymbol{\beta^{S}_{0}}$}
& \multicolumn{1}{c}{$\boldsymbol{\beta^{S}_{d}}$}
& \multicolumn{1}{c}{$\boldsymbol{\beta^{S}_{p}}$}
\\
& & 
\\ \hline
& & 
\\
Between-store Variance & 69.41 & 0.02 & 0.28 & 8.54 & 110.18 & 0.27 & 12.01 \\
& & 
\\
Within-store Variance & 81.01 & 0.02 & 0.07 & 9.78 & 183.81 & 0.11 & 19.35 \\
& & 
\\
\hline \hline
\end{tabular}
\end{center}
\end{table}

\medskip

The parameter estimates imply values for the $(S_{t},s_{t})$ thresholds. 
In section \ref{appendix:ss_estimation} in the Appendix, we investigate heterogeneity across stores in the estimated thresholds. We find strong between-store heterogeneity, especially in the lower threshold $s$.

\section{Structural model \label{sec:structural}}

We propose and estimate a dynamic structural model of inventory management. A price-taking store sells a product and faces uncertain demand. The store manager orders the product from the retail chain's warehouse, and any unsold product rolls over to the next period's inventory. The store-level profit function incorporates four store-specific costs associated with inventory management: per-unit inventory holding cost ($\gamma^{h}_{i,j}$), stockout cost ($\gamma^{z}_{i,j}$), fixed ordering cost ($\gamma^{f}_{i,j}$), and per-unit ordering cost ($\gamma^{c}_{i,j}$). 

\subsection{Non-separability of inventory decisions across products}

At LCBO, store managers are responsible for making inventory decisions for thousands of products. These decisions are not independent of each other due to various factors. Firstly, when a product experiences a stockout, consumers may opt to substitute it with a similar available product. As a result, the cost of a stockout for the store is influenced by the availability of substitute products. Secondly, storage costs are affected by the total volume of items and units in the store, which means the inventory management of one product can impact the storage costs of other products. Lastly, the store's ordering cost is dependent on the cost of filling a truck with units of multiple products and transporting them from the warehouse to the store. The decision to order a truck and the cost associated with it are not separable across products.

We can think of a model that fully accounts for the inter-dependence inventory decisions across products. In this model, a store manager maximizes the aggregate profit from all the products taking into account the substitutability of similar products under stockouts and subject to two overall store-level constraints: the total storage capacity of the store and the capacity of the delivery truck. Our store-product level model aligns with this multiple-product inventory management framework.

By using duality theory, we can show that the marginal conditions for optimality in the multiple-product model are equivalent to those in our single-product model, under appropriate interpretations of our store-product level structural parameters. The inventory holding cost parameter $\gamma^{h}_{i,j}$ represents the shadow price or Lagrange multiplier associated with the storage capacity constraint at store $i$. Similarly, the fixed and unit ordering costs, $\gamma^{f}_{i,j}$ and $\gamma^{c}_{i,j}$ respectively, reflect the shadow prices of the truck's capacity constraint at the extensive and intensive margins. Lastly, the stockout cost parameter $\gamma^{z}_{i,j}$ accounts for the impact of consumer substitution within the store when a stockout occurs for product $j$ at store $i$.

In estimating our model, our approach is valid as long as these Lagrange multipliers do not exhibit significant variation over time. For our counterfactual experiments, we assume that these Lagrange multipliers remain constant in the counterfactual scenario.

To summarize, our approach does not assume the separability of inventory management across products, as this would be an unrealistic assumption. Instead, we employ certain assumptions and shortcuts to address the complexities of the joint inventory management problem, while still maintaining a realistic framework that is consistent with the interdependencies among products.

\subsection{Sequence of events and profit}

For notational simplicity, we omit store and product indexes. Time is indexed by $t$. One period is one day. Every day, the sequence of events is the following.

\medskip

\noindent \textbf{Step (i)}. The day begins with the store manager observing current stock ($k_{t}$), the retail price set by headquarters ($p_{t}$), and her expectation about the mean and variance of the distribution of log-demand: $\ln d^{e}_{t}$ and $\sigma^{2}_{t}$, respectively. Given this information, the store manager orders $y_{t}$ units of inventory from the distribution center. There is \textit{time-to-build} in this ordering decision. More specifically, it takes one day for an order to be delivered to the store and become available to consumers.\footnote{Based on the interviews we conducted with store managers, the most common delivery lag reported is one day. Delivery lags exceeding three days were described as extremely rare.} The ordered amount $y_{t}$ is a discrete variable with support set $\mathcal{Y} \equiv \{0, 1, ..., J\}$.
    
\medskip

\noindent \textbf{Step (ii)}. Demand $d_{t}$ is realized. Demand has a Negative Binomial distribution with log-expected demand and variance:
\begin{equation}
    \begin{cases} 
        \text{ }
        \ln d^{e}_{t} \text{ } = \text{ }
        \boldsymbol{\eta}_{0}^{\prime} \text{ } \mathbf{seas}_{t} + 
        \eta_{p} \text{ }
        \ln p_{t} +
        \eta_{Q} \text{ } 
        \ln Q^{[-7,-1]}_{t}
        \\
        \text{ } \text{ } \text{ } \text{ }
        \sigma^{2}_{t} \text{ } = \text{ }
        d^{e}_{t} \text{ } 
        \left(
            1 + \alpha \text{ } d^{e}_{t} 
        \right) 
    \end{cases} 
\label{eq: demand_eq}
\end{equation}
where $\boldsymbol{\eta}_{0}$, $\eta_{p}$, $\eta_{Q}$, and $\alpha$ are parameters; $\mathbf{seas}_{t}$
is a vector of seasonal dummies (i.e., weekend dummy and main holidays dummy); $Q^{[-7,-1]}_{t}$ is the average daily sales of the product in the store during the last seven days; and $\alpha$ denotes the over-dispersion parameter in the Negative Binomial. We use $F_{d_{t}}$ to represent the distribution of $d_{t}$ conditional on $(p_{t},Q^{[-7,-1]}_{t},\mathbf{seas}_{t})$. Importantly, the stochastic demand shock $u_{t}^{d} \equiv \ln d_{t} - \ln d^{e}_{t}$ is unknown to the store manager at the beginning of the day when she makes her ordering decision.
    
\medskip

\noindent \textbf{Step (iii)}. The store sells $q_{t}$ units of inventory, which is the minimum of supply and demand:
\begin{equation}
        q_{t} 
        \text{ } = \text{ }
        \min
        \{
        \text{ } d_{t}
        \text{ } , 
        \text{ } k_{t}
        \text{ }
        \}
\label{eq:min_dem_sup}
\end{equation}
The store generates flow profits $\Pi_{t}$. The profit function has the following form:
\begin{equation}
        \Pi_{t} 
        \text{ } = \text{ }
        \left(p_{t} - c_{t}\right) \text{ }
        \min
        \{ d_{t}, k_{t} \} 
        + \gamma^{z} \mathbbm{1}
        \{d_{t} > k_{t}\}
        - \gamma^{h} \text{ } k_{t} 
        - \gamma^{c} \text{ } y_{t} 
        - \gamma^{f} \text{ } 
        \mathbbm{1} \{y_{t}>0 \} 
        + \sigma_{\varepsilon} \text{ }
        \varepsilon_{t}(y_{t}) 
\end{equation}
where $c_{t}$ is the wholesale price, and $\gamma^{z}$, $\gamma^{h}$, $\gamma^{c}$, $\gamma^{f}$, and $\sigma_{\varepsilon}$ are store-product-specific structural parameters. 
When $\gamma^{z}>0$, the term $\gamma^{z}\cdot \mathbbm{1}\{d_{t} > k_{t}\}$ captures the situation where the cost of a stockout can be smaller than the revenue loss from excess demand because some consumers substitute the product within the store. On the other hand, if $\gamma^{z}<0$, this term can represent an additional reputational cost of stockouts that goes beyond the lost revenue (see 
\citeauthor{anderson_fitzsimons_2006}, \citeyear{anderson_fitzsimons_2006}). The term $\gamma^{h}\cdot k_{t}$ represents the storage cost associated with holding $k_{t}$ units of inventory at the store. Parameter $\gamma^{c}$ denotes the per-unit cost incurred by the store manager when placing an order, and $\gamma^{f}$ represents the fixed ordering cost, including the transportation cost from the warehouse to the store. The variable $\varepsilon_{t}(y_{t})$ corresponds to a stochastic shock with a mean of zero that affects ordering costs. More specifically, variables $\varepsilon_{t}(0)$, $\varepsilon_{t}(1)$, ..., $\varepsilon_{t}(J)$ are i.i.d. with a Extreme Value type 1 distribution. Parameter $\sigma_{\varepsilon}$ represents the standard deviation of the shocks in ordering costs.

These $\gamma$ parameters are the store manager's perceived costs. For instance, the fixed ordering cost $\gamma^{f}$ and the per-unit holding cost $\gamma^{h}$ can be interpreted as the manager's perception of the shadow prices (or Lagrange multipliers) associated to the capacity constraints of a delivery truck and of the store, respectively.

\medskip

\noindent \textbf{Price-cost margins}. LCBO's retail prices are a constant markup over their respective wholesale prices. There are different markups for Ontario products ($65.5\%$ markup) and non-Ontario products ($71.5\%$ markup) (see \citeauthor{aguirregabiria_ershov_2016}, \citeyear{aguirregabiria_ershov_2016}). A constant markup, say $\tau$, implies that the price-cost margin is proportional to the retail price: $p_{t} - c_{t} = \mathcal{LI} \text{ } p_{t}$, where $\mathcal{LI}$ represents the \textit{Lerner index} that by definition is equal to  $\frac{\tau}{1+\tau}$. The \textit{Lerner index} is equal to 
$\frac{0.655}{1+0.655} = 0.40$ for Ontario products and to $\frac{0.715}{1+0.715} = 0.42$ for non-Ontario products.
    
\medskip

\noindent \textbf{Step (iv)}. Orders placed at the beginning of day $t$, $y_{t}$, arrive to the store at the end of the same day or at the beginning of $t+1$. Inventory is updated according to the following transition rule:
\begin{equation}
        k_{t+1} 
        \text{ } = \text{ }
        k_{t} 
        \text{ } + \text{ }
        y_{t} 
        \text{ } - \text{ }
        q_{t}
\label{eq:trans_k}
\end{equation}
Finally, next period price $p_{t+1}$ is realized according to a first order Markov process with transition distribution function $F_{p}(p_{t+1}|p_{t})$.

\subsection{Dynamic decision problem}

A store manager chooses the order quantity $y_{t}$ to maximize her store's expected and discounted stream of current and future profits. This is a dynamic programming problem with state variables $\mathbf{x}_{t} \equiv (k_{t}, p_{t}, \ln Q^{[-7,-1]}_{t}, \mathbf{seas}_{t})$ and 
$\boldsymbol{\varepsilon}_{t} \equiv (\varepsilon_{t}(0), \varepsilon_{t}(1), ..., \varepsilon_{t}(J))$ and value function  $V(\mathbf{x}_{t},\boldsymbol{\varepsilon}_{t})$. This value function is the unique solution of the following Bellman equation:
\begin{equation}
    V(\mathbf{x}_{t},
    \boldsymbol{\varepsilon}_{t})
    \text{ } = \text{ }
    \max_{y_{t} \in \mathcal{Y}}
    \text{ }
    \{ 
        \text{ }
        \pi(y_{t},\mathbf{x}_{t}) 
        \text{ } + \text{ }
        \sigma_{\varepsilon} \text{ }
        \varepsilon(y_{t})
        \text{ } + \text{ }
        \beta \text{ }
        \mathbb{E}
        \left[
            V(\mathbf{x}_{t+1},
            \boldsymbol{\varepsilon}_{t+1})
            \text{ } | \text{ }
            y_{t}, \mathbf{x}_{t}
        \right]
        \text{ }
    \},
    \label{eq:bellman}
\end{equation}
where $\beta \in (0,1)$ is the store's one-day discount factor; $\pi(y_{t},\mathbf{x}_{t})$ is the expected profit function up to the $\varepsilon_{t}$ shock; and $\mathbb{E}[.| y_{t}, \mathbf{x}_{t}]$ is the expectation over the i.i.d. distribution of $\varepsilon_{t+1}$, and over the distribution of $\mathbf{x}_{t+1}$ conditional on $\mathbf{x}_{t}$. The latter distribution consists of the transition probability $F_{p}(p_{t+1}|p_{t})$ and the distribution of demand $d_{t}$ conditional on $\mathbf{x}_{t}$ which together with equations (\ref{eq:min_dem_sup}) and (\ref{eq:trans_k})
determines the distribution of $(k_{t+1}, p_{t+1}, \ln Q^{[-7,-1]}_{t+1}, \mathbf{seas}_{t+1})$. The solution of this dynamic programming problem implies a time-invariant optimal decision rule: $y_{t} = y^{\ast}(\mathbf{x}_{t}, \boldsymbol{\varepsilon}_{t})$. This optimal decision rule is defined as the \textit{arg max} of the expression within brackets $\{ \}$ in the right-hand-side of equation (\ref{eq:bellman}).

For the solution and estimation of this model, we follow \citeauthor{rust_1987} (\citeyear{rust_1987}, \citeyear{rust_1994}) and use the \textit{integrated value function} 
$V_{\sigma}(\mathbf{x}_{t}) \equiv 
\frac{1}{\sigma_{\varepsilon}} \int     V(\mathbf{x}_{t}, \boldsymbol{\varepsilon}_{t}) d\boldsymbol{\varepsilon}_{t}$ and the corresponding \textit{integrated Bellman equation}. Given the Extreme Value distribution of the $\varepsilon_{t}$ variables, the integrated Bellman equation has the following form:
\begin{equation}
    V_{\sigma}(\mathbf{x}_{t})
    \text{ } = \text{ }
    \ln
    \left[
        \displaystyle{\sum_{y \in \mathcal{Y}}}
        \exp
        \left(    
            \frac{\pi(y,\mathbf{x}_{t})}
            {\sigma_{\varepsilon}}
            \text{ } + \text{ }
            \beta \text{ }
            \mathbb{E}
            \left[
                V_{\sigma}(\mathbf{x}_{t+1})
                \text{ } | \text{ }
                y, \mathbf{x}_{t}
            \right]
        \right)
    \right].
\label{eq:integrated_bellman}
\end{equation}
The expected profit function $\pi(y_{t},\mathbf{x}_{t})$ is linear in the parameters. That is,
\begin{equation}
    \frac{\pi(y_{t},\mathbf{x}_{t})}
    {\sigma_{\varepsilon}}
    \text{ } = \text{ }
    \mathbf{h}(y_{t},\mathbf{x}_{t})' \text{ }
    \mathbf{\gamma},
\end{equation}
where $\mathbf{\gamma}$ is the vector of structural parameters $\mathbf{\gamma} \equiv$ $(1/\sigma_{\varepsilon}$, $\gamma^{h}/\sigma_{\varepsilon}$, $\gamma^{z}/\sigma_{\varepsilon}$, $\gamma^{f}/\sigma_{\varepsilon}$, $\gamma^{c}/\sigma_{\varepsilon})'$, and $\mathbf{h}(y_{t},\mathbf{x}_{t})$ is the following vector of functions of the state variables:
\begin{equation}
    \mathbf{h}(y_{t},\mathbf{x}_{t})'
    \text{ } = \text{ }
    \left(
    \mathcal{LI} \text{ } p_{t} \text{ }
    \mathbb{E}[min\{d_{t},k_{t}\}|\mathbf{x}_{t}], 
    \text{ } -k_{t}, 
    \text{ } \mathbb{E}[\mathbbm{1}\{d_{t} > k_{t}\}|\mathbf{x}_{t}], 
    \text{ } -\mathbbm{1}\{y_{t}>0\}, 
    \text{ } -y_{t}
    \right),
\end{equation}
where the expectation is taken over the distribution of demand conditional on $\mathbf{x}_{t}$.

We consider a discrete space for the state variables $\mathbf{x}_{t}$.\footnote{In the estimation, we discretize the state space using a k-means algorithm.} Let $\mathcal{X} \equiv \{\mathbf{x}^{1}, \mathbf{x}^{2}, ..., \mathbf{x}^{L}\}$ be the support set of $\mathbf{x}_{t}$. We can represent the value function $V_{\sigma}(.)$ as a vector $\mathbf{V}_{\sigma}$ in the Euclidean space $\mathbb{R}^{L}$, and the transition probability functions of $\mathbf{x}_{t}$ for a given value of $y$ as an $L \times L$ matrix $\mathbf{F}_{\mathbf{x}}(y)$. Taking this into account, as well as the linear-in-parameters structure of the expected profit $\pi(y_{t},\mathbf{x}_{t})$, the integrated Bellman equation in vector form is:
\begin{equation}
    \mathbf{V}_{\sigma}
    \text{ } = \text{ }
    \ln
    \left[
        \displaystyle{\sum_{y \in \mathcal{Y}}}
        \exp
        \left(    
            \mathbf{H}(y) \text{ }
            \mathbf{\gamma}
            \text{ } + \text{ }
            \beta \text{ }
            \mathbf{F}_{\mathbf{x}}(y)
            \text{ }
            \mathbf{V}_{\sigma}
        \right)
    \right].
\label{eq:vector_integrated_bellman}
\end{equation}
where $\mathbf{H}(y)$ is a $L \times 5$ matrix that in row $r$ contains vector $\mathbf{h}(y,\mathbf{x}^{r})'$ for $\mathbf{x}^{r} \in \mathcal{X}$. 

The \textit{Conditional Choice Probability (CCP)} function, $P(y|\mathbf{x}_{t})$, is an integrated version of the decision rule $y^{\ast}(\mathbf{x}_{t}, \boldsymbol{\varepsilon}_{t})$. For any $y \in \mathcal{Y}$ and $\mathbf{x}_{t} \in \mathcal{X}$, the CCP $P(y|\mathbf{x}_{t})$ is defined as $\int 1\{ y^{\ast}(\mathbf{x}_{t}, \boldsymbol{\varepsilon}_{t}) = y\} dG(\boldsymbol{\varepsilon}_{t})$, where $G$ is the CDF of $\boldsymbol{\varepsilon}_{t}$. For the Extreme Value type 1 distribution, the CCP function has the Logit form:
\begin{equation}
    P(y|\mathbf{x}_{t}) 
    \text{ } = \text{ }
    \frac
    {
    \exp 
    \{
    \mathbf{h}(y,\mathbf{x}_{t})' 
    \text{ } \mathbf{\gamma}
    \text{ } + \text{ }
    \beta \text{ }
        \mathbb{E}
        \left[
            V_{\sigma}(\mathbf{x}_{t+1})
            \text{ } | \text{ }
            y, \mathbf{x}_{t}
        \right]
    \}
    }
    {
    \sum_{j=0}^{J}
    \exp 
    \{
    \mathbf{h}(j,\mathbf{x}_{t})' 
    \text{ } \mathbf{\gamma}
    \text{ } + \text{ }
    \beta \text{ }
        \mathbb{E}
        \left[
            V_{\sigma}(\mathbf{x}_{t+1})
            \text{ } | \text{ }
            j, \mathbf{x}_{t}
        \right]
    \}
    }
    \text{ },
\end{equation}

Following \citeauthor{aguirregabiria_mira_2002} (\citeyear{aguirregabiria_mira_2002}), we can represent the vector of CCPs, $\mathbf{P} \equiv \{P(y|\mathbf{x}): (y,\mathbf{x}) \in \mathcal{Y} \times \mathcal{X} \}$, as the solution of a fixed-point mapping in the probability space: $\mathbf{P} = \psi(\mathbf{P})$. Mapping $\psi$ is denoted the \textit{policy iteration} mapping, and it is the composition of two mappings: $\psi(\mathbf{P}) \equiv \lambda(\upsilon(\mathbf{P}))$. Mapping $\lambda(\mathbf{V})$ is the \textit{policy improvement}. It takes as given a vector of values $\mathbf{V}$ and obtains the optimal CCPs as "best responses" to these values. Mapping $\upsilon(\mathbf{P})$ is the \textit{valuation} mapping. It takes as given a vector of CCPs $\mathbf{P}$ and obtains the corresponding vector of values if the agent behaves according to these CCPs.\footnote{See \citeauthor{puterman_2014} (\citeyear{puterman_2014}) for a description of these three mappings in the context of a general dynamic programming problem.} In our Logit model, the \textit{policy improvement} mapping has the following vector form, for any $y \in \mathcal{Y}$:
\begin{equation}
    \mathbf{P}(y)
    \text{ } = \text{ }
    \lambda(y,\mathbf{V})
    \text{ } = \text{ }
    \frac
    {
    \exp 
    \{
        \mathbf{H}(y) \text{ }
        \mathbf{\gamma}
        \text{ } + \text{ }
        \beta \text{ }
        \mathbf{F}_{\mathbf{x}}(y)
        \text{ }
        \mathbf{V}
    \}
    }
    {
    \sum_{j=0}^{J}
    \exp 
    \{
        \mathbf{H}(j) \text{ }
        \mathbf{\gamma}
        \text{ } + \text{ }
        \beta \text{ }
        \mathbf{F}_{\mathbf{x}}(j)
        \text{ }
        \mathbf{V}
    \}
    }
    \text{ }.
\label{eq:policy_improvement}
\end{equation}
The \textit{valuation} mapping has the following form:
\begin{equation}
    \mathbf{V}
    \text{ } = \text{ }
    \upsilon(\mathbf{P})
    \text{ } = \text{ }
    \left[
        \mathbf{I} - \beta
        \sum_{y=0}^{J} 
        \mathbf{P}(y) \ast \mathbf{F}_{x}(y)
    \right]^{-1}
    \left[
        \sum_{y=0}^{J} 
        \mathbf{P}(y) \ast 
        \left(
        \mathbf{H}(y) \text{ }
        \mathbf{\gamma}
        + euler
        - \ln \mathbf{P}(y)
        \right)
    \right],
\label{eq:valuation}
\end{equation}
where $euler$ is Euler's constant, and $\ast$ is the element-by-element vector product.

\subsection{Parameter estimates \label{sec: estimates}}

For every LCBO store and product in our working sample, we estimate the store-product specific parameters in vector $\mathbf{\gamma}$ using a \textit{Two-Step Pseudo Likelihood (2PML)} estimator (\citeauthor{aguirregabiria_mira_2002} (\citeyear{aguirregabiria_mira_2002})). Given a dataset $\{y_{t}, \mathbf{x}_{t}: t=1,2, ..., T\}$  and arbitrary vectors of CCPs and structural parameters $(\mathbf{P},\mathbf{\gamma})$, define the pseudo (log) likelihood function:
\begin{equation}
    Q(\mathbf{P},\mathbf{\gamma})
    \text{ } = \text{ }
    \sum_{t=1}^{T} \text{ }
    \ln \psi
    \left( 
        y_{t}, \text{ }
        \mathbf{x}_{t}
        \text{ } ; \text{ }
        \mathbf{P},\mathbf{\gamma}
    \right),
\end{equation}
where $\psi(.)$ is the policy iteration mapping defined by the composition of equations (\ref{eq:policy_improvement}) and (\ref{eq:valuation}). Note that the likelihood function $Q(\mathbf{P},\mathbf{\gamma})$ is a function of the store's one-day discount factor $\beta$. In the estimation, we fix the value\footnote{We could eventually relax this assumption and treat $\beta$ as a parameter to be estimated, and allow it to vary across store managers in order to potentially capture different degrees of myopia or impatience.} of this discount factor equal to $0.95^{\frac{1}{365}}$. In the first step of the 2PML method, we obtain a reduced form estimation of the vector of CCPs $\mathbf{P}$ using a Kernel method. In the second step, the 2PML estimator is the vector $\widehat{\mathbf{\gamma}}$ that maximizes the pseudo-likelihood function when $\mathbf{P}=\widehat{\mathbf{P}}$. That is:
\begin{equation}
        \text{ }
        \widehat{\mathbf{\gamma}}
        \text{ } = \text{ }
        \textit{arg max}_{\mathbf{\gamma}}
        \text{ }
        Q(\widehat{\mathbf{P}},\mathbf{\gamma})
\label{eq:npl_estimator}
\end{equation}
\citeauthor{aguirregabiria_mira_2002} (\citeyear{aguirregabiria_mira_2002}) show that this estimator is consistent and asymptotically normal with the same asymptotic variance as the full maximum likelihood estimator. In section \ref{appendix_estimation_method} in the Appendix, we provide further details on the implementation of this estimator.

In a similar vein to the estimation of $(S,s)$ thresholds in section \ref{sec:s_s_rules}, a portion of the variation in parameter estimates $\widehat{\gamma}_{i,j}$ can be attributed to estimation error rather than genuine heterogeneity. To address this issue and mitigate the excessive dispersion or spurious heterogeneity resulting from estimation error, we employ a \textit{shrinkage estimator}. The details of this estimator can be found in section \ref{appendix_shrinkage_estimator} in the Appendix.

Table \ref{tab: struc_estimates} presents the medians from the empirical
distributions (across stores and products)
of our estimates of the four structural parameters, measured in dollar amounts.\footnote{More specifically, we first obtain the two-step PML estimate of the vector $\mathbf{\gamma} \equiv$ $(1/\sigma_{\varepsilon}$, $\gamma^{h}/\sigma_{\varepsilon}$, $\gamma^{z}/\sigma_{\varepsilon}$, $\gamma^{f}/\sigma_{\varepsilon}$, $\gamma^{c}/\sigma_{\varepsilon})'$, and then we divide elements 2 to 5 of this vector by the first element to obtain estimates of costs in dollar amount. We use the delta method to obtain standard errors.} The median values of the estimates are $\$0.0036$ for the per-unit inventory holding cost, $\$0.0219$ for the stockout cost, $\$2.9658$ for the fixed ordering cost, and $\$0.0341$ for the per-unit ordering cost. To have an idea of the importance of these dollar amounts, in Section \ref{sec:contribution_costs} below we provide measures of the implied magnitude of each cost relative to revenue. These magnitudes are consistent with other cost estimates in the inventory management literature (see \cite{aguirregabiria_1999}, \cite{bray_yao_2019}). Median standard errors and t-statistics in Table \ref{tab: struc_estimates} show that the inventory holding cost and the fixed ordering cost are very precisely estimated (median t-ratios of $5.32$ and $12.65$, respectively), while a substantial fraction of the estimates of the stockout cost are quite imprecise (median t-ratio of $0.27$).

\medskip

\begin{table}[ht]
\begin{center}
\caption{Structural Estimates of Cost Parameters (in Canadian Dollars) \label{tab: struc_estimates}}
\resizebox{0.9\textwidth}{!}{\begin{tabular}{r|cccc}
\hline \hline
&
\multicolumn{1}{c}{\textit{Median}}
&
\multicolumn{1}{c}{\textit{Std. Dev.}}
&
\multicolumn{1}{c}{\textit{Median}}
&
\multicolumn{1}{c}{\textit{Median}}

\\
&
\multicolumn{1}{c}{\textit{Estimate}}
&
\multicolumn{1}{c}{\textit{Estimate}}
&
\multicolumn{1}{c}{\textit{S.e.}}
&
\multicolumn{1}{c}{\textit{t-stat.}}

\\ \hline
& 
\\
$\gamma^{h}$ : \textit{Per-unit Inventory Holding Cost} 
& $0.0036$  & 0.0029 & $0.0007$  & 5.3259  \\
&
\\
$\gamma^{z}$: \textit{Stockout Cost} 
& $0.0219$  & 0.3094 & $0.1465$  & 0.2672  \\
&
\\
$\gamma^{f}$: \textit{Fixed Ordering Cost} 
& $2.9658$ & 1.0847 & $0.2356$ & 12.6557  \\
&
\\
$\gamma^{c}$: \textit{Per-Unit Ordering Cost} 
& $0.0341$ & 0.0607 & $0.0274$ & 1.4763  \\ 
&
\\
\hline 
\# of observed store-product pairs & 3,076 & & \\
\# of store-product pairs with structural estimates & 2,589 & & \\
\hline \hline
\end{tabular}}
\end{center}
\end{table}

\medskip

\begin{figure}[ht]
\begin{center}
\caption{$\boldsymbol{\gamma}$ Estimates: Shrinkage Estimator
\label{fig:gamma_histograms_shrinkage}}
\begin{subfigure}{.4\textwidth}
    \caption{$\gamma^{h}$}
    \centering
    \includegraphics[width=6cm]{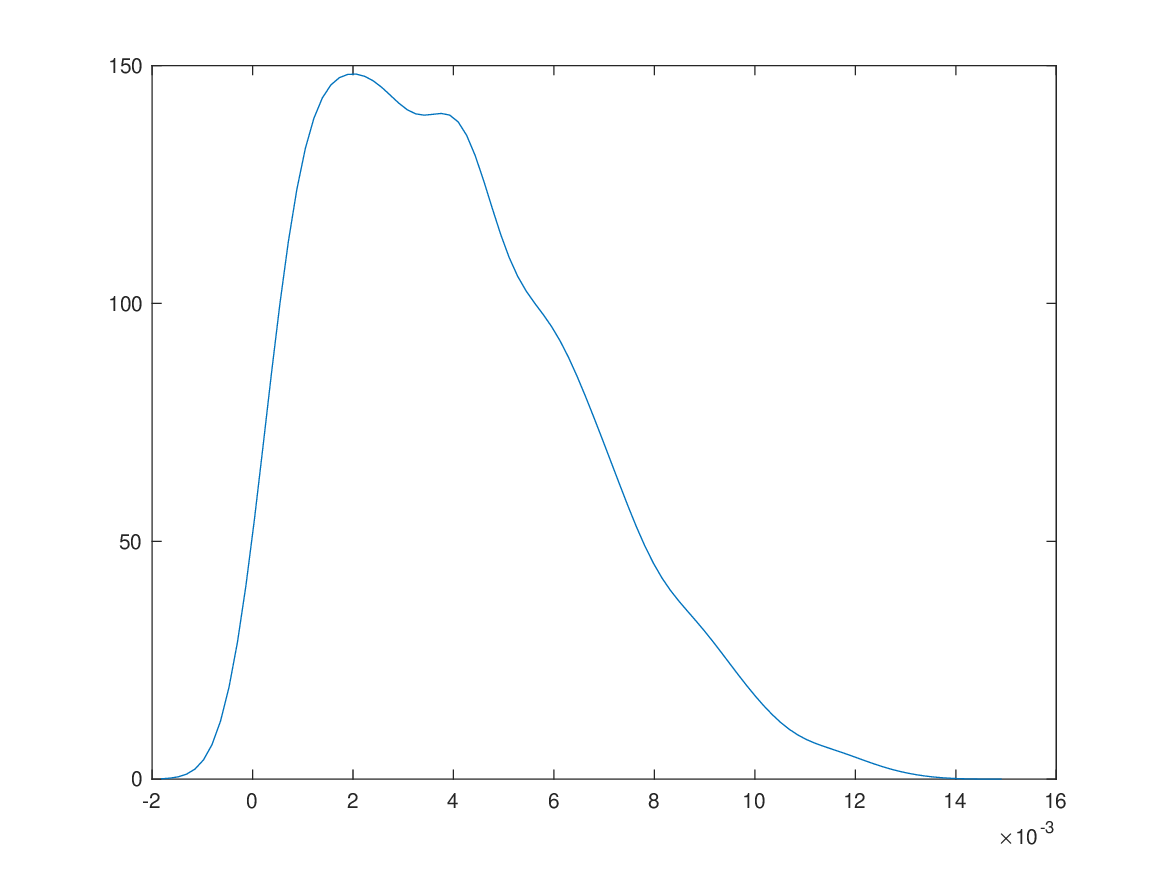}
\end{subfigure}
\begin{subfigure}{.4\textwidth}
    \caption{$\gamma^{z}$}
    \centering
    \includegraphics[width=6cm]{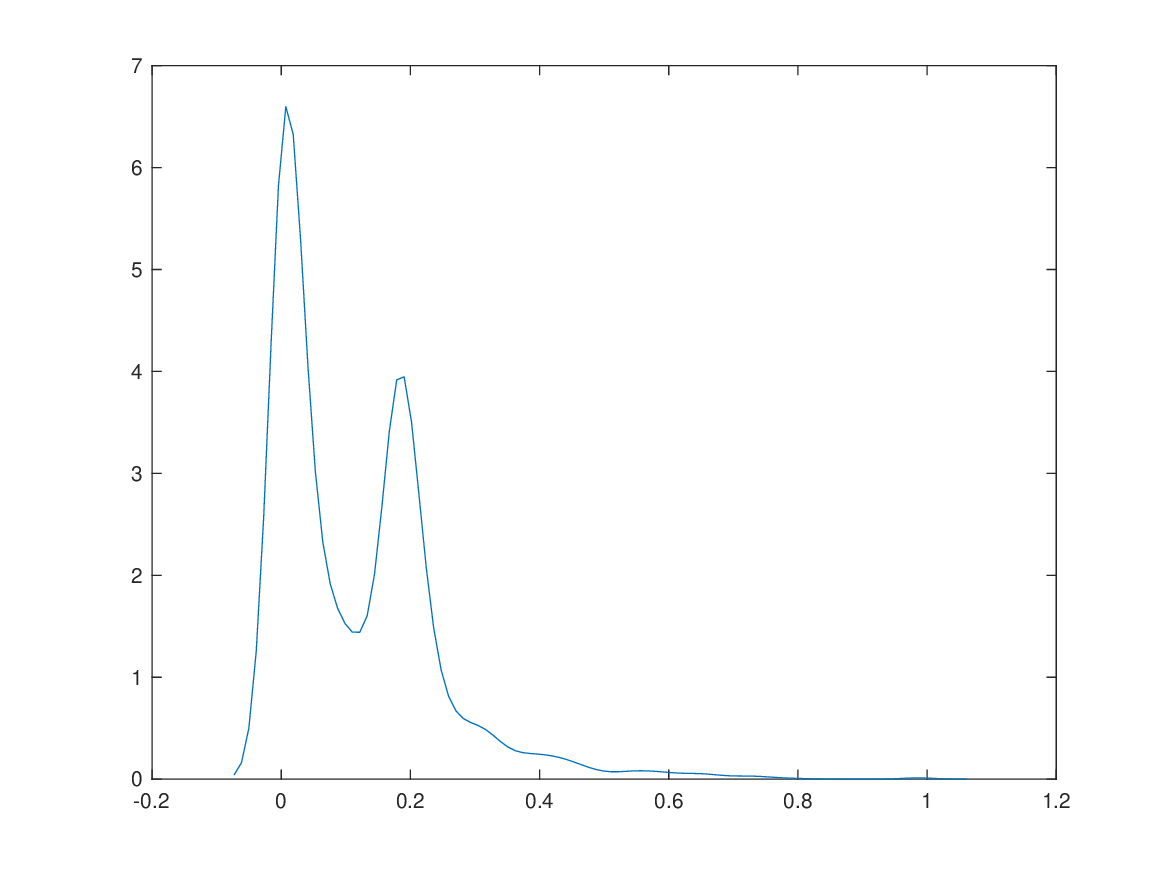}
    \vspace*{2mm}
\end{subfigure}
\begin{subfigure}{.4\textwidth}
    \caption{$\gamma^{f}$}
    \centering
    \includegraphics[width=6cm]{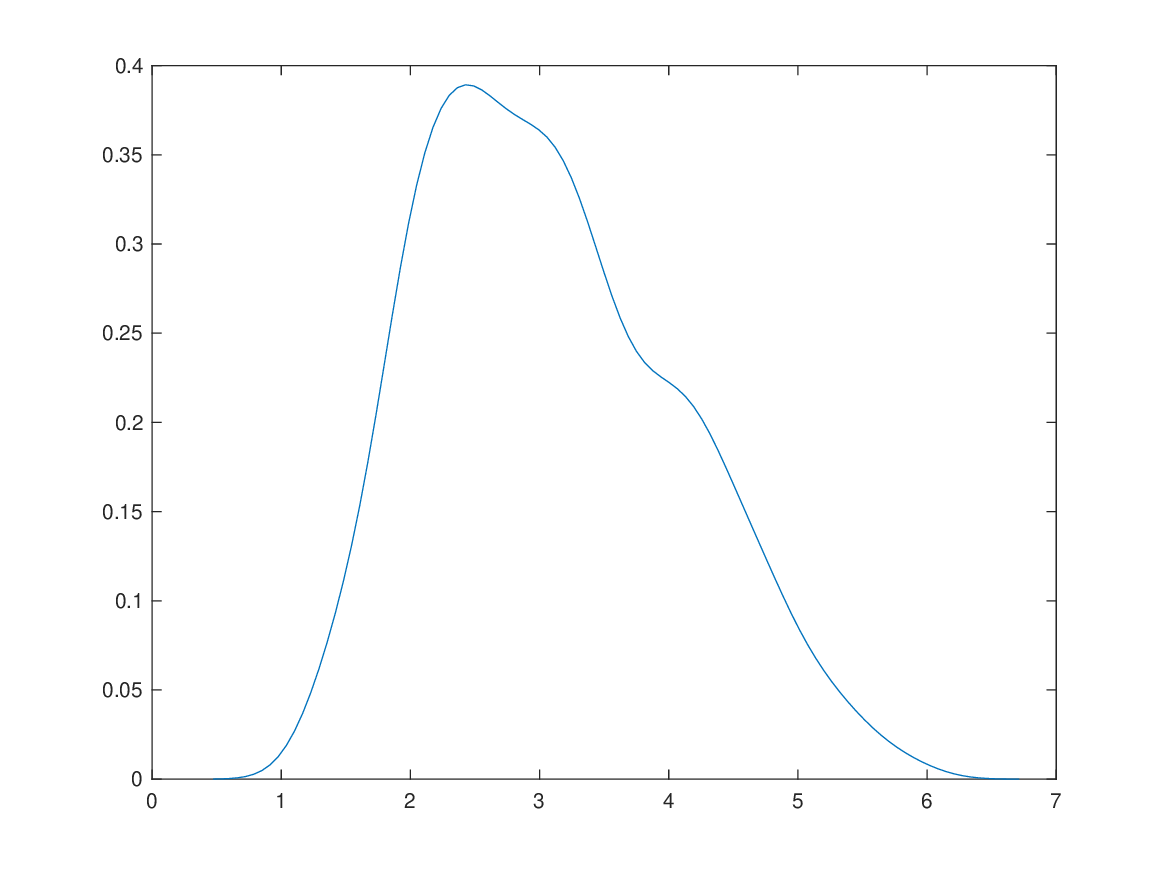}
\end{subfigure}
\begin{subfigure}{.4\textwidth}
    \caption{$\gamma^{c}$}
    \centering
    \includegraphics[width=6cm]{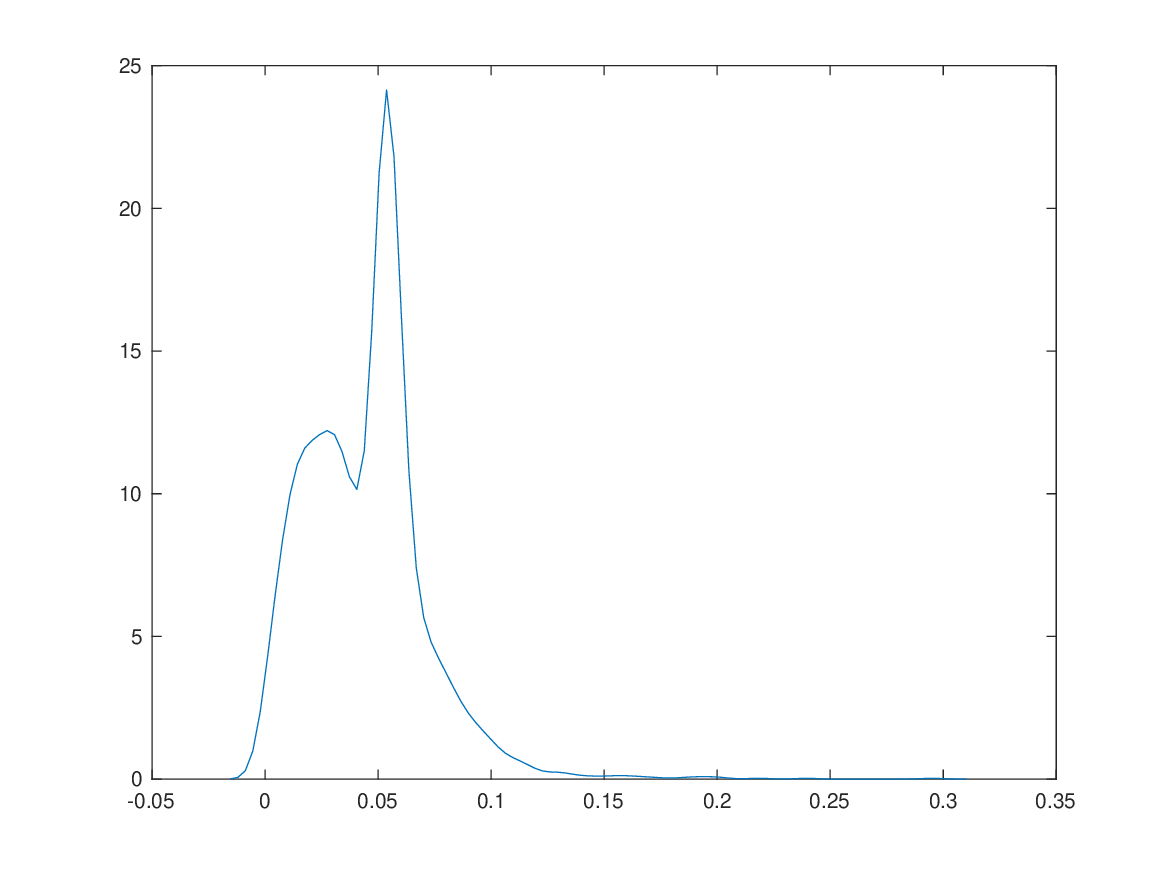}
\end{subfigure}
\end{center}
\end{figure}

\medskip

\subsection{Relative contribution of the different costs \label{sec:contribution_costs}}

In this section we assess the magnitude of the different inventory management costs relative to store revenues. The purpose of this exercise is twofold. First, we want to evaluate whether our parameter estimates imply realistic magnitudes for the realization of these costs. And second, it is relevant to measure to what extent the heterogeneity in cost parameters that we have presented above generates heterogeneity in profits across stores. Conditional on their perception of cost parameters, store managers' optimal behavior should compensate -- at least partly -- for the differences in cost parameters such that heterogeneity in realized costs should be smaller. We want to measure the extent of this compensating effect.

For each component of the inventory management cost, and for every store-product, we calculate the ratio between the realized value of the cost during our sample period and the realized value of revenue during the same period. More specifically, we calculate the following ratios for every store-product: inventory holding cost to revenue; stockout cost to revenue; fixed ordering cost to revenue; variable ordering cost to revenue; and total inventory management cost to revenue. We have an empirical distribution over store-products for each of these ratios. 

Table \ref{tab: inv_ratios} presents the median and the standard deviation in these distributions. To evaluate the magnitude of these ratios, it is useful taking into account that -- according to the LCBO's annual reports -- the total expenses to sales ratio of the retail chain is consistently around $16\%$ each year.\footnote{Of course, these expenses do not include the cost of merchandise.} According to our estimate, the total inventory cost-to-revenue ratio for the median store is approximately $1.37\%$. This would imply that the retail chain's cost of managing the inventories of their stores would represent around $10\%$ of total costs, which entails that non-inventory related costs would account for approximately $90\%$ of total costs (e.g. labor costs, fixed capital costs, delivery costs). This seems to be of the right order of magnitude. Table \ref{tab: inv_ratios} shows that the fixed ordering cost is the largest realized cost for store managers at LCBO, followed by storage costs. Realized stockout costs are negligible. This is due to a combination of a small parameter that captures the stockout cost, and infrequent stockouts in our working sample.

\medskip

\begin{table}[ht]
\begin{center}
\caption{Realized Inventory Management Costs to Revenue Ratios \label{tab: inv_ratios}}
\begin{tabular}{r|cc}
\hline \hline
& 
\multicolumn{1}{c}{\textit{Median}}
& \multicolumn{1}{c}{\textit{St. Dev.}}
\\ \hline
\\
\textit{Inventory Holding Cost to Revenue Ratio (\%)}
& 0.2863
& 0.2182 \\
\textit{Stockout Cost to Revenue Ratio (\%)}
& 0.0005
& 0.0197 \\
\textit{Fixed Ordering Cost to Revenue Ratio (\%)}
& 0.8731
& 0.7728 \\
\textit{Variable Ordering Cost to Revenue Ratio (\%)}
& 0.2045
& 0.1895 \\
\\
\hline
\\
\textit{Total Inventory Cost to Revenue Ratio (\%)}
& 1.3749
& 0.9235 \\
\\
\hline \hline
\end{tabular}
\end{center}
\end{table}

\medskip

In section \ref{appendix_realized_costs} in the Appendix, we present the empirical distribution across stores and products of each of the four cost-to-revenue ratios. We also show the extent to which managers' inventory decisions compensate for the heterogeneity in the structural parameters. 

\medskip

\subsection{Heterogeneity in cost parameters}

Below, we investigate two potential sources for the large heterogeneity in our cost parameters: (i) differences across stores, such as store type according to LCBO's classification of stores, physical area, total product assortment, distance to the warehouse, and consumer socioeconomic characteristics; and (ii) differences across local managers. Our goal in this section is to separate the heterogeneity attributable to store characteristics, and the heterogeneity stemming from the managers themselves. We proceed using a sequential approach. First, we regress our parameter estimates on a set of store characteristics. Then, we take the residual components from the first step and regress them on manager characteristics.

\medskip

\def\sym#1{\ifmmode^{#1}\else\(^{#1}\)\fi}
\begin{table}[ht]
\begin{center}
\caption{Regressions on Store Characteristics \label{tab_store_reg}}
\resizebox{\textwidth}{!}{\begin{tabular}{lcccc}
\hline \hline 
& ($\gamma^{h}$) & ($\gamma^{z}$)
& ($\gamma^{f}$) & ($\gamma^{c}$)
\\ \hline 
& \textit{Est.} & \textit{Est.} 
& \textit{Est.} & \textit{Est.} 
\\
& \textit{(s.e.)} & \textit{(s.e.)} 
& \textit{(s.e.)} & \textit{(s.e.)} 
\\ \hline
\addlinespace
Store Class & & & & \\
\multicolumn{1}{c}{\textit{AA}}        &    0.000554         &     -0.0216         &      -0.164         &     0.00302         \\
                    &  (0.000411)         &    (0.0194)         &     (0.185)         &   (0.00587)         \\
\addlinespace
\multicolumn{1}{c}{\textit{A}}        &    0.000511         &     -0.0204         &      -0.203         &     0.00816         \\
                    &  (0.000397)         &    (0.0181)         &     (0.178)         &   (0.00572)         \\
\addlinespace
\multicolumn{1}{c}{\textit{B}}        &   -0.000415         &     -0.0237         &       0.108         &     0.00761         \\
                    &  (0.000434)         &    (0.0204)         &     (0.190)         &   (0.00590)         \\
\addlinespace
\multicolumn{1}{c}{\textit{C}}         &    -0.00209\sym{***}&     -0.0531\sym{**} &       0.785\sym{***}&     0.00477         \\
                    &  (0.000520)         &    (0.0243)         &     (0.221)         &   (0.00657)         \\
\addlinespace
\multicolumn{1}{c}{\textit{D}}         &    -0.00334\sym{***}&     -0.0379         &       1.223\sym{***}&     0.00479         \\
                    &  (0.000564)         &    (0.0262)         &     (0.244)         &   (0.00693)         \\
\addlinespace
ln(Product Assortment Size)       &    0.000336         &    -0.00932         &      -0.364\sym{***}&     0.00388\sym{*}  \\
                    &  (0.000205)         &   (0.00973)         &    (0.0775)         &   (0.00201)         \\
\addlinespace
ln(Population in City)      &  -0.0000801\sym{**} &    -0.00136         &     0.00948         &    0.000413         \\
                    & (0.0000381)         &   (0.00220)         &    (0.0186)         &  (0.000483)         \\
\addlinespace
ln(Median Income in City)    &    0.000362         &      0.0493\sym{*}  &     -0.0577         &     0.00795         \\
                    &  (0.000520)         &    (0.0255)         &     (0.212)         &   (0.00646)         \\
\addlinespace
\hline
Location dummies (25 regions, 4 districts) &  YES  & YES   & YES   & YES \\
Product dummies (5 products)  &  YES  & YES   & YES   & YES \\
\addlinespace
\hline
R-squared & 0.3954 & 0.1345 & 0.5265 & 0.0598 \\
Observations        &        2,589         &        2,589         &        2,589         &        2,589         \\
\hline \hline
\multicolumn{5}{l}{\footnotesize{(1) Location dummies based on LCBO's own division of Ontario into 25 regional markets and 4 districts.}}\\
\multicolumn{5}{l}{\footnotesize{(2) Robust standard errors clustered at the store level in parentheses}}\\
\multicolumn{5}{l}{\footnotesize{(3) * means p-value<0.10, ** means p-value<0.05, *** means p-value<0.01}}\\
\end{tabular}}
\end{center}
\end{table}

\medskip

\noindent \textbf{First step: store characteristics.} Table \ref{tab_store_reg} presents estimation results from the first-step regressions of each estimated cost parameter against store and location characteristics: LCBO's store type dummies (6 types); LCBO's regional market dummies (25 regions); logarithm of the number of unique products offered by the store; logarithm of population in the store's city; and logarithm of median income level in the store's city. As we have cost estimates at the store-product level, we also include product fixed effects.\footnote{Note that the store location dummies capture various factors, including the effect of the distance between the store and the warehouse.}

These store and location characteristics can explain an important part of the variation across stores in inventory holding costs and fixed ordering costs: the R-squared coefficients for these regressions are $0.39$, and $0.53$, respectively. Fixed ordering costs decline significantly with the number of products in the store, which is consistent with economies of scope in ordering multiple products. Inventory holding costs increase with assortment size and are significantly higher for $AAA$ stores relative to $D$ stores. In contrast, only $6\%$ of the variation in unit ordering costs and $13\%$ of the variation in stockout costs can be explained by these store and location characteristics. These results are robust to other specifications of the regression equation based on transformations of explanatory or/and dependent variables.

\medskip

\noindent \textbf{Second step: manager characteristics.} Table \ref{tab_man_reg} presents the estimation results from the second-step regressions of cost parameters on manager characteristics (educational attainment, years of experience at the LCBO, and other industry experience), after controlling for the variation explained by store characteristics. The overall finding is that managers' education and experience have non-significant effects in these regressions. There are two main reasons that can explain these negligible effects. 

First, there is a substantial correlation between store characteristics and managers' skills. More skilled managers tend to be allocated to higher-class stores (positive assortative matching). Therefore, in the first-step regression, where store characteristics are included, these characteristics are also capturing the effect of managers' skills. As a result, the direct effect of managers' skills in the second-step regressions becomes less apparent.

Second, the insignificant effect of managers' skills on the estimated cost parameters aligns with the interpretation that the residual component of these parameters is associated with biased perceptions. More skilled managers may have a better measure of these costs, while less skilled managers may have noisier estimates. However, this does not imply a larger or smaller effect of managers' skills on the mean value of cost parameters. Instead, the effect would appear in the variance of the cost parameters, indicating differences in the precision of their estimates. Indeed, when we regress the variance of the cost parameters on managers' skills, we find evidence supporting this interpretation.

\clearpage

\begin{table}[ht]
\begin{center}
\caption{Regressions on Manager Characteristics \label{tab_man_reg}}
\begin{tabular}{lcccc}
\hline \hline 
& ($res(\gamma^{h})$) & ($res(\gamma^{z})$)
& ($res(\gamma^{f})$) & ($res(\gamma^{c})$)
\\ \hline 
& \textit{Est.} & \textit{Est.} 
& \textit{Est.} & \textit{Est.} 
\\
& \textit{(s.e.)} & \textit{(s.e.)} 
& \textit{(s.e.)} & \textit{(s.e.)} 
\\ \hline
\addlinespace
Educational Attainment & & & & \\
\multicolumn{1}{c}{\textit{High School}}         &  -0.0000165         &    0.000752         &  -0.0000456         &    0.000420         \\
                    &  (0.000237)         &    (0.0122)         &    (0.0769)         &   (0.00271)         \\
\addlinespace
\multicolumn{1}{c}{\textit{University}}          &  -0.0000801         &     0.00474         &      0.0194         &   -0.000213         \\
                    &  (0.000170)         &   (0.00794)         &    (0.0577)         &   (0.00183)         \\
\addlinespace
LCBO Experience &   $0.0000128^{**}$ &   -0.000204         &     0.00228         &  -0.0000967         \\
                    &(0.00000627)         &  (0.000270)         &   (0.00223)         & (0.0000669)         \\
\addlinespace
Other Experience    &   0.0000156         &    0.000576         &     0.00471         &  -0.0000587         \\
                    & (0.0000125)         &  (0.000475)         &   (0.00437)         &  (0.000115)         \\
\addlinespace
\hline
R-squared & 0.0095 & 0.0051 & 0.0058 & 0.0038 \\
Observations        &        2,589         &        2,589         &        2,589         &        2,589         \\
\hline \hline
\multicolumn{5}{l}{\footnotesize{(1) Robust standard errors clustered at the store level in parentheses}}\\
\multicolumn{5}{l}{\footnotesize (2) * means p-value<0.10, ** means p-value<0.05, *** means p-value<0.01}\\
\multicolumn{5}{l}{\footnotesize (3) We use multiple imputation to account for missing values of our explanatory variables}\\
\end{tabular}
\end{center}
\end{table}

\medskip

In Section \ref{appendix_cost_dispersion} of the Appendix, we also examine how the variance of the cost parameters depends on store characteristics. We find that the dispersion of the (second-step) manager component of costs is larger on average for lower-class stores. Since managers in these stores generally have lower levels of human capital (i.e. education and experience), we interpret the second-step manager component as a biased perception of the true cost from the point of view of store managers. That is, the (first-step) store component of the costs will be interpreted as the true cost, and the manager component will be interpreted as deviations from this true cost. In order to illustrate the interpretation of the residual component as manager bias, we present in Section \ref{appendix_cost_dispersion} of the Appendix two granular examples in which pairs of stores -- located in close proximity to each other -- are similar in size, sales, store classification, but have very different levels of manager experience and estimates of the cost parameters.

We explore the interpretations of our cost parameters, and their impact on store-level inventory outcomes, in the subsequent counterfactual experiments of Section \ref{sec:counterfactuals}.

\medskip

\section{Counterfactual experiments \label{sec:counterfactuals}}

This section presents two sets of counterfactual experiments based on the model that we have estimated in the previous section. First, we study the contribution to inventory management outcomes from the heterogeneity in store managers' perceptions of costs. Second, we evaluate the effects of a counterfactual centralization of inventory management decisions at LCBO headquarters. We present this counterfactual experiment under different scenarios on the information that headquarters has about demand and costs at the store level.

\subsection{Removing store managers' idiosyncratic effects \label{sec:counterfactuals_shutdown_res}}

Let $\widehat{\boldsymbol{\gamma}}_{i,j}$ be the vector of estimates of cost parameters for product $j$ and store $i$. Based on the regressions in Tables \ref{tab_store_reg} and \ref{tab_man_reg}, we decompose this vector into two additive and orthogonal components: the part explained by store and location characteristics, that we represent as $\widehat{\boldsymbol{\gamma}}^{sto}_{i,j}$; and the part explained by local managers, $\widehat{\boldsymbol{\gamma}}^{man}_{i,j}$. Below, we construct a counterfactual scenario that removes the idiosyncratic component $\widehat{\boldsymbol{\gamma}}^{man}_{i,j}$ from the inventory decision problem for store $i$ and product $j$. For every store-product $(i,j)$ in our working sample, we implement a separate counterfactual experiment for each of the four cost parameters, and one experiment that shuts down together the manager component of the four cost parameters. This implies a total number of $15,850$ experiments.

We implement each of these experiments by solving the dynamic programming problem and obtaining the corresponding CCPs under the counterfactual values of the structural parameters. We use this vector of CCPs to calculate the corresponding ergodic distribution of the state variables for the store-product.\footnote{Note that this ergodic distribution incorporates the seasonal effects in the demand part of the model, as seasonal dummies are a component of the vector of state variables of the model.} Finally, we use the vector of CCPs and the ergodic distribution to calculate mean values of relevant outcome variables related to inventory management. We compare these average outcomes with their corresponding values under the factual values of structural parameters. In terms of outcome variables, we look at the same descriptive statistics as those reported in Table \ref{tab:summary} and Figure \ref{fig:inv_het}: stockout frequency, ordering frequency, inventory to sales ratio, inventory to sales ratio after an order (i.e. \textit{S} threshold), and inventory to sales ratio before an order (i.e. \textit{s} threshold). 

\begin{figure}[ht]
\begin{center}
\caption{Counterfactual Outcome: Stockout Frequency \label{fig: counterfactual_stockout}}
\begin{subfigure}{.4\textwidth}
    \centering
    \caption{$\gamma^{h,man}$}
    \includegraphics[width=6cm]{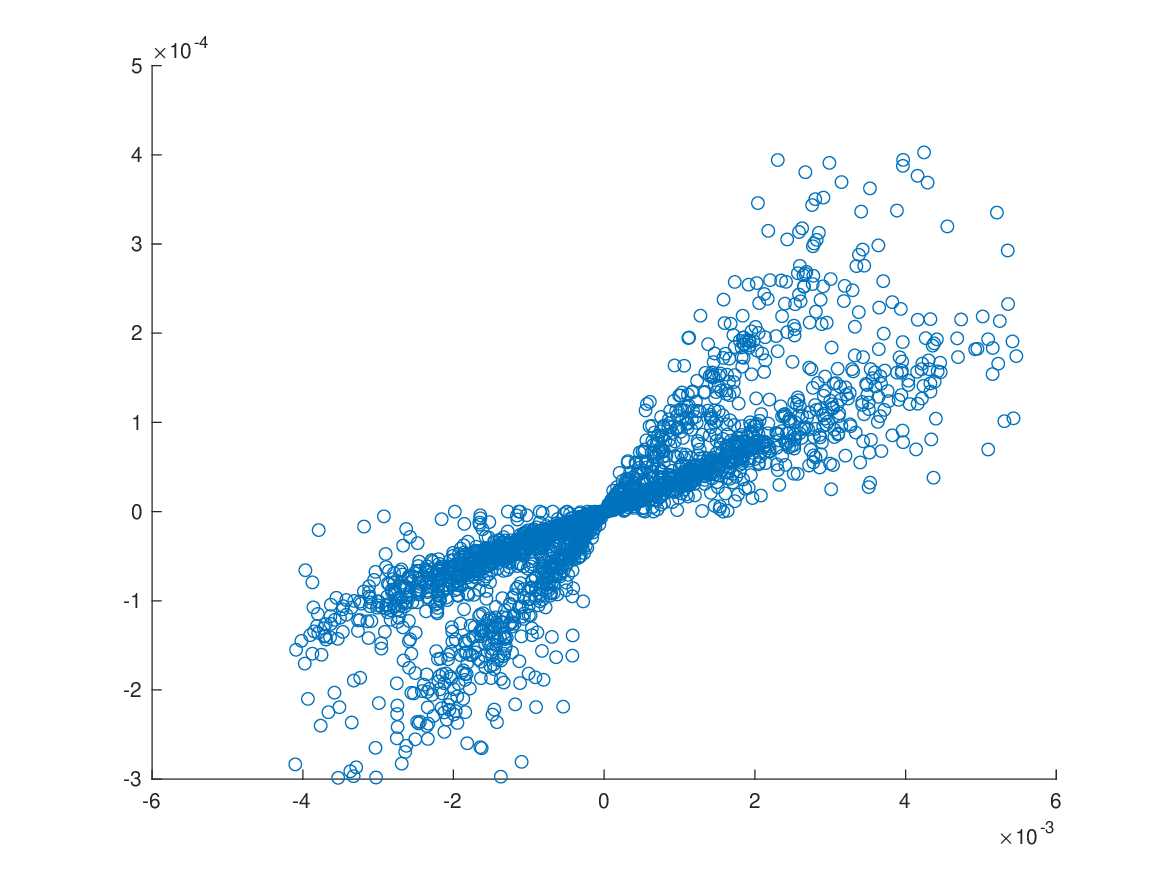}
\end{subfigure}
\begin{subfigure}{.4\textwidth}
    \centering
    \caption{$\gamma^{z,man}$}
    \includegraphics[width=6cm]{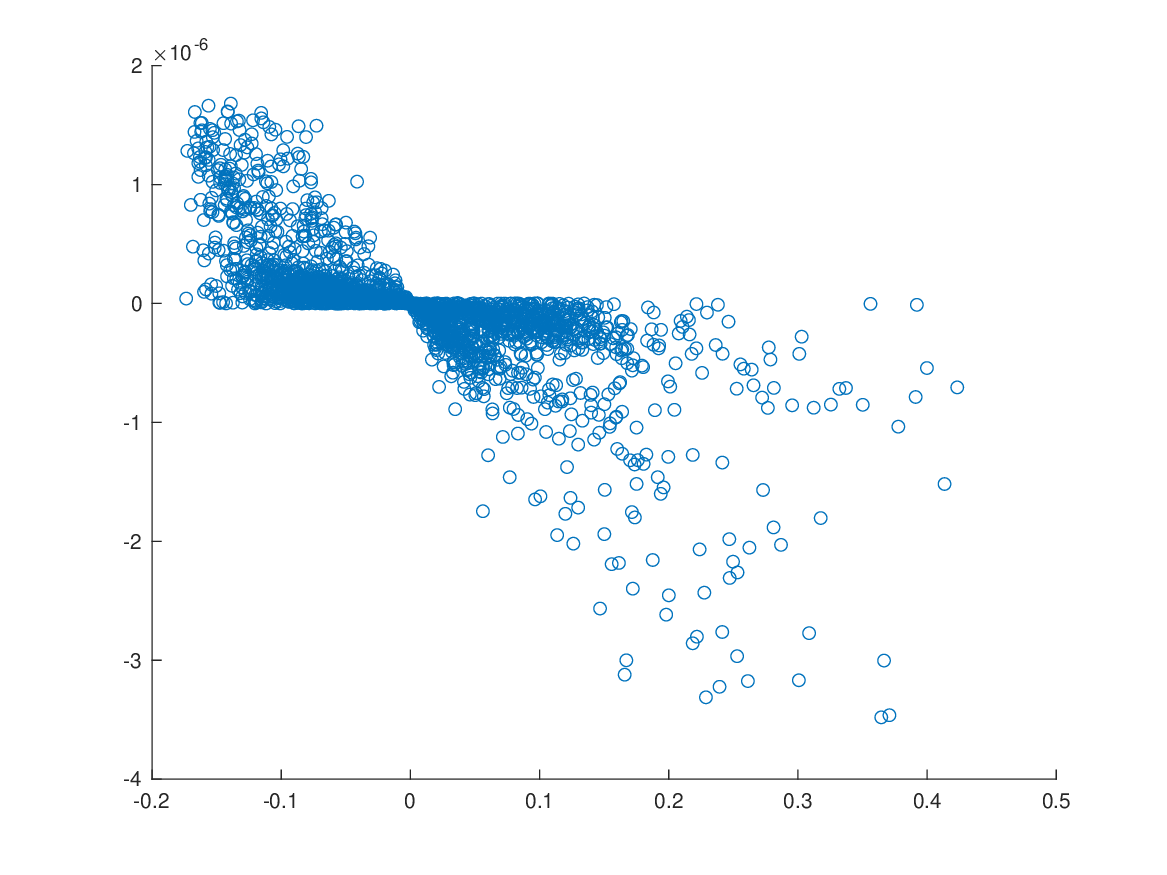}
\end{subfigure}
\begin{subfigure}{.4\textwidth}
    \centering
    \caption{$\gamma^{f,man}$}
    \includegraphics[width=6cm]{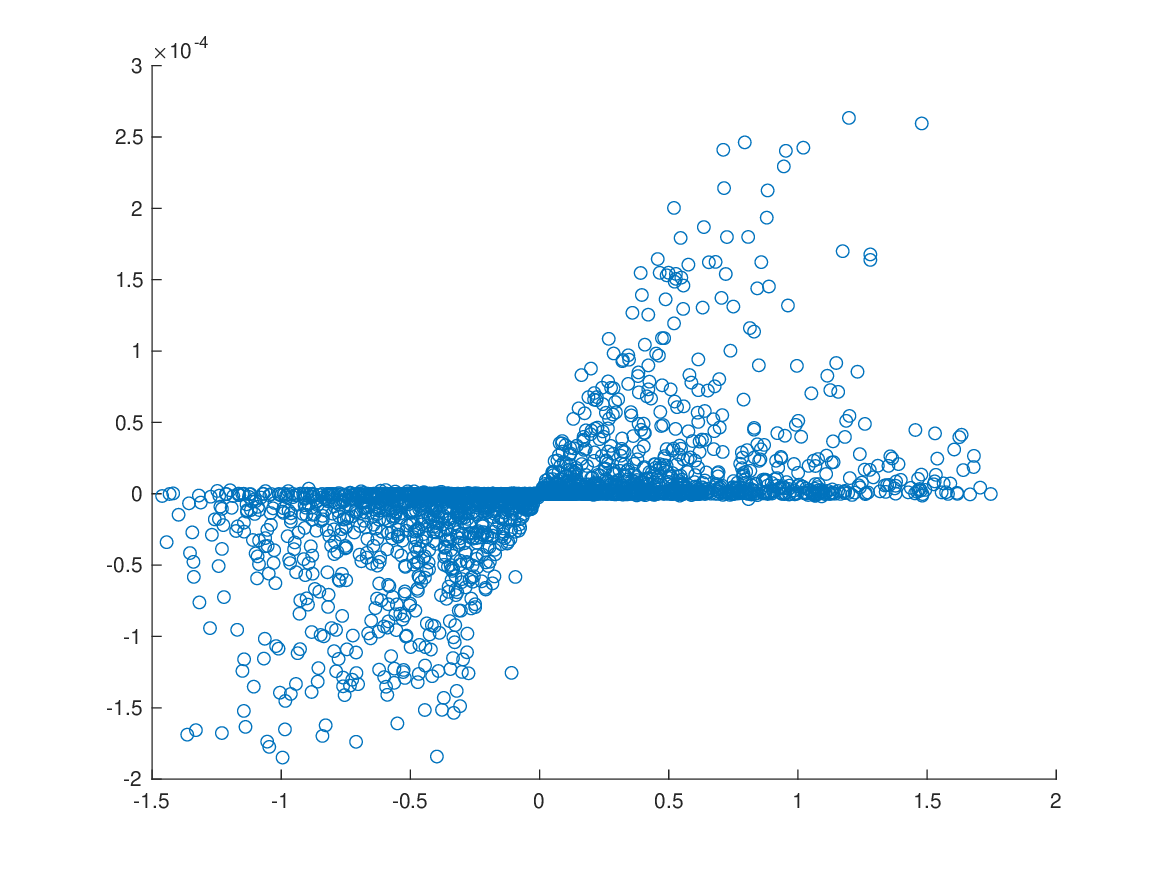}
\end{subfigure}
\begin{subfigure}{.4\textwidth}
    \centering
    \caption{$\gamma^{c,man}$}
    \includegraphics[width=6cm]{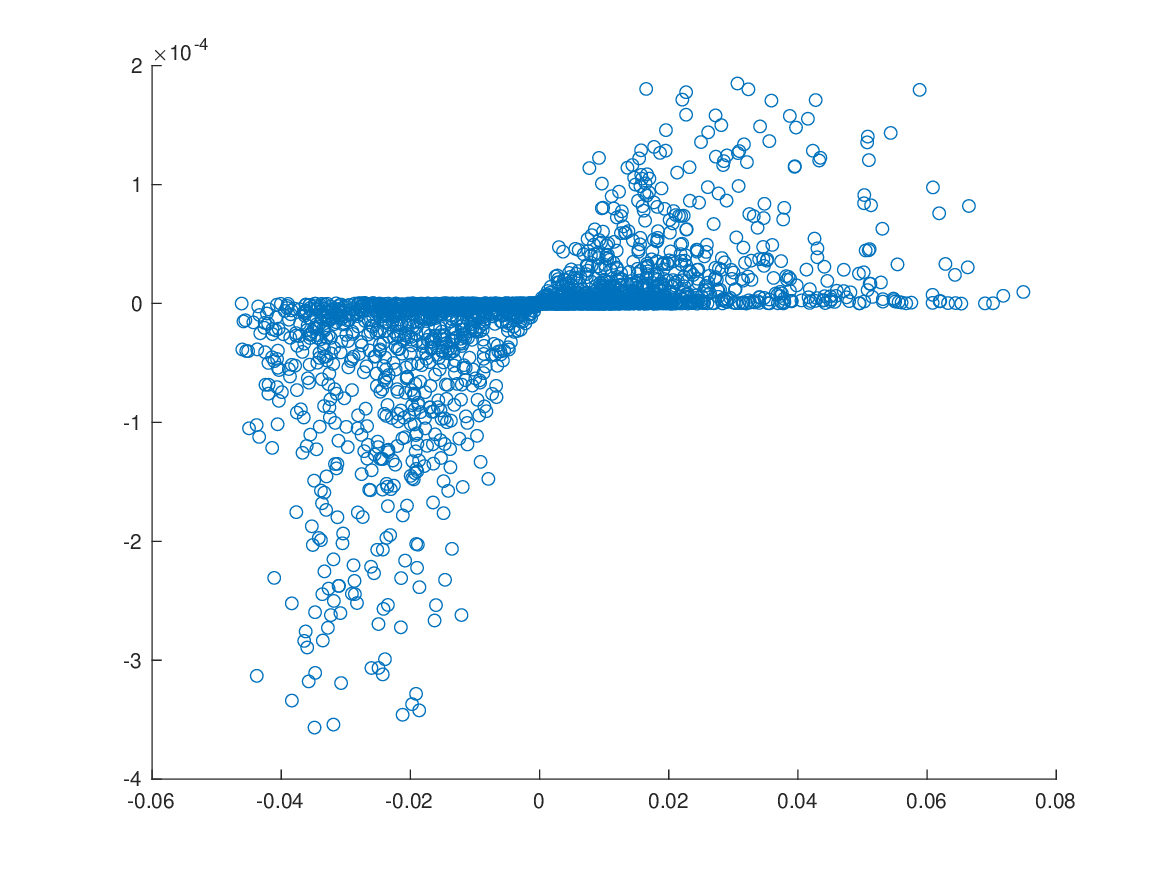}
\end{subfigure}
\begin{center}
\footnotesize{$\Delta$ Stockout Frequency on y-axis, $\gamma^{man}$ on x-axis}
\end{center}
\end{center}
\end{figure}

\begin{figure}[ht]
\begin{center}
\caption{Counterfactual Outcome: Ordering Frequency \label{fig: counterfactual_order}}
\begin{subfigure}{.4\textwidth}
    \caption{$\gamma^{h,man}$}
    \centering
    \includegraphics[width=6cm]{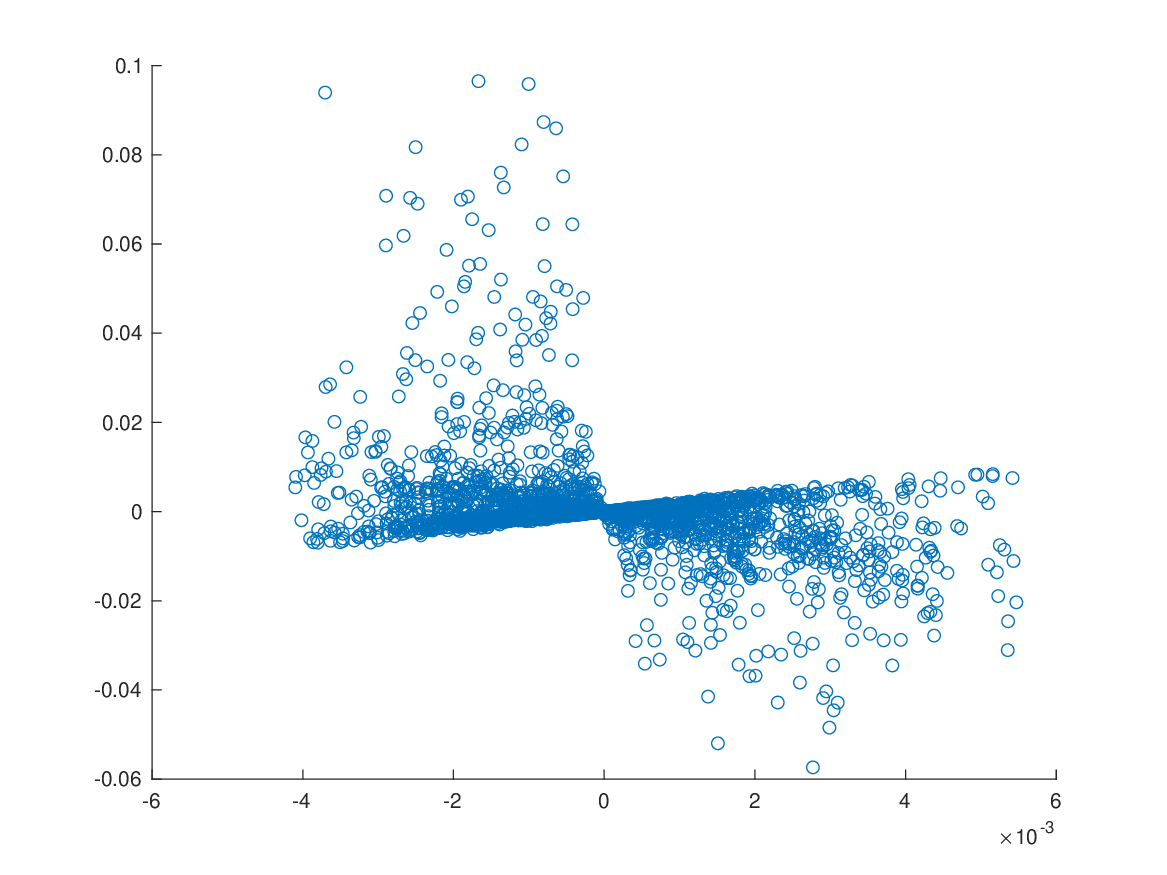}
\end{subfigure}
\begin{subfigure}{.4\textwidth}
    \caption{$\gamma^{z,man}$}
    \centering
    \includegraphics[width=6cm]{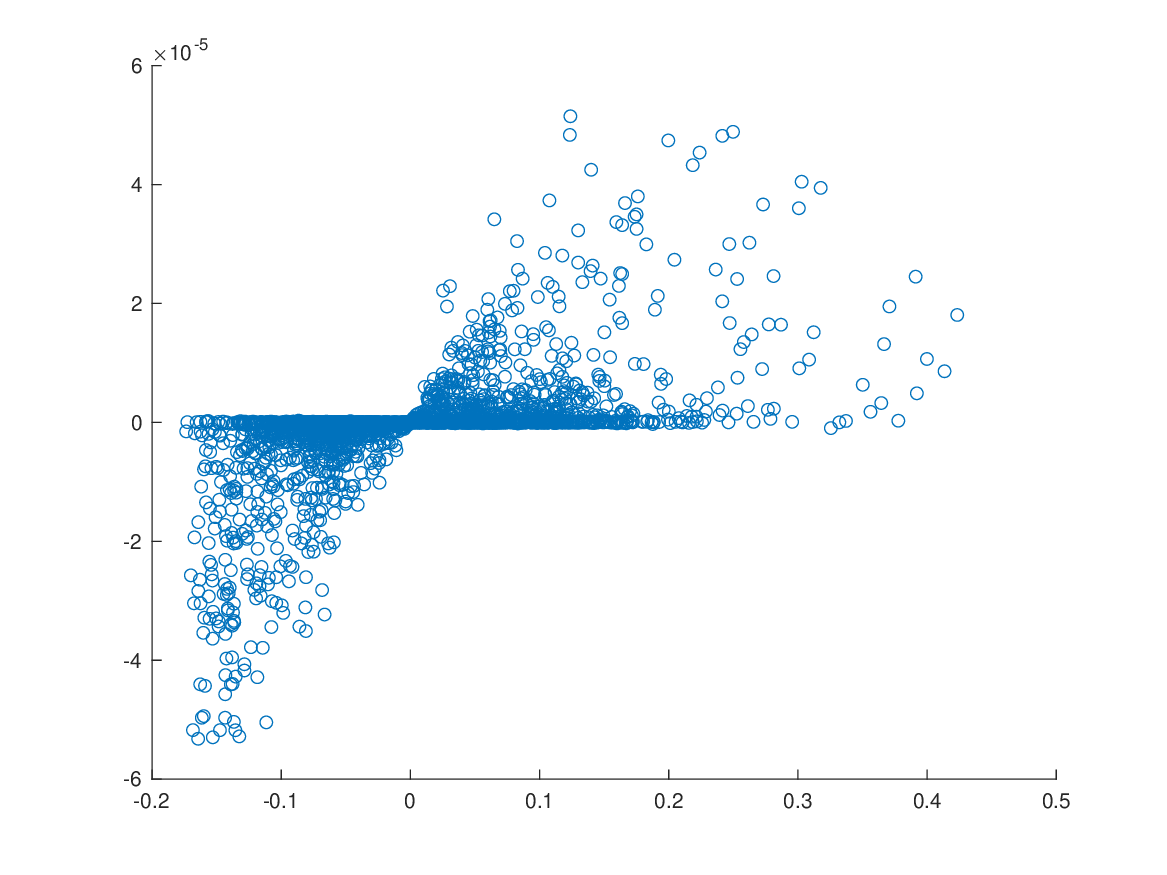}
\end{subfigure}
\begin{subfigure}{.4\textwidth}
    \centering
    \caption{$\gamma^{f,man}$}
    \includegraphics[width=6cm]{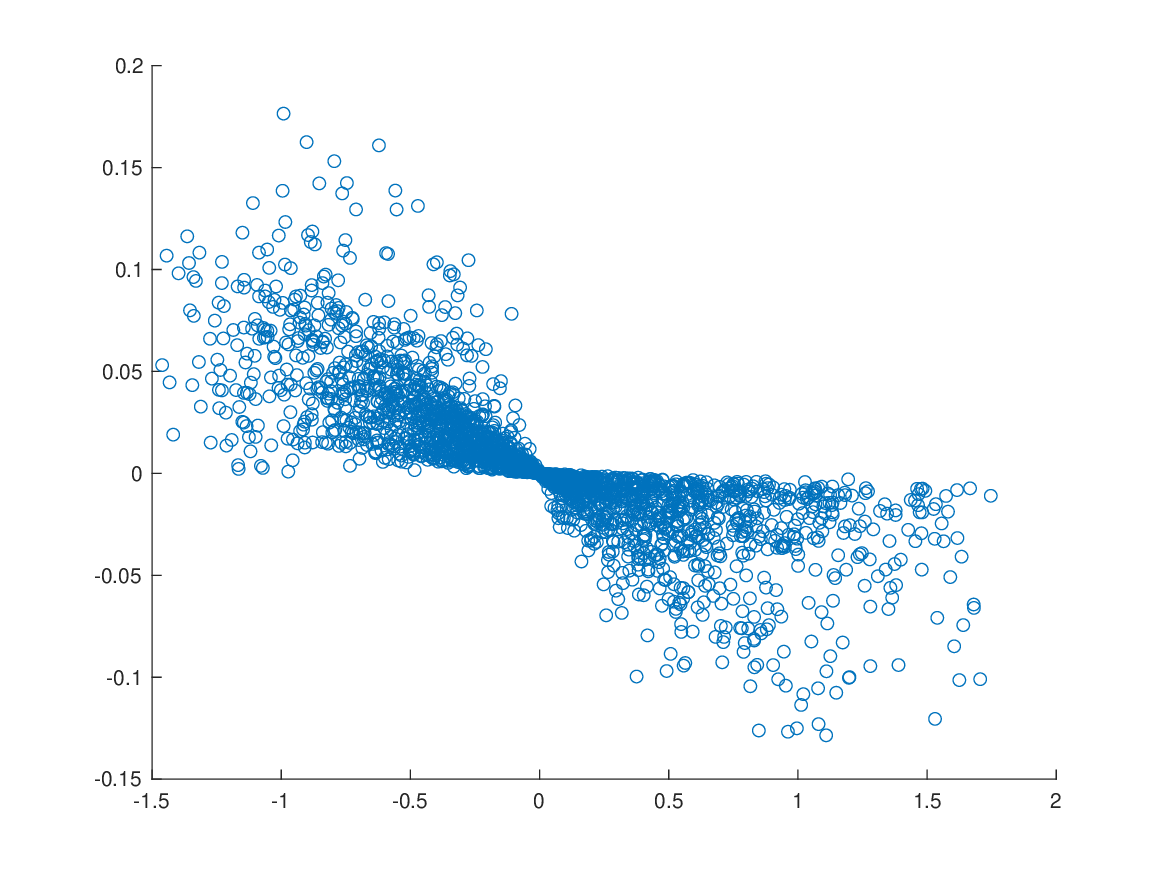}
\end{subfigure}
\begin{subfigure}{.4\textwidth}
    \centering
    \caption{$\gamma^{c,man}$}
    \includegraphics[width=6cm]{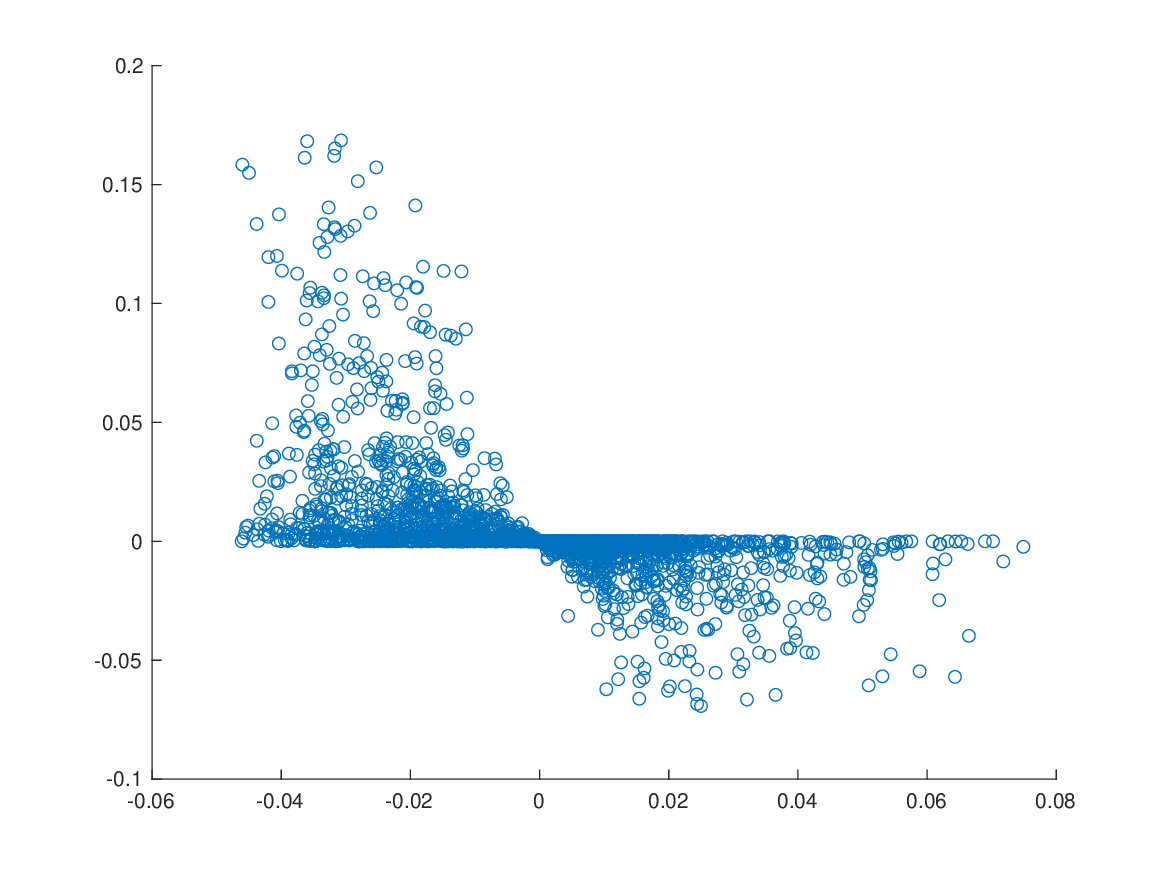}
\end{subfigure}
\begin{center}
\footnotesize{$\Delta$ Ordering Frequency on y-axis, $\gamma^{man}$ on x-axis}
\end{center}
\end{center}
\end{figure}

\begin{figure}[ht]
\begin{center}
\caption{Counterfactual Outcome: Inventory-to-sales Ratio \label{fig: counterfactual_invent}}
\begin{subfigure}{.4\textwidth}
    \centering
    \caption{$\gamma^{h,man}$}
    \includegraphics[width=6cm]{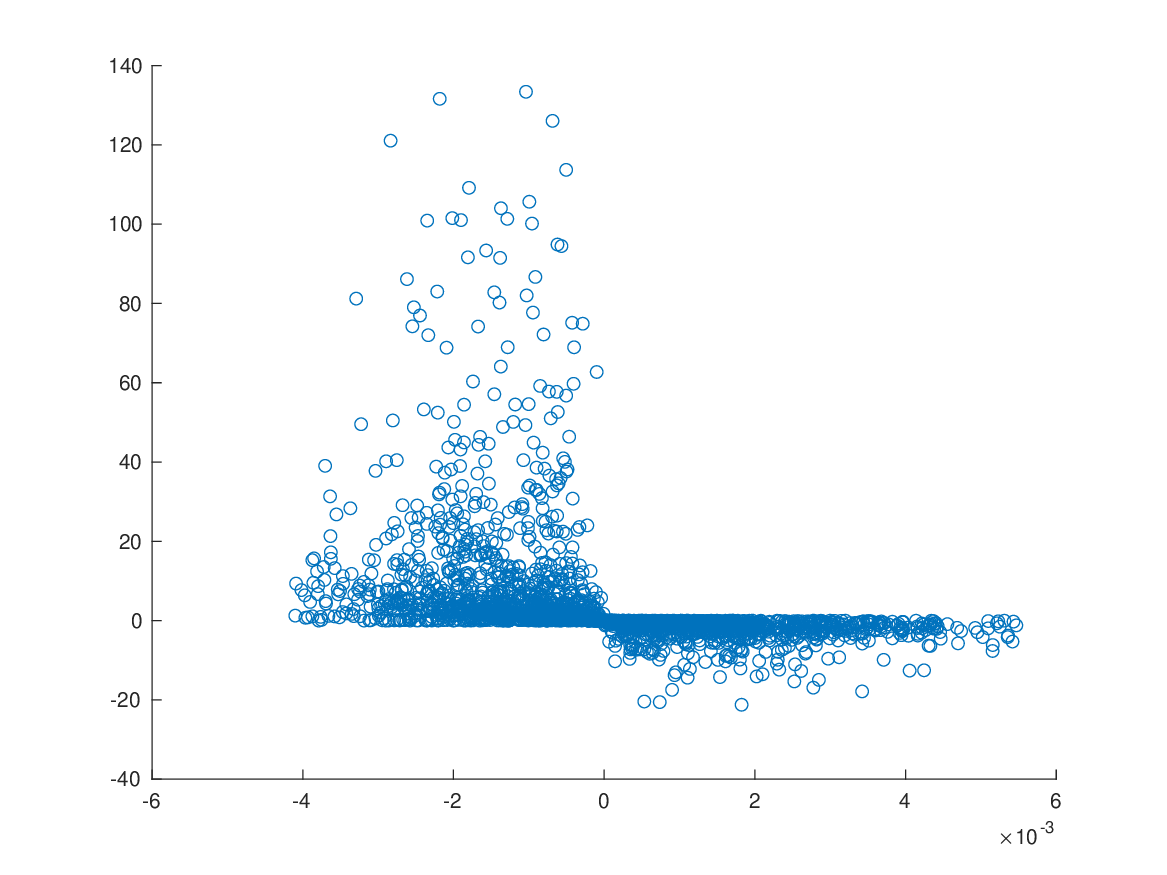}
\end{subfigure}
\begin{subfigure}{.4\textwidth}
    \centering
    \caption{$\gamma^{z,man}$}
    \includegraphics[width=6cm]{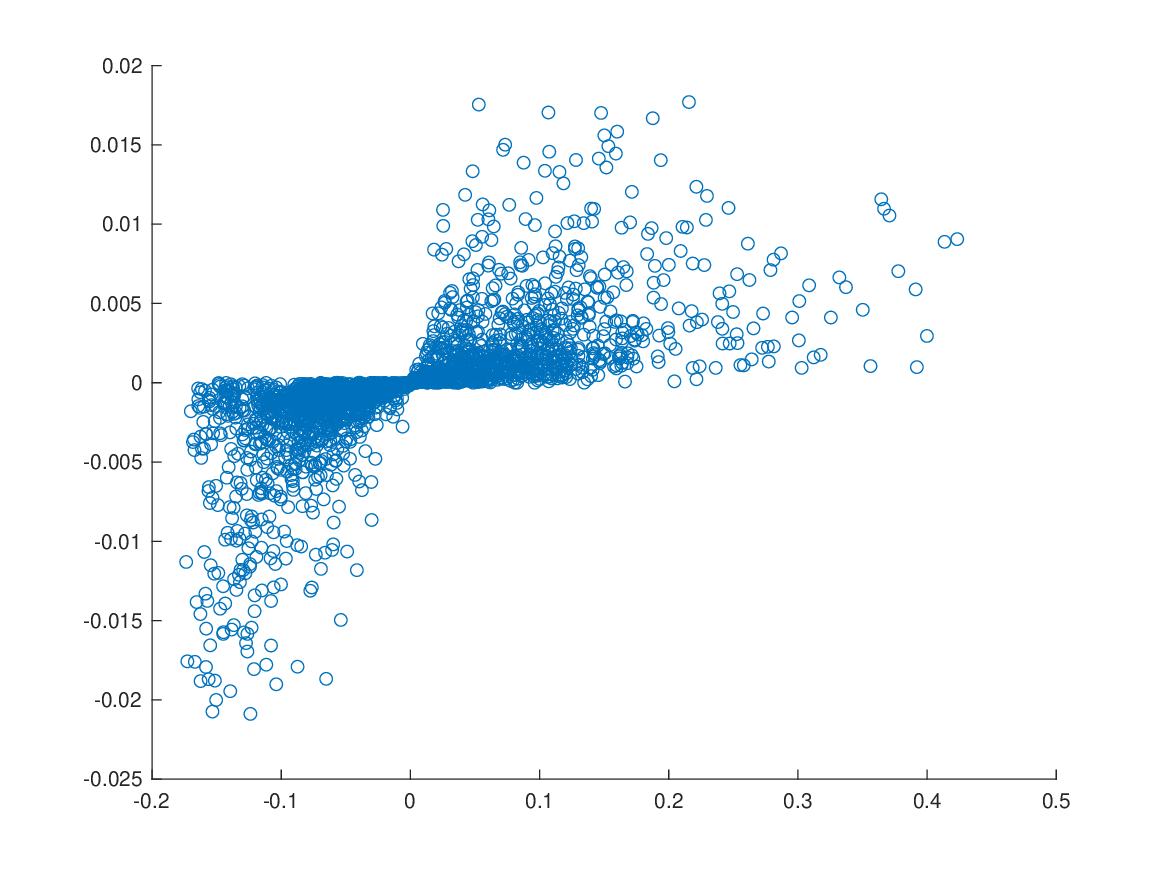}
\end{subfigure}
\begin{subfigure}{.4\textwidth}
    \centering
    \caption{$\gamma^{f,man}$}
    \includegraphics[width=6cm]{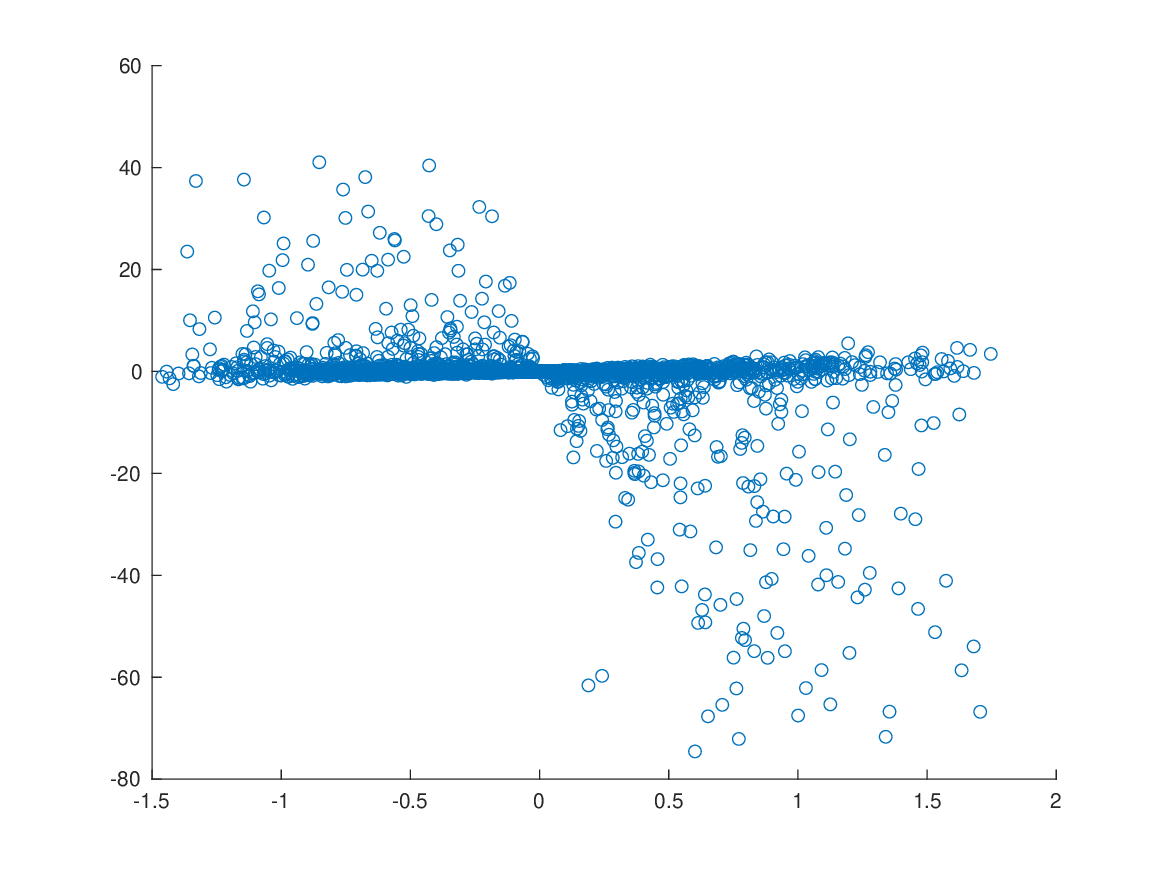}
\end{subfigure}
\begin{subfigure}{.4\textwidth}
    \centering
    \caption{$\gamma^{c,man}$}
    \includegraphics[width=6cm]{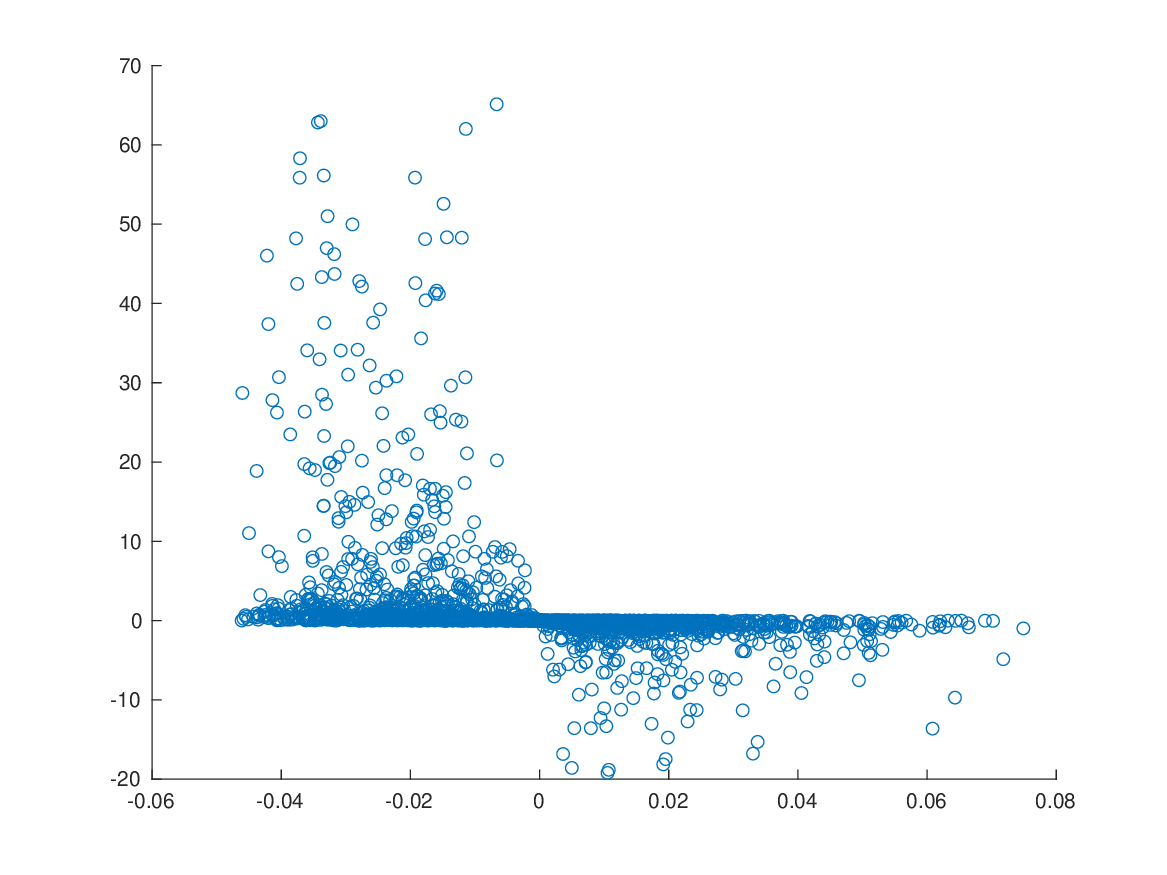}
\end{subfigure}
\begin{center}
\footnotesize{$\Delta$ Inventory to Sales Ratio on y-axis, $\gamma^{man}$ on x-axis}
\end{center}
\end{center}
\end{figure}

Figures \ref{fig: counterfactual_stockout} (for stockout frequency), \ref{fig: counterfactual_order} (for ordering frequency), and \ref{fig: counterfactual_invent} (for inventory-to-sales ratio) summarize the results from these experiments. In each figure, the horizontal axis measures the value of the corresponding parameter $\gamma_{i,j}^{man}$, and the vertical axis measures the difference in the mean value of the outcome variable between the factual and the counterfactual scenario. For instance, in Figure \ref{fig: counterfactual_stockout}(a), the horizontal axis represents $\gamma_{i,j}^{h,man}$, and the vertical axis measures $\Delta SOF_{i,j} = SOF_{i,j}^{factual} - SOF_{i,j}^{counter}$, where $SOF$ is stockout frequency.

Note that the counterfactual experiment of shutting down $\gamma_{i,j}^{h,man}$ to zero is equivalent to a change in parameter $\gamma_{i,j}^{h}$ from the counterfactual value $\gamma_{i,j}^{h,sto}$ to the factual value $\gamma_{i,j}^{h,sto} + \gamma_{i,j}^{h,man}$. Therefore, we can see the cloud of points in, say, Figure \ref{fig: counterfactual_stockout}(a) as the results of many comparative statics exercises, all of them consisting in changes in the value of parameter $\gamma^{h}$. In these figures, there are multiple curves relating a change in $\gamma^{h}$ with a change in the outcome variable because store-products have different values of the other structural parameters. However, each of these figures shows a monotonic relationship between a change in a cost parameter and the corresponding change in an outcome variable. 

\medskip

\medskip

We compare the relationship between parameters and outcomes implied by these figures with the theoretical predictions from the model as depicted in equation (\ref{eq:compstatics_ss}). More specifically, note that $S-s$ is negatively related to the ordering frequency (the larger the $S-s$, the smaller the ordering frequency); $s$ is negatively related to the stockout rate (the larger the $s$, the smaller the stockout rate); and the levels of both $S$ and $s$ are positively related to the inventory to sales ratio (the larger the $S$ and $s$, the larger the ratio). The pattern in our figures is fully consistent with Blinder's theoretical predictions for this class of models.

Figure \ref{fig: counterfactual_stockout} depicts the relationship between cost parameters and the stockout frequency. According to  Blinder's formula, the lower threshold $s$ depends negatively on $\gamma^{h}$ and $\gamma^{f}$, and positively on $\gamma^{z}$, while the effect of $\gamma^{c}$ is ambiguous. Panels (a) to (d) in Figure \ref{fig: counterfactual_stockout} confirm the signs of these effects on the stockout frequency.

\medskip

\medskip

In Figure \ref{fig: counterfactual_order}, we present the relationship between cost parameters and the ordering frequency. Blinder's formula says that $S-s$ depends negatively on $\gamma^{h}$ and positively on $\gamma^{f}$. Panels (a) and (c) confirm the sign of these effects on the ordering frequency. According to Blinder, the sign of the effects of $\gamma^{z}$ and $\gamma^{c}$ on ordering frequency is ambiguous because they affect the two thresholds $S$ and $s$ in the same direction. In Panel (b), we find a positive relationship between the stockout cost $\gamma^{z}$ and ordering frequency. Panel (d) shows that the frequency of placing an order falls when the unit ordering cost increases.

\clearpage

Figure \ref{fig: counterfactual_invent} illustrates the relationship between cost parameters and the inventory-to-sales ratio. Blinder's formula establishes that the two thresholds $S$ and $s$ depend negatively on $\gamma^{h}$ and positively $\gamma^{z}$. Panels (a) and (b) confirm these signs for the inventory-to-sales ratio. However, according to Blinder's formula, the sign of the effects of $\gamma^{f}$ and $\gamma^{c}$ on the inventory-to-sales ratio is ambiguous. Panels (c) and (d) show negative effects of $\gamma^{f}$ and $\gamma^{c}$ on the inventory-to-sales ratio.

\clearpage

\begin{table}[ht]
\begin{center}
\caption{Removing the Manager Component of Cost Parameters \label{tab: counterfactual_inv}}
\resizebox{\textwidth}{!}{\begin{tabular}{r|cccccc}
\hline \hline 
& \multicolumn{6}{c}{\textit{Parameters Shut Down}}
\\
&
\multicolumn{1}{c}{\textit{None}}
&
\multicolumn{1}{c}{$\gamma^{h}$}
&
\multicolumn{1}{c}{$\gamma^{z}$}
&
\multicolumn{1}{c}{$\gamma^{f}$}
&
\multicolumn{1}{c}{$\gamma^{c}$}
&
\multicolumn{1}{c}{\textit{All}}
\\ \hline
{\textbf{Store-product level inventory outcomes}}
& \textit{Mean} & \textit{Mean} 
& \textit{Mean} & \textit{Mean} 
& \textit{Mean} & \textit{Mean}
\\
& (\textit{st.dev}) & (\textit{st.dev}) 
& (\textit{st.dev}) & (\textit{st.dev}) 
& (\textit{st.dev}) & (\textit{st.dev})
\\ \hline
\\
\textit{Stockout Frequency} 
& 0.0003 & 0.0003 & 0.0003 & 0.0003 & 0.0003 & 0.0003 \\
& (0.0004) & (0.0003) & (0.0004) & (0.0004) & (0.0004) & (0.0003) \\
\\
\textit{Ordering Frequency} 
& 0.1776 & 0.1770 & 0.1776 & 0.1714 & 0.1729  & 0.1618 \\
& (0.1329) & (0.1353) & (0.1329) & (0.1260) & (0.1313) & (0.1240) \\
\\
\textit{Inventory to Sales Ratio} 
& 26.8373 & 23.3465 & 26.8376 & 27.8804 & 25.7753 & 22.1358 \\
& (21.1587) & (13.3625) & (21.1589) & (25.1945) & (19.1622) & (11.8119) \\
\\
\textit{Inventory to Sales Ratio After Order}
& 31.6917 & 26.7060 & 31.6920 & 33.7313 & 29.6416 & 24.2525 \\ 
& (29.7575) & (17.9471) & (29.7576) & (36.5440) & (25.5906) & (13.9094) \\
\\
\textit{Inventory to Sales Ratio Before Order}
& 22.6063 & 18.3920 & 22.6066 & 24.6034 & 20.9836 & 16.6467 \\ 
& (22.0222) & (12.2635) & (22.0224) & (28.0980) & (18.7917) & (8.7038) \\
\\
\textit{Change in Total Inventory Cost (\%)}
& -- & -4.1913 & 0.0005 & 1.5256 & -4.5866 & -12.0838 \\  
& & (11.7293) & (0.0048) & (24.1220) & (11.4978) & (17.5204) \\
\\
\hline \hline
\end{tabular}}
\end{center}
\end{table}

\medskip

It is of interest to measure the average effect across stores and products of shutting down the store manager-specific component in costs. Table \ref{tab: counterfactual_inv} presents these average effects for each of the four cost parameters and for the combination of the four.\footnote{Note that, by construction, the manager component $\boldsymbol{\gamma}^{man}_{i,j}$ has mean zero and is orthogonal to the store component $\boldsymbol{\gamma}^{sto}_{i,j}$. Therefore, if the model implied a linear relationship between outcome variables and structural parameters, then the average effect of shutting down the residual component would be zero. For the same reason, a first-order linear approximation to this average effect is zero. However, the model implies a nonlinear relationship between outcomes and structural parameters such that it is a relevant empirical question to look at these average effects. In fact, we find that the effect is not negligible at all.} Removing the manager component in all four inventory costs generates a decrease in the mean ordering frequency of $1.6$ percentage points, from $17.8\%$ to $16.2\%$; a decrease in the inventory-to-sales ratio of $4.7$ days of average sales, from $26.8$ to $22.1$ days; a decrease in the lower $s$ threshold of $6$ days, from $22.6$ to $16.6$ days; and a decrease in the $S-s$ gap of $1.5$ days, from $9.1$ to $7.6$ days.

Store managers' idiosyncratic perception of costs has a substantial effect on inventory management at the aggregate firm level. It entails a $6$-day decrease in waiting time between two orders, an increase in the average order amount of $1.5$ days of average sales, and a $21\%$ increase in the inventory-to-sales ratio, but a negligible effect on the frequency of stockouts. Accordingly, if this idiosyncratic component is a biased perception, then it has a substantial negative impact on the firm’s profit as it increases storage and ordering costs with almost no effect on stockouts and revenue. The bottom row in Table \ref{tab: counterfactual_inv} presents the effect of removing $\boldsymbol{\gamma}^{man}_{i,j}$ on total inventory management cost calculated using $\boldsymbol{\gamma}^{sto}_{i,j}$ but not $\boldsymbol{\gamma}^{man}_{i,j}$. We find that, on average, this cost declines by $12.1\%$. This substantial effect plays an important role in the counterfactual experiment on centralization that we present in the next section.

\subsection{Centralizing inventory decision-making}

We now address the main question that motivates this paper: would the LCBO retail chain benefit from managing the stores' inventories at the headquarter level, as opposed to allowing heterogeneous store managers to have autonomy in their inventory decisions? To answer this question, we need to establish some conditions on the headquarters' information about store-level demand, inventories, and cost parameters. The experiments that we present below are based on the following conditions.

First, based on the institutional details we describe in Section \ref{sec:lcbo}, we consider that headquarters process store-product level transactions data with a one-week delay. Transmission of information from stores to headquarters occurs in real time, without any substantial delay. However, it takes time to process that information to generate headquarters demand predictions and ordering recommendations. Though a fully automated inventory management system is possible, human supervision can add value by accounting for soft information (\citeauthor{cimini_lagorio_2019}, \citeyear{cimini_lagorio_2019}). Accordingly, in the counterfactual centralized system, we replace state variable $Q^{[-7,-1]}_{t}$ with the one-week lag of this variable, i.e., $Q^{[-14,-8]}_{t}$.

Second, to compare profits between the centralized and decentralized structures, we must take a stance on what are the "true" cost parameters. We assume that the true cost parameters are $\boldsymbol{\gamma}^{sto}_{i,j}$ which are determined by store and location characteristics. Under the centralized system, the headquarters know these costs and take inventory decisions for every store based on these costs. In contrast, we interpret $\boldsymbol{\gamma}^{man}_{i,j}$ as store managers' behavioral biases and not as "true" costs. Under the decentralized system, store managers make decisions as if the cost parameters were $\boldsymbol{\gamma}^{sto}_{i,j}+\boldsymbol{\gamma}^{man}_{i,j}$, but our measure of their profits is based only on $\boldsymbol{\gamma}^{sto}_{i,j}$. The evaluation of profits under this assumption provides an upper bound for the (profit) gains from centralization. Alternatively, we could assume that $\boldsymbol{\gamma}^{man}_{i,j}$ is a true component of profit that is known to the store manager but unknown to the headquarters. This alternative assumption would provide a lower bound for the gains from centralization.

Based on these assumptions, this counterfactual experiment measures the following trade-off in the choice between centralized and decentralized inventory management. A negative aspect of decentralization is that store managers have different skills and behavioral biases as captured by the idiosyncratic components $\boldsymbol{\gamma}^{man}_{i,j}$. These biases should have a negative effect on LCBO profits. The positive aspect of decentralization is that store managers have just-in-time information about demand, sales, and inventories, while the firm's headquarters process this information with one week delay. This just-in-time information should have a positive effect on LCBO profits.

\medskip

\begin{table}[ht]
\caption{Decentralized vs. Centralized Profits: Average Daily Profit Per-Store Per-Product \label{tab: centralization}}
\centering
\resizebox{\textwidth}{!}{\begin{tabular}{r|cccccc}
\hline \hline 
& \textit{Mean} & \textit{Pct. 10\%} 
& \textit{Pct. 25\%} & \textit{Median} 
& \textit{Pct. 75\%} & \textit{Pct. 90\%}
\\ \hline
\\
\textit{Centralized Solution (\$)}
& 53.72 & 10.54 & 20.29 & 44.06 & 80.34 & 109.21
\\
\\
\textit{Decentralized Solution (\$)} 
& 52.81 & 10.16 & 19.64 & 43.34 & 79.66 & 108.04
\\
\\
\textit{Gains in Profit from Decentralization (\$)} 
& -0.91 & -3.10 & -1.58 & -0.66 & -0.05 & 0.80
\\
\\
\textit{Gains in Profit from Decentralization (\%)} 
& -1.97 & -6.09 & -3.81 & -2.09 & -0.19 & 1.85
\\
\\
\textit{Change in Inventory Cost from Decentralization (\%)}
& 22.99 & -8.06 & -2.59 & 3.72 & 21.99 & 62.66 \\ 
\\
\hline \hline
\multicolumn{7}{l}{\textit{$1\%$ change in profit per store-product is approximately $\$17$ million in total annual profit for LCBO.}}
\end{tabular}}
\end{table}

\bigskip

\begin{figure}[ht]
\caption{Change in Daily Profit From Decentralization (\$) \label{fig: centralization profits}}
    \centering
    \includegraphics[width=10cm]{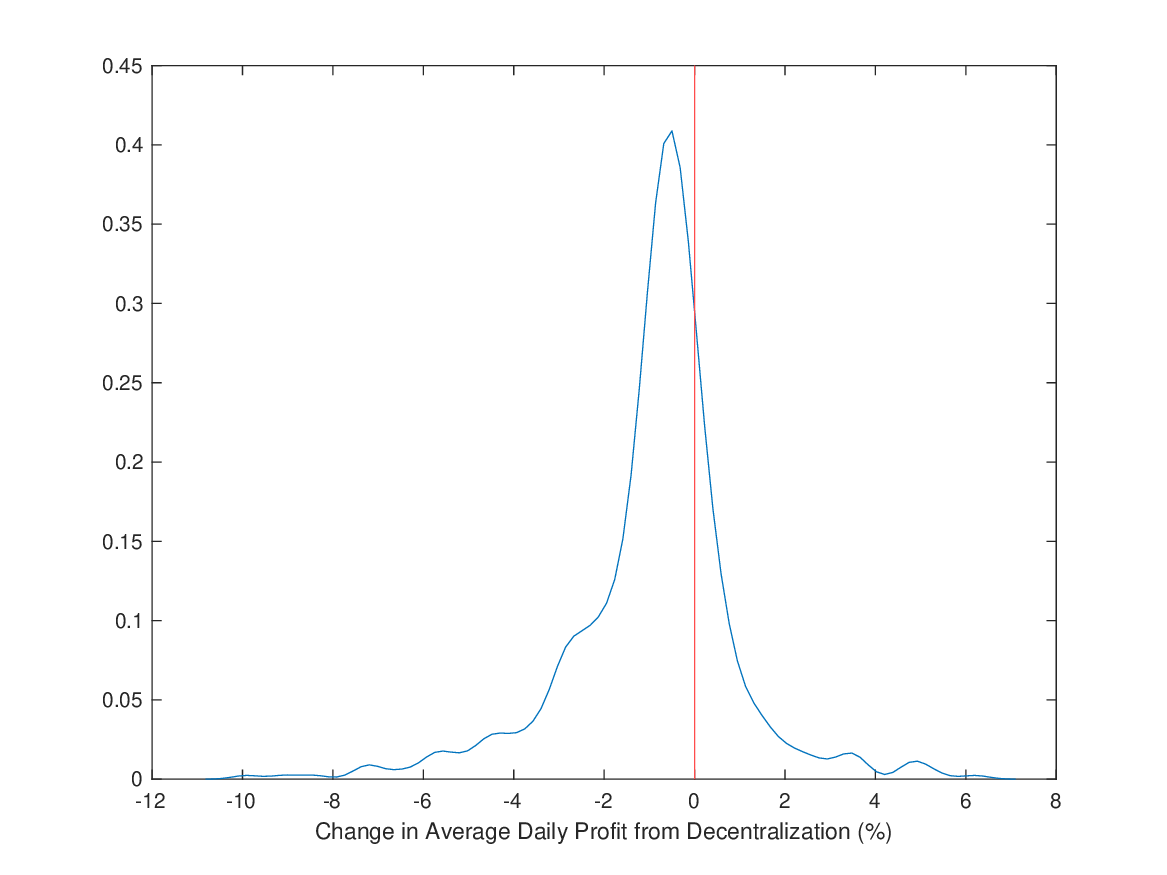}
\end{figure}

\clearpage

\begin{figure}[ht]
\caption{Change in Daily Inventory Cost From Decentralization (\%) \label{fig: centralization costs}}
    \centering
    \includegraphics[width=10cm]{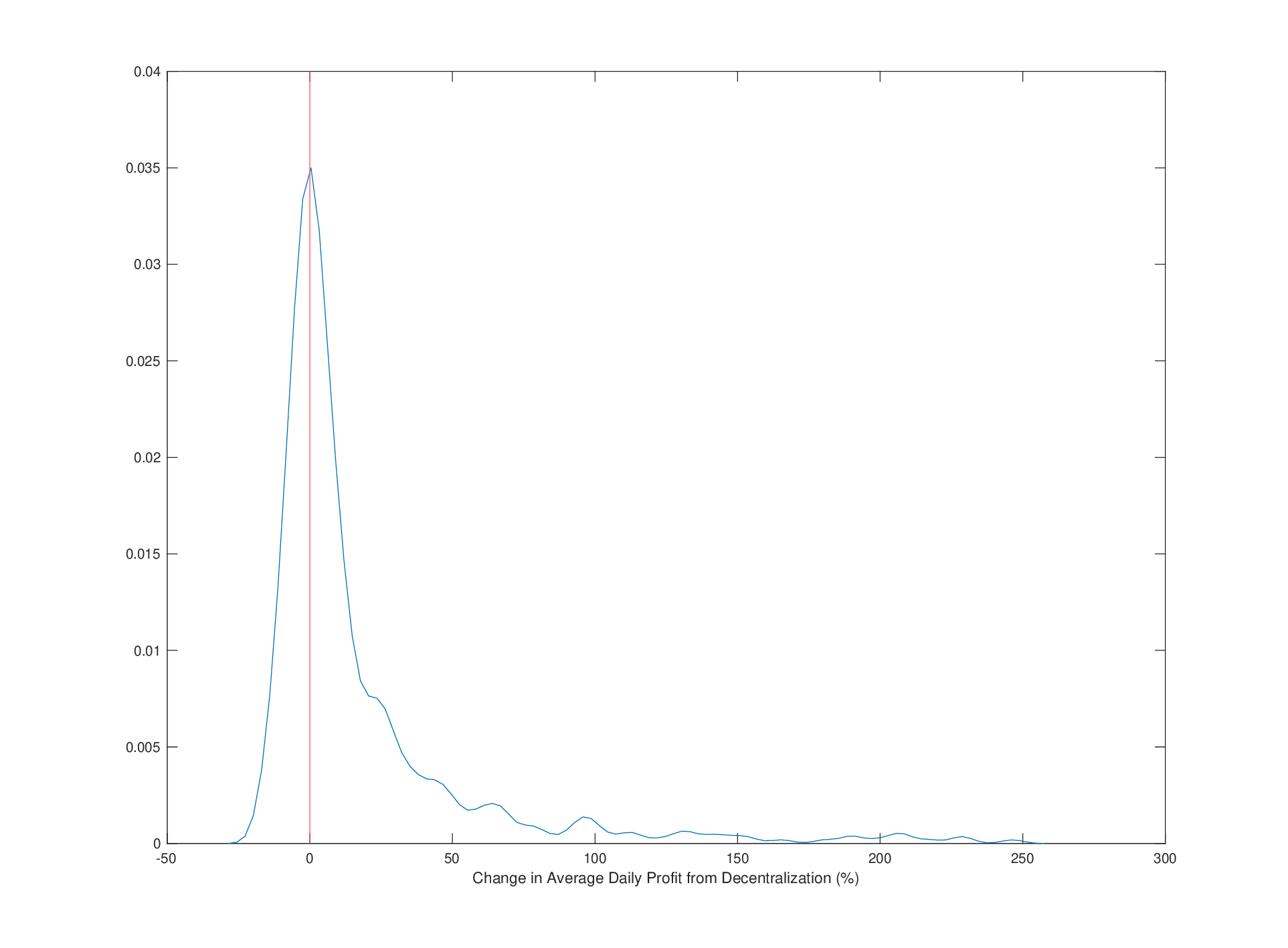}
\end{figure}

We depict the results of this experiment in Table \ref{tab: centralization}, and Figures \ref{fig: centralization profits} and \ref{fig: centralization costs}. Similarly as for the counterfactuals in section \ref{sec:counterfactuals_shutdown_res}, we evaluate the effects using the ergodic distributions of the state variables under the factual and counterfactual scenarios. Table \ref{tab: centralization} presents means, medians, and several percentiles for the profit per store-product under the centralized and decentralized systems and for the gains from decentralization. To have a better perspective of the implications of these effects, it is useful to take into account that a $1\%$ change in profit per store-product represents approximately $\$17$ million in total annual profit for LCBO in year 2012.\footnote{According to the 2012-2013 LCBO annual report, the annual profit (net income) of the company was $\$ 1.7$ billion. Therefore, $1\%$ of this profit is $\$17$ million.} At the aggregate level, decentralization has a negative impact on LCBO profits. It implies a $2\%$ decline in profits, which represents $\$34$ million in annual profits for the retail chain. The effect on the median store is also negative: $-2.1\%$. This relatively modest effect is the result of combining two large effects with opposite signs. The one-week delay in the processing of information in the centralized system has a non-negligible negative impact on profits at every store. However, this negative effect of the centralized system is more than compensated by the large increase in profits due to reducing ordering and storage costs when removing store managers' biased perceptions of costs. This is illustrated in the bottom row of Table \ref{tab: centralization}: on average, decentralization increases total inventory costs by $23\%$.

The evidence on the mean and median effects in Table \ref{tab: centralization} is not necessarily sufficient for a retail chain to adopt a centralized inventory management system. A retail chain may need to assess the distributional effects of the gains/losses across its stores before adopting a substantial organizational change, and not simply rely on the average effect. Table \ref{tab: centralization} and Figure \ref{fig: centralization profits} show significant heterogeneity in the impact of decentralization, with a considerable amount of stores benefiting from the decentralized structure: the $90^{\text{th}}$ percentile store has a gain in profit of $1.8\%$, which is not negligible. Therefore, although centralization would generate positive gains in total profits for the retail chain relative to the existing decentralized structure, the distributional effects of these gains are important to assess.

Figure \ref{fig: centralization costs} provides a closer look at the heterogeneous effect from (de)centralization. It presents the empirical distribution of the percentage change in average daily inventory costs. The median of this distribution presents a positive increase in costs (i.e., $3.7\%$), but most striking is the long right tail of this distribution, which implies a $23\%$ increase in average inventory costs.

\section{Conclusion \label{sec:conclusion}}

Retail chains are complex organizations with various divisions and teams, each having its own decision-making authority. Store managers, in particular, play a crucial role within certain retail chains. These managers have the advantage of collecting and processing timely information specific to their individual stores. This store-level information is more consistent and manageable compared to information at the chain-wide level. However, the transfer and processing of this information from stores to headquarters can introduce delays ranging from days to weeks, which can negatively impact decision-making and overall profitability. 

On the other hand, store managers exhibit heterogeneous skills, motivations, and levels of effort. A centralized decision-making system that selectively utilizes the most competent managers within the organization can help mitigate the negative effects stemming from the variation in managers' skills.

In this paper, we examine the trade-off between centralized and decentralized decision-making in the context of inventory management within a large retail chain. Leveraging a unique dataset containing daily information on inventories, sales, prices, and stockouts at the individual store and product level, we estimate a dynamic structural model to capture store managers' inventory decisions. Using revealed preference as a guiding principle, we obtain separate estimates for each store and product regarding four cost parameters: per unit inventory holding cost, stockout cost, fixed ordering costs, and per unit ordering costs. Our analysis reveals significant heterogeneity across stores in these cost parameters. While observable store and location characteristics account for part of this heterogeneity, a substantial portion can be attributed to idiosyncratic information and perceptions of store managers.

We utilize the estimated model to conduct a counterfactual experiment, evaluating the impact of centralizing inventory management at LCBO. In this experiment, we assume that the idiosyncratic cost component associated with store managers represents a behavioral bias rather than true costs. This assumption allows us to provide an upper-bound estimate for the gains from centralization. Our findings indicate that a centralized inventory management system would lead to a modest $2\%$ increase in LCBO's annual profit. This outcome arises from the combination of two opposing effects. The negative effect on profits due to the loss of just-in-time information from store managers is outweighed by the significant reduction in ordering and storage costs resulting from the elimination of behavioral biases and skill heterogeneity among store managers (averaging at $23\%$ reduction overall and $3.7\%$ reduction for the median store). 

Furthermore, the effects of centralization are highly heterogeneous across stores within the retail chain, with a substantial number of stores experiencing significant losses from adopting centralization. This distributional effect has important implications for the decision-making process when considering organizational changes aimed at maximizing overall company profit (\citeauthor{inderst_muller_2007}, \citeyear{inderst_muller_2007}).

Our empirical findings highlight the advantages of a hybrid inventory management system that combines decentralized decision-making with centralized control. By assigning decision rights to high-skilled store managers and utilizing a centralized system for stores where skill levels are lower, we can eliminate subjective biases while retaining the benefits of just-in-time local information for some of the stores. The structural model presented in this paper provides a useful tool for determining the specific allocation of decision rights across stores.

\newpage

\baselineskip 18pt

\bibliography{references}

\begin{thebibliography}{39}
\providecommand{\natexlab}[1]{#1}
\providecommand{\url}[1]{\texttt{#1}}
\expandafter\ifx\csname urlstyle\endcsname\relax
  \providecommand{\doi}[1]{doi: #1}\else
  \providecommand{\doi}{doi: \begingroup \urlstyle{rm}\Url}\fi

\bibitem[Adda and Cooper(2000)]{adda_cooper_2000}
J.~Adda and R.~Cooper.
\newblock Balladurette and juppette: A discrete analysis of scrapping
  subsidies.
\newblock \emph{Journal of Political Economy}, 108\penalty0 (4):\penalty0
  778--806, 2000.

\bibitem[Aghion et~al.(2021)Aghion, Bloom, Lucking, Sadun, and
  Van-Reenen]{aghion_bloom_2021}
P.~Aghion, N.~Bloom, B.~Lucking, R.~Sadun, and J.~Van-Reenen.
\newblock Turbulence, firm decentralization, and growth in bad times.
\newblock \emph{American Economic Journal: Applied Economics}, 13\penalty0
  (1):\penalty0 133--69, 2021.

\bibitem[Aguirregabiria(1999)]{aguirregabiria_1999}
V.~Aguirregabiria.
\newblock The dynamics of markups and inventories in retailing firms.
\newblock \emph{The Review of Economic Studies}, 66\penalty0 (2):\penalty0
  275--308, 1999.

\bibitem[Aguirregabiria and Magesan(2020)]{aguirregabiria_magesan_2020}
V.~Aguirregabiria and A.~Magesan.
\newblock Identification and estimation of dynamic games when players’
  beliefs are not in equilibrium.
\newblock \emph{The Review of Economic Studies}, 87\penalty0 (2):\penalty0
  582--625, 2020.

\bibitem[Aguirregabiria and Mira(2002)]{aguirregabiria_mira_2002}
V.~Aguirregabiria and P.~Mira.
\newblock Swapping the nested fixed point algorithm: A class of estimators for
  discrete markov decision models.
\newblock \emph{Econometrica}, 70\penalty0 (4):\penalty0 1519--1543, 2002.

\bibitem[Aguirregabiria et~al.(2016)Aguirregabiria, Ershov, and
  Suzuki]{aguirregabiria_ershov_2016}
V.~Aguirregabiria, D.~Ershov, and J.~Suzuki.
\newblock Estimating the effects of deregulation in the ontario wine retail
  market.
\newblock \emph{Working Paper}, 2016.

\bibitem[Anderson et~al.(2006)Anderson, Fitzsimons, and
  Simester]{anderson_fitzsimons_2006}
E.~Anderson, G.~Fitzsimons, and D.~Simester.
\newblock Measuring and mitigating the costs of stockouts.
\newblock \emph{Management Science}, 52\penalty0 (11):\penalty0 1751--1763,
  2006.

\bibitem[Arrow et~al.(1951)Arrow, Harris, and Marschak]{arrow_harris_1951}
J.~Arrow, T.~Harris, and J.~Marschak.
\newblock Optimal inventory policy.
\newblock \emph{Econometrica}, 19\penalty0 (3):\penalty0 250--272, 1951.

\bibitem[Arthur and Vassilvitskii(2007)]{arthur_vassilvitskii_2007}
D.~Arthur and S.~Vassilvitskii.
\newblock k-means++: The advantages of careful seeding.
\newblock \emph{Proceedings of the 18th ACM-SIAM Symposium on Discrete
  Algorithms (SODA)}, page 1027–1035, 2007.

\bibitem[Attanasio(2000)]{attanasio_2000}
O.~Attanasio.
\newblock Consumer durables and inertial behaviour: Estimation and aggregation
  of (s, s) rules for automobile purchases.
\newblock \emph{The Review of Economic Studies}, 67\penalty0 (4):\penalty0
  667--696, 2000.

\bibitem[Bertola et~al.(2005)Bertola, Guiso, and
  Pistaferri]{bertola_guiso_2005}
G.~Bertola, L.~Guiso, and L.~Pistaferri.
\newblock Uncertainty and consumer durables adjustment.
\newblock \emph{Review of Economic Studies}, 72\penalty0 (4):\penalty0
  973--1007, 2005.

\bibitem[Blinder(1981)]{blinder_1981}
A.~Blinder.
\newblock Retail inventory behavior and business fluctuations.
\newblock \emph{Brooking Papers of Economic Activity}, 2\penalty0 (1):\penalty0
  443--505, 1981.

\bibitem[Bray et~al.(2019)Bray, Yao, Duan, and Huo]{bray_yao_2019}
R.~Bray, Y.~Yao, Y.~Duan, and J.~Huo.
\newblock Ration gaming and the bullwhip effect.
\newblock \emph{Operations Research}, 67\penalty0 (2):\penalty0 453--467, 2019.

\bibitem[Cho and Rust(2010)]{cho_rust_2010}
S.~Cho and J.~Rust.
\newblock The flat rental puzzle.
\newblock \emph{The Review of Economic Studies}, 77\penalty0 (2):\penalty0
  560--594, 2010.

\bibitem[Cimini et~al.(2019)Cimini, Lagorio, Pirola, and
  Pinto]{cimini_lagorio_2019}
C.~Cimini, A.~Lagorio, F.~Pirola, and R.~Pinto.
\newblock Exploring human factors in logistics 4.0: Empirical evidence from a
  case study.
\newblock \emph{IFAC-PapersOnLine}, 52\penalty0 (13):\penalty0 2183--2188,
  2019.

\bibitem[DellaVigna and Gentzkow(2019)]{dellavigna_gentzkow_2019}
S.~DellaVigna and M.~Gentzkow.
\newblock Uniform pricing in us retail chains.
\newblock \emph{NBER Working Paper 23996}, 2019.

\bibitem[Denardo(1981)]{denardo_1981}
E.~Denardo.
\newblock \emph{Dynamic Programming: Models and Applications}.
\newblock Prentice Hall PTR, USA, 1981.
\newblock ISBN 0132215071.

\bibitem[Doraszelski et~al.(2018)Doraszelski, Lewis, and
  Pakes]{doraszelski_lewis_2018}
U.~Doraszelski, G.~Lewis, and A.~Pakes.
\newblock Just starting out: Learning and equilibrium in a new market.
\newblock \emph{American Economic Review}, 108\penalty0 (3):\penalty0 565--615,
  2018.

\bibitem[Eberly(1994)]{eberly_1994}
J.~Eberly.
\newblock Adjustment of consumers' durables stocks: Evidence from automobile
  purchases.
\newblock \emph{Journal of Political Economy}, 102\penalty0 (3):\penalty0
  403--436, 1994.

\bibitem[Ellison et~al.(2018)Ellison, Snyder, and Zhang]{ellison_snyder_2018}
S.~Ellison, C.~Snyder, and H.~Zhang.
\newblock Costs of managerial attention and activity as a source of sticky
  prices: Structural estimates from an online market.
\newblock \emph{NBER Working Paper 24680}, 2018.

\bibitem[Goldfarb and Xiao(2011)]{goldfarb_xiao_2011}
A.~Goldfarb and M.~Xiao.
\newblock Who thinks about the competition? managerial ability and strategic
  entry in us local telephone markets.
\newblock \emph{American Economic Review}, 101\penalty0 (7):\penalty0 3130--61,
  2011.

\bibitem[Gu and Koenker(2017)]{gu_koenker_2017}
J.~Gu and R.~Koenker.
\newblock Empirical bayesball remixed: Empirical bayes methods for longitudinal
  data.
\newblock \emph{Journal of Applied Econometrics}, 32\penalty0 (3):\penalty0
  575--599, 2017.

\bibitem[Hadley and Whitin(1963)]{hadley_whitin_1963}
G.~Hadley and T.~Whitin.
\newblock \emph{Analysis of inventory systems}.
\newblock Prentice-Hall, 1963.

\bibitem[Hall and Rust(2000)]{hall_rust_2000}
G.~Hall and J.~Rust.
\newblock An empirical model of inventory investment by durable commodity
  intermediaries.
\newblock \emph{Carnegie-Rochester Conference Series on Public Policy},
  52:\penalty0 171--214, 2000.

\bibitem[Hitsch et~al.(2021)Hitsch, Horta{\c{c}}su, and
  Lin]{hitsch_hortacsu_2021}
G.~Hitsch, A.~Horta{\c{c}}su, and X.~Lin.
\newblock Prices and promotions in us retail markets.
\newblock \emph{Quantitative Marketing and Economics}, pages 1--80, 2021.

\bibitem[Horta{\c{c}}su and Puller(2008)]{hortacsu_puller_2008}
A.~Horta{\c{c}}su and S.~Puller.
\newblock Understanding strategic bidding in multi-unit auctions: a case study
  of the texas electricity spot market.
\newblock \emph{The RAND Journal of Economics}, 39\penalty0 (1):\penalty0
  86--114, 2008.

\bibitem[Horta{\c{c}}su et~al.(2019)Horta{\c{c}}su, Luco, Puller, and
  Zhu]{hortacsu_luco_2019}
A.~Horta{\c{c}}su, F.~Luco, S.~Puller, and D.~Zhu.
\newblock Does strategic ability affect efficiency? evidence from electricity
  markets.
\newblock \emph{American Economic Review}, 109\penalty0 (12):\penalty0
  4302--42, 2019.

\bibitem[Horta{\c{c}}su et~al.(2021)Horta{\c{c}}su, Natan, Parsley, Schwieg,
  and Williams]{hortacsu_natan_2021}
A.~Horta{\c{c}}su, O.~Natan, H.~Parsley, T.~Schwieg, and K.~Williams.
\newblock Organizational structure and pricing: Evidence from a large us
  airline.
\newblock \emph{National Bureau of Economic Research, Working Paper}, 29508,
  2021.

\bibitem[Huang et~al.(2022)Huang, Ellickson, and Lovett]{huang_ellickson_2022}
Y.~Huang, P.~B. Ellickson, and M.~J. Lovett.
\newblock Learning to set prices.
\newblock \emph{Journal of Marketing Research}, 59\penalty0 (2):\penalty0
  411--434, 2022.

\bibitem[Inderst et~al.(2007)Inderst, M{\"u}ller, and
  W{\"a}rneryd]{inderst_muller_2007}
R.~Inderst, H.~M. M{\"u}ller, and K.~W{\"a}rneryd.
\newblock Distributional conflict in organizations.
\newblock \emph{European Economic Review}, 51\penalty0 (2):\penalty0 385--402,
  2007.

\bibitem[Kryvtsov and Midrigan(2013)]{kryvtsov_midrigan_2013}
O.~Kryvtsov and V.~Midrigan.
\newblock Inventories, markups, and real rigidities in menu cost models.
\newblock \emph{Review of Economic Studies}, 80\penalty0 (1):\penalty0
  249--276, 2013.

\bibitem[{Liquor Control Board of Ontario}(2012)]{lcbo_2011_2012}
{Liquor Control Board of Ontario}.
\newblock Annual report 2011-2012, 2012.
\newblock
  \url{https://www.lcbo.com/content/dam/lcbo/corporate-pages/about/pdf/2011_2012.pdf}.

\bibitem[{Liquor Control Board of Ontario}(2013)]{lcbo_2012_2013}
{Liquor Control Board of Ontario}.
\newblock Annual report 2012-2013, 2013.
\newblock
  \url{https://www.lcbo.com/content/dam/lcbo/corporate-pages/about/pdf/LCBO_AR12-13-english.pdf}.

\bibitem[{Liquor Control Board of Ontario}(2016)]{lcbo_2016}
{Liquor Control Board of Ontario}.
\newblock An introduction to web based ordering \& sos (store order services),
  2016.
\newblock LCBO PSM Trade Symposium.

\bibitem[Miravete et~al.(2020)Miravete, Seim, and Thurk]{miravete_seim_2020}
E.~Miravete, K.~Seim, and J.~Thurk.
\newblock One markup to rule them all: Taxation by liquor pricing regulation.
\newblock \emph{American Economic Journal: Microeconomics}, 12\penalty0
  (1):\penalty0 1--41, 2020.

\bibitem[Puterman(2014)]{puterman_2014}
M.~Puterman.
\newblock \emph{Markov decision processes: discrete stochastic dynamic
  programming}.
\newblock John Wiley \& Sons, 2014.

\bibitem[Rust(1987)]{rust_1987}
J.~Rust.
\newblock Optimal replacement of gmc bus engines: An empirical model of harold
  zurcher.
\newblock \emph{Econometrica}, 55\penalty0 (5):\penalty0 999--1033, 1987.

\bibitem[Rust(1994)]{rust_1994}
J.~Rust.
\newblock Structural estimation of markov decision processes.
\newblock \emph{Handbook of Econometrics}, 4:\penalty0 3081--3143, 1994.

\bibitem[Scarf(1959)]{scarf_1959}
H.~Scarf.
\newblock The optimality of (s,s) policies in the dynamic inventory problem.
\newblock In K.~Arrow, S.~Karlin, and P.~Suppes, editors, \emph{Mathematical
  Methods in the Social Sciences}, chapter~13. Stanford University Press,
  Stanford, 1959.

\end{thebibliography}

\clearpage

\baselineskip 18pt

\appendix

\section{Appendix \label{sec:appendix}}

\subsection{Correlations between inventory outcomes \label{appendix_correlations}}

Figure \ref{fig:inv_het_scatter} below, presents five scatter plots at the store level: (a) stockout frequency against ordering frequency; (b) stockout frequency against inventory-to-sales ratio; (c) stockout frequency against inventory-to-sales ratio after an order is received; (d) stockout frequency against inventory-to- sales ratio before an order is placed; and (e) inventory-to-sales ratio after an order is received against inventory-to-sales ratio before an order is placed. The simple correlations in these figures provide preliminary descriptive evidence on the possible sources of structural heterogeneity, such as heterogeneity across stores in storage cost, stockout cost, ordering cost, or demand uncertainty, which are structural parameters in our model in Section \ref{sec:structural}.

The strongest correlation appears in Panel (e), for the relationship between our measures of the thresholds \textit{S} and \textit{s}.\footnote{For the interpretation of this empirical evidence, it is useful to take into the comparative statics of the $(S,s)$ thresholds as functions of the structural parameters in the profit function. We present these comparative statics in equation (\ref{eq:compstatics_ss}) in Section     \ref{sec:ss_model}, based on results in \citeauthor{hadley_whitin_1963} (\citeyear{hadley_whitin_1963}) and \citeauthor{blinder_1981} (\citeyear{blinder_1981}).} This positive correlation can be explained by store heterogeneity in stockout costs and/or storage costs: a higher stockout cost (storage cost) implies higher (lower) values of both $s$ and $S$. In contrast, a higher lump-sum ordering cost implies a lower $s$ but a negligible effect on $S$. Therefore, the positive correlation between the lower and upper thresholds we observe seems more compatible with store heterogeneity in stockout and/or storage costs rather than with heterogeneity in ordering costs. We confirm this conjecture in the estimation of the structural model in Section \ref{sec:structural}.

Panel (a) shows a negative relationship between the stockout frequency and the ordering frequency. As one would expect, stores placing orders more frequently tend to have lower stockout rates. Panels (b) and (c) show a small negative relationship between stockout rates and the inventory to sales ratio overall and after an order is received, respectively. These findings are what we would expect: stores that have lower inventory-on-hand on average experience higher stockout rates, and stores that order up to a smaller threshold \textit{S} also experience higher stockout rates. Relatedly, panel (d) shows a small negative relationship between stockout rates and our measure of the threshold \textit{s}. Again, this is what we would expect, as stores with a lower safety stock level are more likely to experience higher stockout rates.

\clearpage

\begin{figure}[ht]
\begin{center}    
\caption{Inventory Scatter Plots\label{fig:inv_het_scatter}}
\begin{subfigure}{.4\textwidth}
    \caption{Stockouts and Orders}
    \centering
    \includegraphics[width=6cm]{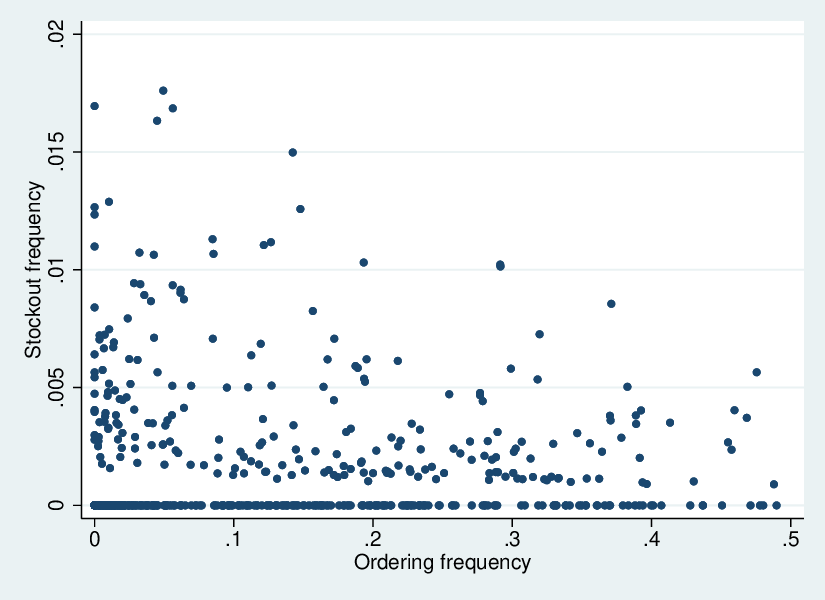}
\end{subfigure}
\vspace{5mm}
\begin{subfigure}{.4\textwidth}
    \caption{Stockouts and Inventory}
    \centering
    \includegraphics[width=6cm]{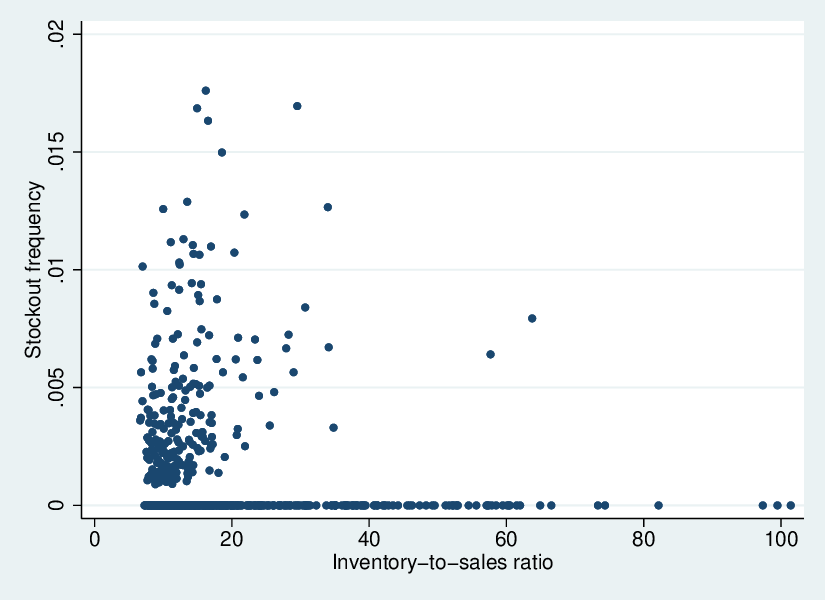}
    \end{subfigure}
\vspace{5mm}
\begin{subfigure}{.4\textwidth}
    \captionsetup{justification=centering}
    \caption{Stockouts and Inventory,
    \\ After Order}
    \centering
    \includegraphics[width=6cm]{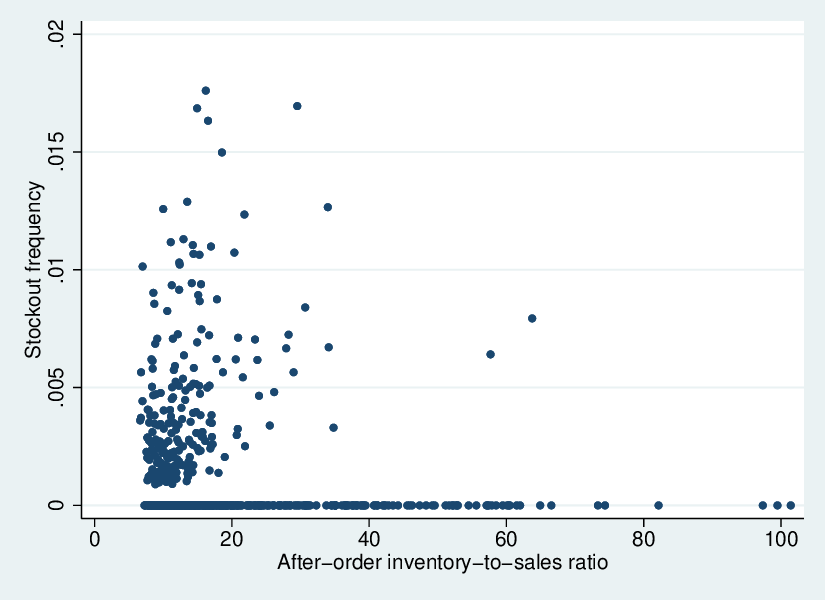}
\end{subfigure}
\begin{subfigure}{.4\textwidth}
    \captionsetup{justification=centering}
    \caption{Stockouts and Inventory, 
    \\ Before Order}
    \centering
    \includegraphics[width=6cm]{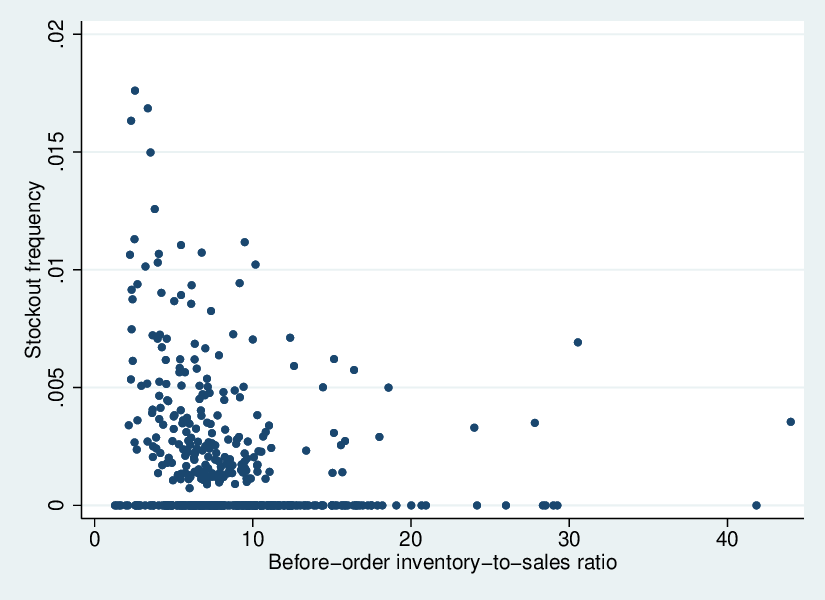}
    \end{subfigure}
\begin{center}
\begin{subfigure}{.4\textwidth}
    \captionsetup{justification=centering}
    \caption{Inventory Before and
    \\ After Order}
    \centering
    \includegraphics[width=6cm]{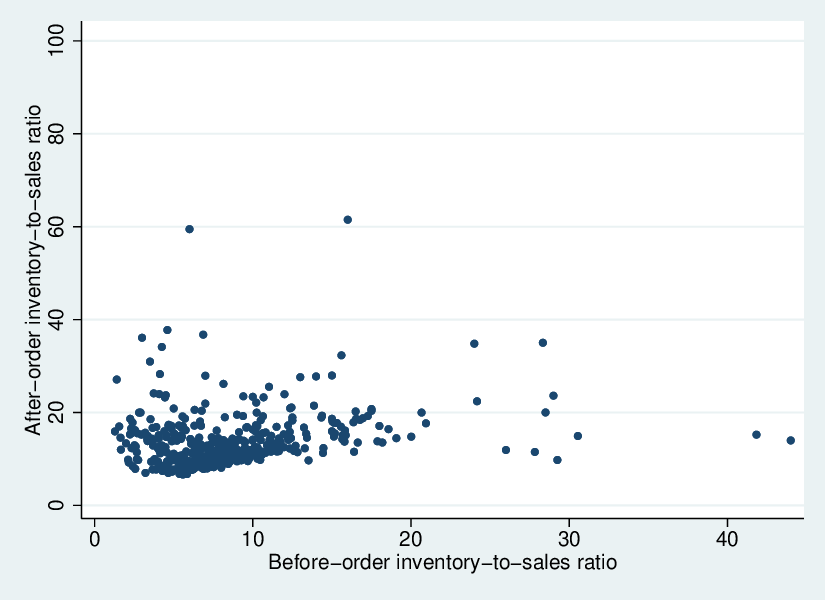}
\end{subfigure}
\vspace{5mm}
\end{center}
\end{center}
\end{figure}

\clearpage

\subsection{Correlations between manager and store characteristics \label{sec_man_sto_correlations}}

\begin{figure}[ht]
\begin{center}
\caption{Manager Characteristics and Store Classification \label{fig:man_sto_correlations}}
\begin{subfigure}{.4\textwidth}
    \caption{LCBO Experience}
    \centering
    \includegraphics[width=6cm]{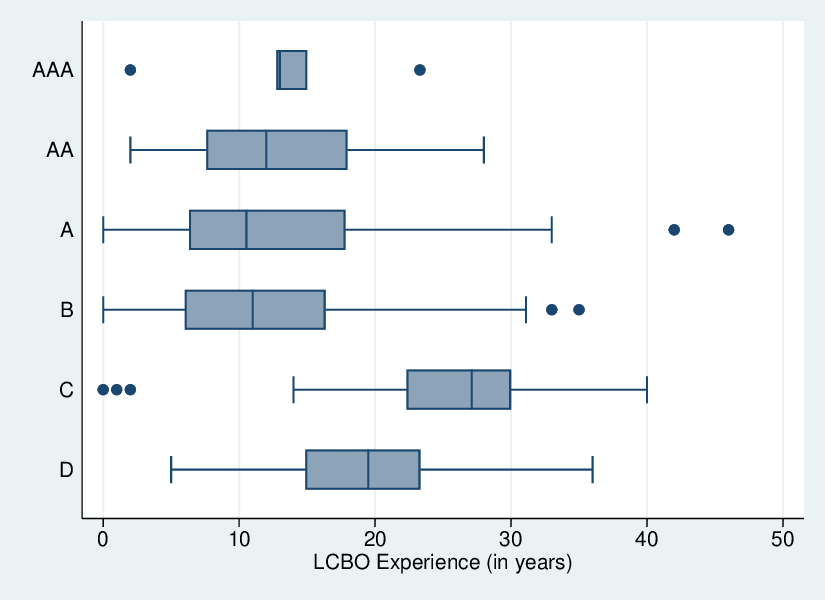}
\end{subfigure}
\vspace{5mm}
\begin{subfigure}{.4\textwidth}
    \caption{Other Retail Experience}
    \centering
    \includegraphics[width=6cm]{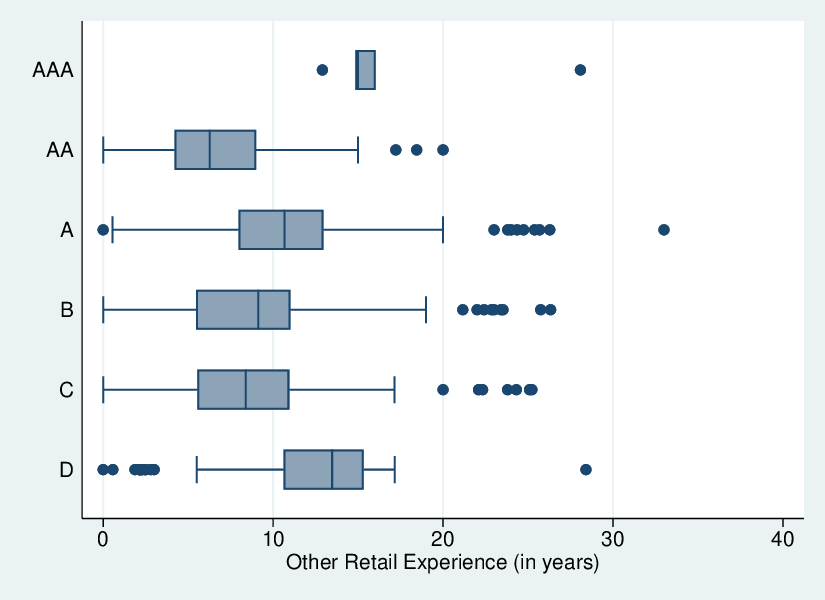}
    \end{subfigure}
\vspace{5mm}
\begin{center}
\begin{subfigure}{.4\textwidth}
    \caption{Educational Attainment}
    \centering
    \includegraphics[width=6cm]{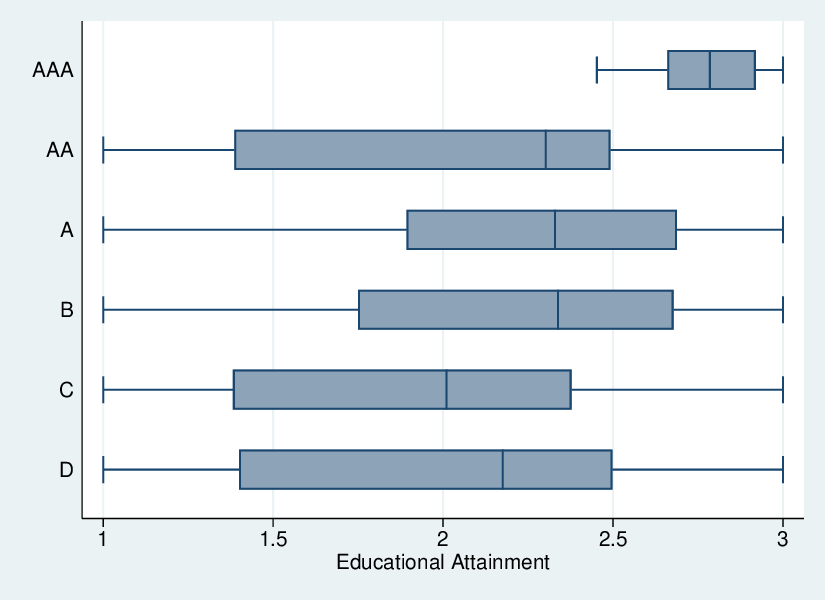}
\end{subfigure}
\vspace{5mm}
\end{center}
\end{center}
\end{figure}

Figure \ref{fig:man_sto_correlations} presents correlations between education and experience of managers and store classification. Across all three panels, there seems to be a small positive relationship between store classification and manager characteristics. However the strongest relationship is in Panel (c), where higher educational attainment is associated with higher store classification.

\medskip

\clearpage

\subsection{Estimates of Sales Forecasting Equation \label{sec:appendix_sales_equation}}

Table \ref{tab: sales_forecast} summarizes our estimation results of the sales forecasting function. For each store and product, we estimate a Negative Binomial regression function using Maximum Likelihood. The set of explanatory variables includes the logarithm of retail price, the logarithm of the store-product sales in the last week, and two seasonal dummies: a weekend dummy, and a holiday dummy for a major holiday. For each product, Table \ref{tab: sales_forecast} reports the three quartiles in the distribution across stores of parameter estimates and their respective standard errors for the coefficient of log-price, the coefficient of log-lagged-weekly-sales, and the over-dispersion parameter in Negative Binomial model. Given our interest in the forecasting power of this equation, we also report the three quartiles of McFadden's  Pseudo R-squared coefficient (i.e., one minus the ratio between the log-likelihoods of the estimated model and a model only with a constant term).

The estimates of for the lagged-sales coefficient show very substantial time persistence for all products and most stores. Standard errors show that this parameter is estimated with enough precision. The estimates for the log-price coefficient are mostly negative and large in absolute value, though they are not precisely estimated as LCBO changes prices quite infrequently. The estimate of the over-dispersion parameter is substantially smaller than one for almost all the stores and products, which implies evidence of over-dispersion and the rejection of the Poisson regression model. The magnitude of the Pseudo R-squared coefficient is on average around $6\%$ for the median store, which seems small. However, it is important to note that the uncertainty about daily sales of a single product and store can be substantially larger than the uncertainty about aggregate sales at monthly level or over products or/and stores.

\newpage

\begin{table}[ht]
\begin{center}
\caption{Sales Forecasting Equation -- Negative Binomial Model\label{tab: sales_forecast}}
\resizebox{0.9\textwidth}{!}{\begin{tabular}{r|ccc}
\hline \hline 
&
\multicolumn{1}{c}{\textit{$1^{\text{st}}$ Quartile}}
&
\multicolumn{1}{c}{\textit{$2^{\text{nd}}$ Quartile}}
&
\multicolumn{1}{c}{\textit{$3^{\text{rd}}$ Quartile}}
\\ \hline
& \textit{Est.} & \textit{Est.} & \textit{Est.} 
\\
& (\textit{st.err.}) & (\textit{st.err.}) & (\textit{st.err.}) 
\\ \hline
\multicolumn{1}{r|}
{{\textit{SKU \#67 -- Smirnoff Vodka}}}
& & &
\\
ln(\textit{p})
& -1.9740 & -0.6215 & 0.3442 \\
& (2.1803) & (1.0068) & (1.1901) \\
ln(\textit{Q})
& 0.3168 & 0.5202 & 0.6352 \\
& (0.0784) & (0.0687) & (0.0567) \\
$\alpha$
& 0.2569 & 0.3344 & 0.5130 \\
& (0.0336) & (0.1439) & (0.2490) \\
$\text{Pseudo-R}^{2}$
& 0.0633 & 0.0852 & 0.1034 \\
\hline
\multicolumn{1}{r|}
{{\textit{SKU \#117 -- Bacardi Superior White Rum}}}
& & &
\\
ln(\textit{p})
& -4.9425 & -2.4639 & -0.7968 \\
& (3.5171) & (2.6876) & (1.4862) \\
ln(\textit{Q})
& 0.2901 & 0.4855 & 0.6233 \\
& (0.0925) & (0.0811) & (0.0552) \\
$\alpha$
& 0.2246 & 0.2967 & 0.4746 \\
& (0.0532) & (0.0401) & (0.0795) \\
$\text{Pseudo-R}^{2}$
& 0.0565 & 0.0840 & 0.1015 \\
\hline
\multicolumn{1}{r|}
{{\textit{SKU \#340380 -- Two Oceans Sauvignon Blanc}}}
& & &
\\
ln(\textit{p})
& -6.1316 & -4.7974 & -3.3746 \\
& (0.9763) & (0.6034) & (0.6559) \\
ln(\textit{Q})
& 0.1208 & 0.2760 & 0.4100 \\
& (0.0801) & (0.0802) & (0.0882) \\
$\alpha$
& 0.4366 & 0.6963 & 1.2437 \\
& (0.0421) & (0.1397) & (0.1145) \\
$\text{Pseudo-R}^{2}$
& 0.0361 & 0.0534 & 0.0710 \\
\hline
\multicolumn{1}{r|}
{{\textit{SKU \#550715 -- Forty Creek Select Whisky}}}
& & &
\\
ln(\textit{p})
& -4.4488 & -2.2356 & -0.5021 \\
& (3.4604) & (3.9153) & (2.3409) \\
ln(\textit{Q})
& 0.1275 & 0.2772 & 0.4409 \\
& (0.0890) & (0.0848) & (0.0723) \\
$\alpha$
& 0.1507 & 0.2112 & 0.3286 \\
& (0.0182) & (0.0577) & (0.0291) \\
$\text{Pseudo-R}^{2}$
& 0.0469 & 0.0727 & 0.0880 \\
\hline
\multicolumn{1}{r|}
{{\textit{SKU \#624544 -- Yellow Tail Shiraz Red}}}
& & &
\\
ln(\textit{p})
& -4.5672 & -3.1739 & -2.2505 \\
& (1.1090) & (1.6732) & (1.5370) \\
ln(\textit{Q})
& 0.2080 & 0.4003 & 0.5749 \\
& (0.0762) & (0.0772) & (0.0625) \\
$\alpha$
& 0.3849 & 0.5730 & 0.9906 \\
& (0.0333) & (0.0425) & (0.1443) \\
$\text{Pseudo-R}^{2}$
& 0.0299 & 0.0491 & 0.0642 \\
\hline \hline
\end{tabular}}
\end{center}
\end{table}

\clearpage

\subsection{Stockouts at the Warehouse Level \label{appendix:warehouse_stockouts}}

\medskip

\begin{table}[ht]
\caption{Warehouse Stockout Events \label{tab: warehouse_stockouts}}
\centering
\begin{tabular}{r|cccccc}
\hline \hline 
& \textit{Days with a} & \textit{\% of total days} \\
& \textit{warehouse stockout} & \textit{in sample} \\
\hline
\\
\textit{SKU \# 67}
& 3 & 0.4 \%
\\
\textit{SKU \# 117}
& 1 & 0.1 \%
\\
\textit{SKU \# 340380}
& 1 & 0.1 \%
\\
\textit{SKU \# 550715}
& 1 & 0.1 \%
\\
\textit{SKU \# 624544}
& 1 & 0.1 \%
\\ 
\\
\hline \hline
\end{tabular}
\end{table}

\medskip

\begin{figure}[ht]
    \caption{Total Daily Deliveries}
    \centering
    \includegraphics[width=12cm]{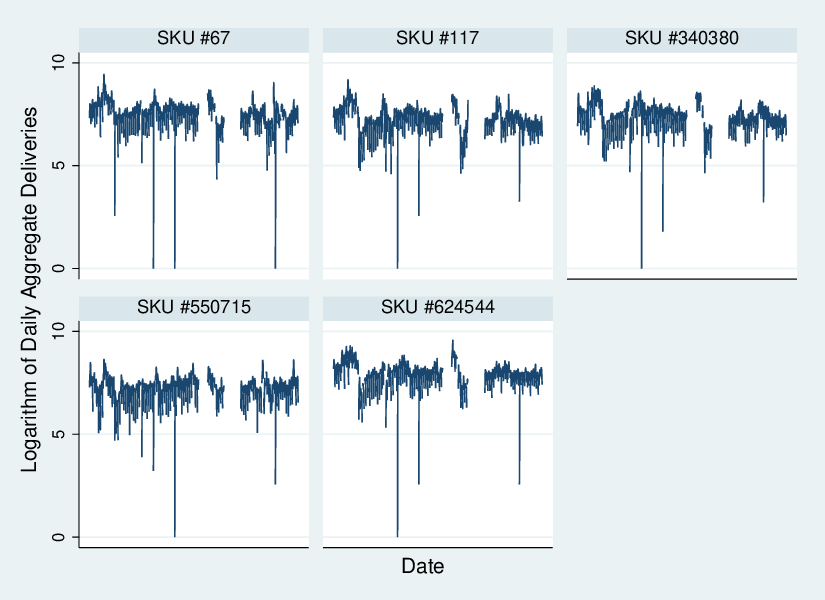}
\end{figure}

\clearpage

\begin{figure}[ht]
    \caption{Daily Number of Stores Receiving Deliveries}
    \centering
    \includegraphics[width=12cm]{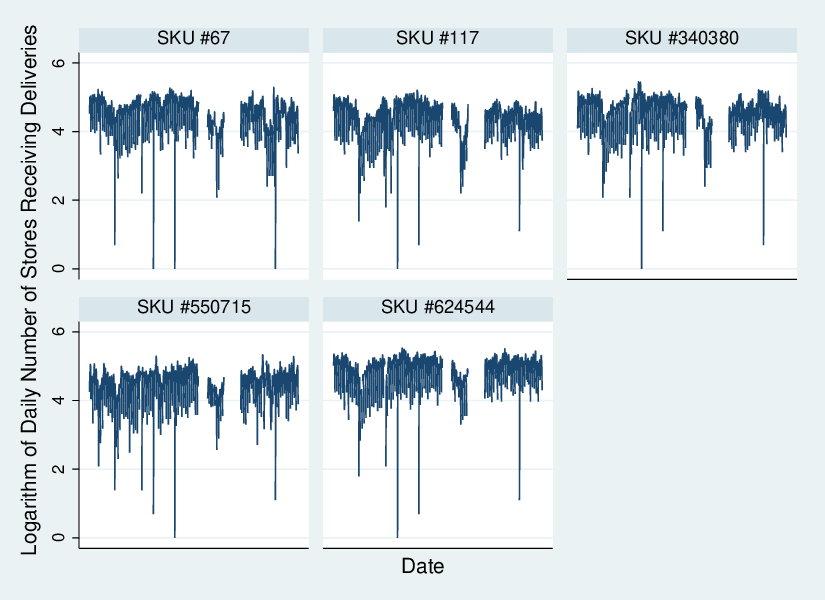}
\end{figure}

\medskip

\subsection{Store heterogeneity in estimated (\textit{S,s}) thresholds \label{appendix:ss_estimation}}

\begin{figure}[ht]
\begin{center}
\caption{Optimal Thresholds \label{fig: threshold_cdf}}
\begin{subfigure}{.4\textwidth}
    \caption{\textit{Log-}$s_{0}$}
    \centering
    \includegraphics[width=6cm]{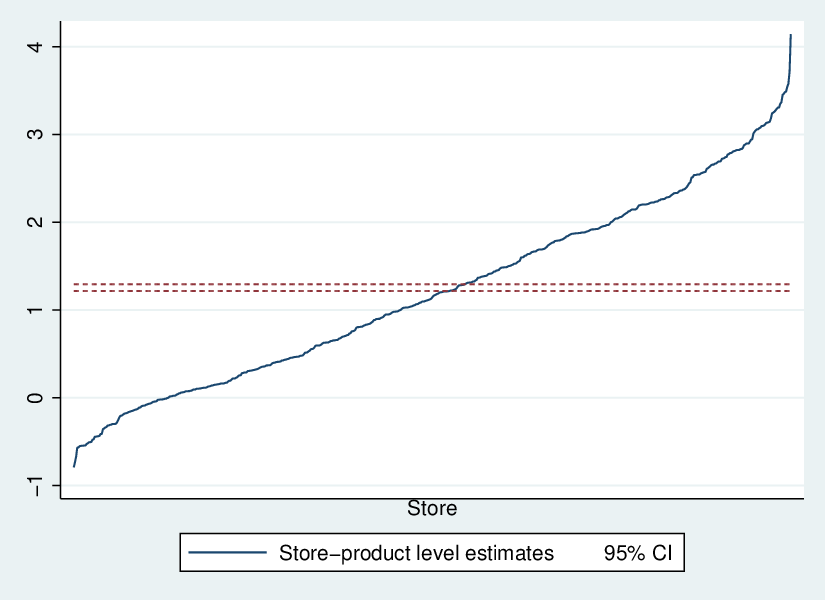}
\end{subfigure}
\begin{subfigure}{.4\textwidth}
    \caption{\textit{Log-}$S_{0}$}
    \centering
    \includegraphics[width=6cm]{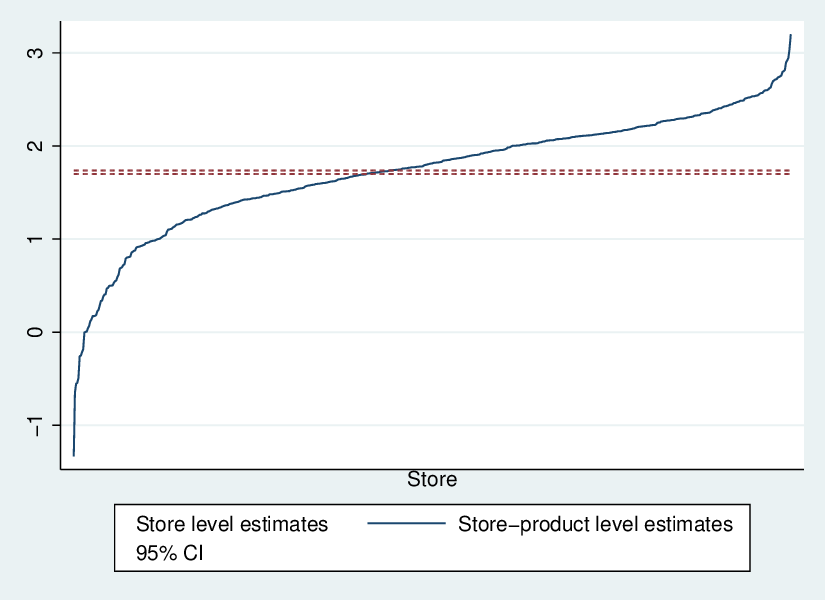}
    \end{subfigure}
\end{center}
\end{figure}

We investigate the heterogeneity across stores in the estimated thresholds. For each store-product pair, we begin by obtaining the \textit{log-lower-threshold} and the \textit{log-upper-threshold} evaluated at the mean value of log-price and retail-specific mean value of log-expected-demand. We denote these log-thresholds as \textit{log-}$s_{0}$ and \textit{log-}$S_{0}$, respectively. Figure \ref{fig: threshold_cdf} presents the inverse CDF of the store-specific average estimates of \textit{log-}$s_{0}$ and \textit{log-}$S_{0}$ and the Bonferroni $95\%$ confidence band under the null hypothesis of store homogeneity.  These distributions show very significant differences in store level estimates. For the log-lower-threshold, 98\% of stores lie outside the confidence bands and therefore have different values of store level log-thresholds. For the upper-log-threshold, only 3\% of stores lie within the confidence bands, which entails that 97\% of stores have different values of store level log-thresholds. 

In addition to this between-store heterogeneity, we also observe significant positive correlation between the two log-thresholds. This confirms our previous conjecture from Figure \ref{fig:inv_het_scatter}, in which we observed a positive correlation between the inventory to sales ratio before and after an order is placed. Again, this correlation can be explained by differences across stores in stockout costs or/and inventory holding costs. 

Given that we have estimates at the store-product level, we can also explore within-store heterogeneity. Table \ref{tab: threshold_decomp} below presents a variance decomposition of the log-thresholds $s_{0}$ and $S_{0}$. More specifically, we are interested in disentangling how much of the differences we observe in Figure \ref{fig: threshold_cdf} is attributable to variation across stores, and how much is because of differences across products. Table \ref{tab: threshold_decomp} presents an interesting finding: for the lower threshold, between-store variance is significantly larger than within-store variance, while the opposite is true for the upper threshold. That is, the order-up-to quantity seems to be relatively homogeneous across stores, while the safety stock level seems to vary significantly.

\medskip 

\medskip

\begin{table}[ht]
\begin{center}
\caption{Variance Decomposition of Log-thresholds \label{tab: threshold_decomp}}
\begin{tabular}{c c c}
\hline \hline 
& & 
\\
\multicolumn{1}{c}{\textbf{Variance }} 
& \multicolumn{1}{c}{$\boldsymbol{\textit{\textbf{Log-}}s_{0}}$} 
& \multicolumn{1}{c}{$\boldsymbol{\textit{\textbf{Log-}}S_{0}}$}
\\
& & 
\\ \hline
& & 
\\
Between-store Variance & 1.03 & 0.38 \\
& & 
\\
Within-store Variance & 0.48 & 0.66 \\
& & 
\\
\hline \hline
\end{tabular}
\end{center}
\end{table}

\subsection{Details on estimation method of structural parameters \label{appendix_estimation_method}}

\noindent \textbf{(i) Nonparametric estimation of CCP function}. In the first step of the 2PML method, we use the following Kernel method for the estimation of the CCP function. For every $(y,\mathbf{x}) \in \mathcal{Y} \times \mathcal{X}$:

\begin{equation}
    \widehat{P}(y|\mathbf{x}) \text{ } = \text{ }
    \frac{ \sum_{t=1}^{T} \mathbbm{1} \{y_{t} = y\} K_{T}(\textbf{x}_{t} - \textbf{x})}{\sum_{t=1}^{T} K_{T}(\textbf{x}_{t} - \textbf{x})}
\end{equation}

\medskip

where $K_{T}(\mathbf{u})$ is the Kernel function $1/(1 + \sqrt{T} \lvert\lvert \mathbf{u} \rvert\rvert)$ with $\lvert\lvert . \rvert\rvert$ being the Euclidean distance. 

\medskip 

\noindent \textbf{(ii) Discretization of state variables}. This estimation method applies to models where the vector of state variables $\textbf{x}$ has discrete support. In principle, our state variables have continuous support. We have applied a \textit{K-means clustering} method for the discretization of the exogenous state variables. We apply this method separately for each store-product. More specifically, we apply K-means to discretize variables $p_{t}$ and $\ln Q^{[-7,-1]}_{t}$. For every store and product pair, we cluster these variables using a k-means algorithm with a squared Euclidean distance metric, and using \citeauthor{arthur_vassilvitskii_2007} (\citeyear{arthur_vassilvitskii_2007})'s  \textit{k-means}++ cluster initialization. For both variables, we impose the number of clusters to be $2$. For the endogenous state variable $k_{t}$, along with the choice variable $y_{t}$, we choose a set of fixed grid points. Specifically, we allow $k_{t}$ to take values between 0 and 100 with an interval of 2, and $y_{t}$ to take values between 0 and 48 with an interval of 6. The latter preserves an important aspect of the nature of orders being placed by store managers at LCBO: most orders are placed in multiples of 6, and most order sizes are smaller or equal to 48 \footnote{Note that $\ln d^{e}$ and $\sigma^{2}$ are indirectly clustered through $\ln Q$ and $p$, as the sales forecasting equation determines the space of the variables $\ln d^{e}$ and $\sigma^{2}$}. Table \ref{tab: order_freq} below presents the frequency of orders in the choice set. Overall, the grid points in the choice space represent approximately 98\% of orders that we observe in the data.

\medskip

\begin{table}[ht]
\begin{center}
\caption{Frequency of Orders \label{tab: order_freq}}
\begin{tabular}{r|ccccccccc}
\hline \hline
&
\multicolumn{1}{c}{0}
&
\multicolumn{1}{c}{6}
&
\multicolumn{1}{c}{12}
&
\multicolumn{1}{c}{18}
&
\multicolumn{1}{c}{24}
&
\multicolumn{1}{c}{30}
&
\multicolumn{1}{c}{36}
&
\multicolumn{1}{c}{42}
&
\multicolumn{1}{c}{48}
\\ \hline
& 
\\
Frequency of orders (in \%)
& $84.29$  & $0.17$ & $10.05$  &  $0.04$ &  $2.63$ &  $0.01$ &  $0.91$ &  $0.01$ &  $0.41$ \\
\\
\hline \hline
\end{tabular}
\end{center}
\end{table}

Finally, in Table \ref{tab: state_space_decomp} below, we present a variance decomposition of the state space. Our goal is to assess whether the discretization of the state space is such that it captures most of the variation we observe in the data. The discretization of variables $p$ and $\text{ln}Q$ preserves most of their sample variation, with discretized variance representing approximately 99\% and 89\% of overall variance, respectively. However, for variable $k$, the discretization is significantly restrictive. The variance of the discretized variable represents only approximately 20\% of the overall variance.

\begin{table}[ht]
\begin{center}
\caption{State Space Variance Decomposition \label{tab: state_space_decomp}}
\resizebox{\textwidth}{!}{\begin{tabular}{r|cccc}
\hline \hline
&
\multicolumn{1}{c}{\textit{Overall Variance}}
&
\multicolumn{1}{c}{\textit{Between (Discretized) Variance}}
&
\multicolumn{1}{c}{\textit{Within (Residual) Variance}}
&
\multicolumn{1}{c}{\textit{Between Variance \%}}

\\ \hline
& 
\\
$k$
& $4,055.56$  & $835.20$ & $2,256.29$  &  $20.59$ \\
&
\\
$p$
& $51.03$  & $51.00$ & $0.03$  & $99.93$  \\
&
\\
$\text{ln}Q$
& $1.30$ & $1.16$ & $0.14$ & $89.36$  \\
&
\\
\hline \hline
\end{tabular}}
\end{center}
\end{table}

\noindent \textbf{(iii) Computing time}.
Most of the computing time in the implementation of this two-step estimator comes from the calculation of present values, and more specifically from the inversion of matrix $\mathbf{I} - \beta \sum_{y=0}^{J}        \mathbf{P}(y) \ast \mathbf{F}_{x}(y)$ that has dimension $|\mathcal{X}| \times |\mathcal{X}|$. Nevertheless, the computing time to obtain the 2PML for one store-product -- using standard computer equipment -- was around $20$ seconds, and the total computing time for the approximately $634 \times 5 = 3,160$ store-products in our working sample was less than $18$ hours.

\subsection{Empirical distribution of parameter estimates \label{appendix_empirical_distribution_estimates}}

\begin{figure}[ht]
\begin{center}    
\caption{$\boldsymbol{\gamma}$ Estimates
\label{fig:gamma_histograms}}
\begin{subfigure}{.4\textwidth}
    \caption{$\gamma^{h}$}
    \centering
    \includegraphics[width=6cm]{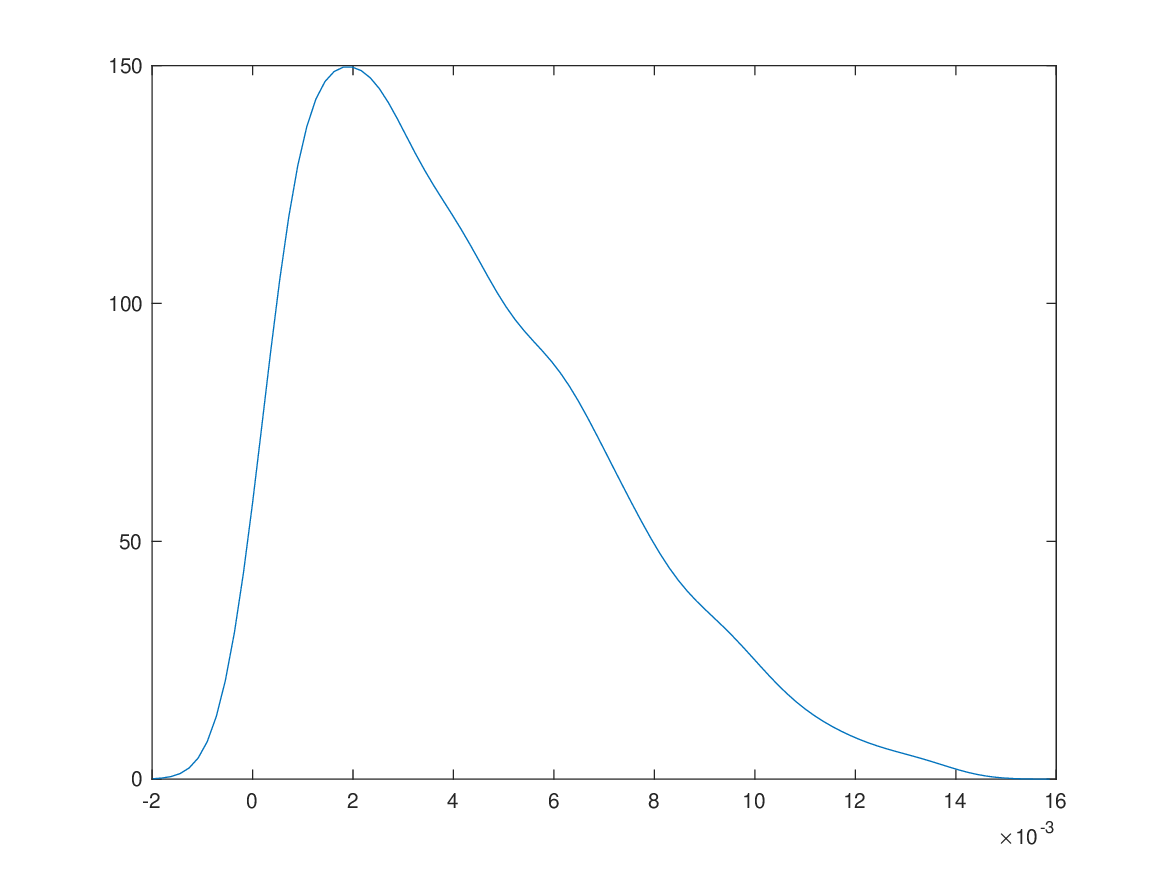}
\end{subfigure}
\begin{subfigure}{.4\textwidth}
    \caption{$\gamma^{z}$}
    \centering
    \includegraphics[width=6cm]{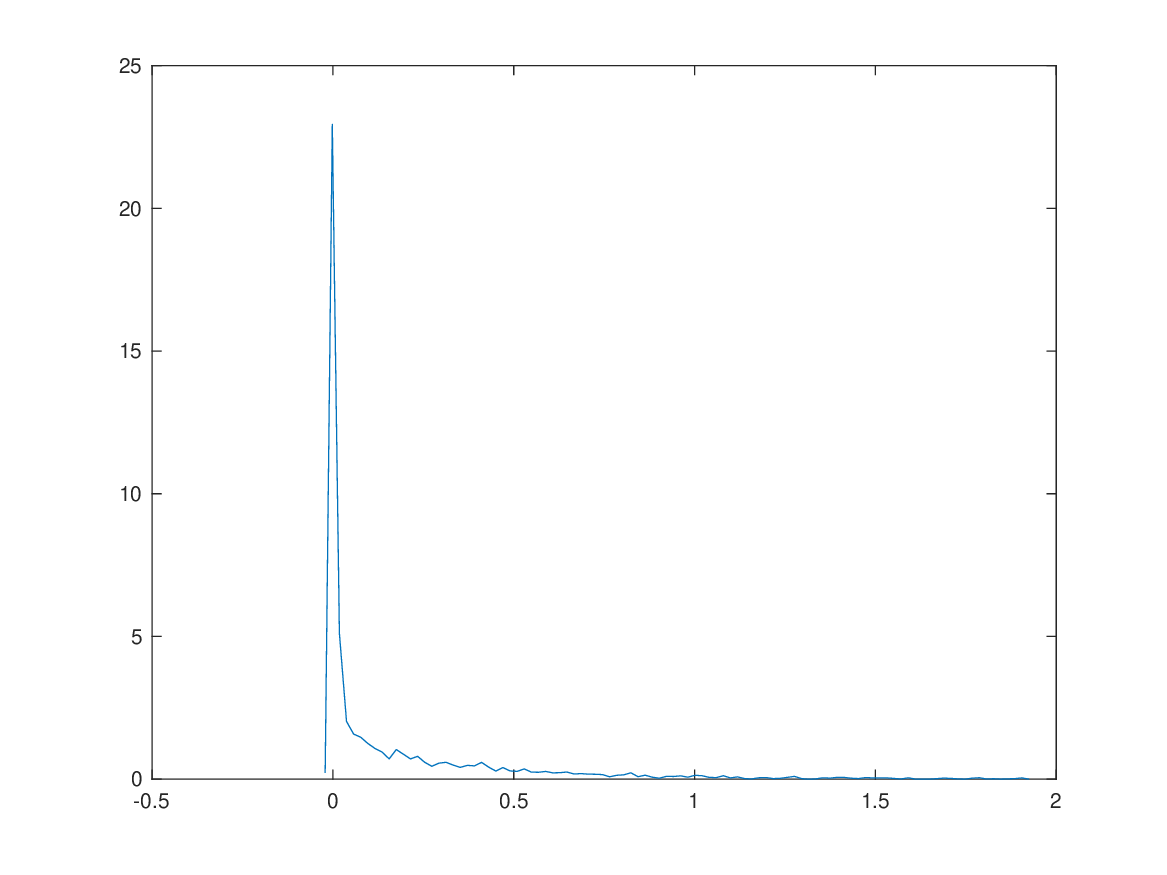}
    \vspace*{2mm}
\end{subfigure}
\begin{subfigure}{.4\textwidth}
    \caption{$\gamma^{f}$}
    \centering
    \includegraphics[width=6cm]{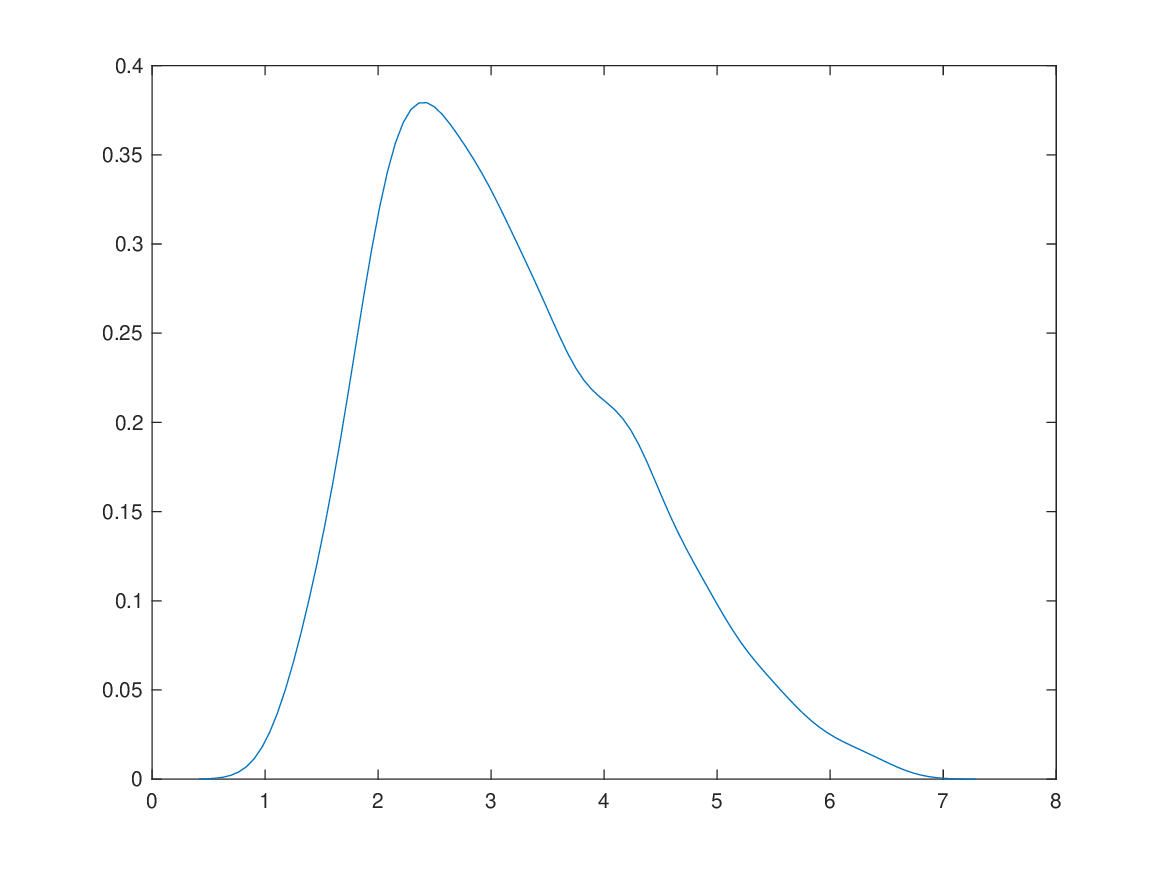}
\end{subfigure}
\begin{subfigure}{.4\textwidth}
    \caption{$\gamma^{c}$}
    \centering
    \includegraphics[width=6cm]{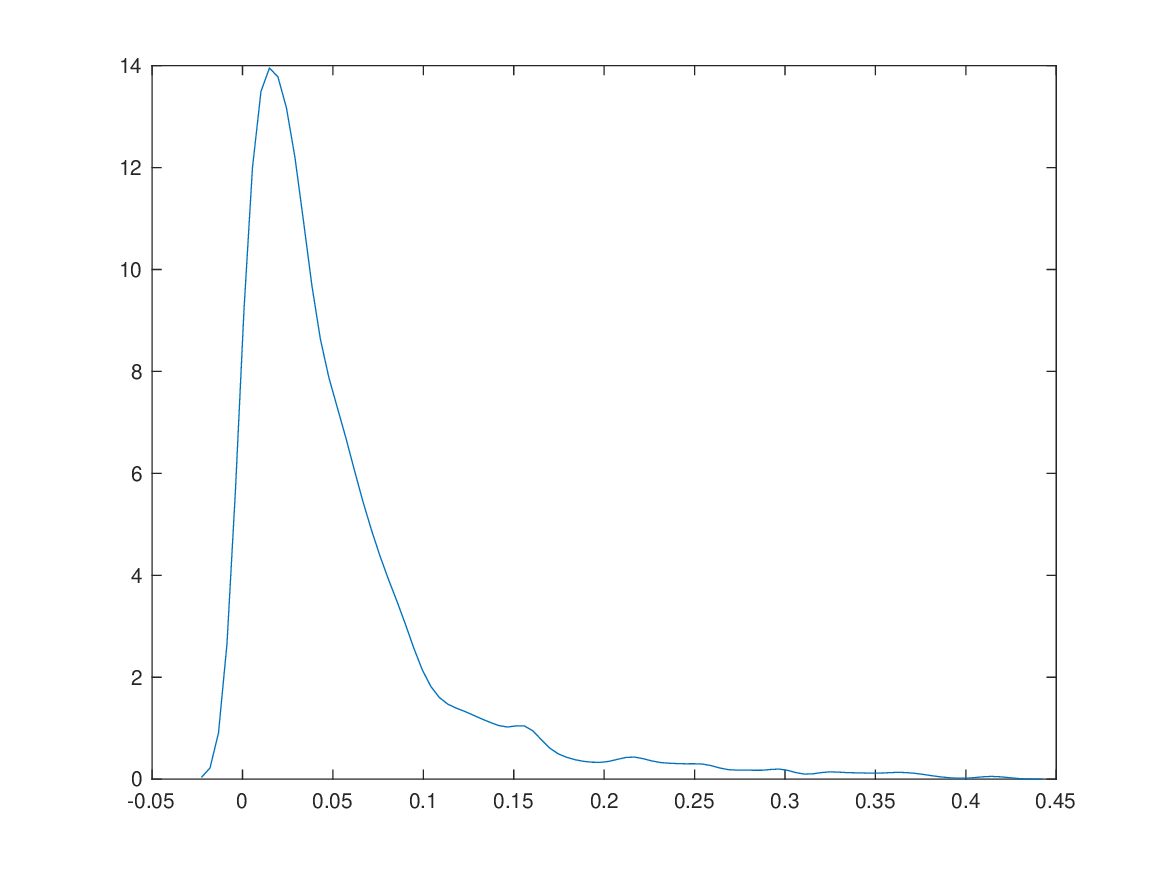}
\end{subfigure}
\end{center}
\end{figure}

Figure \ref{fig:gamma_histograms} plots the empirical density across stores and products of our raw estimates of the four structural parameters, measured in dollar amounts. These empirical densities show substantial heterogeneity across stores and products in the four parameter estimates. 

\subsection{Shrinkage estimator \label{appendix_shrinkage_estimator}}

To correct for excess dispersion, we consider the following \textit{shrinkage estimator} (see \cite{gu_koenker_2017}):
\begin{equation*}
    \widehat{\gamma}^{\ast}_{i,j} \text{ } = \text{ }
    \bar{\widehat{\gamma}} \text{ } + \text{ }
    \left(
        1 - \frac{\widehat{\sigma}^{2}_{i,j}}
        {Var(\widehat{\gamma})}
    \right)^{1/2}
    \text{ }
    \left(
    \widehat{\gamma}_{i,j} -
    \bar{\widehat{\gamma}}
    \right)
\end{equation*}
where $\widehat{\gamma}_{i,j}$ is the original parameter estimate; $\widehat{\sigma}_{i,j}$ is its standard error, and $\bar{\widehat{\gamma}}$ and  $Var(\widehat{\gamma})$ represent the mean and variance, respectively, in the empirical distribution of $\widehat{\gamma}_{i,j}$ across stores and products. This estimator generates a distribution of estimates across stores-products that corrects for the spurious heterogeneity due to estimation error. By construction, we have that $Var(\hat{\gamma}^{*}_{ij}) = Var(\hat{\gamma}_{ij}) - E(\sigma^{2}_{ij})$.

\subsection{Heterogeneity in realized inventory management costs \label{appendix_realized_costs}}

\begin{figure}[ht]
\begin{center}
\caption{Empirical CDFs of Realized Inventory Management Costs to Revenue Ratios \label{fig: inv_ratios}}
\begin{subfigure}{.4\textwidth}
    \caption{Holding Cost to Revenue}
    \centering
    \includegraphics[width=6cm]{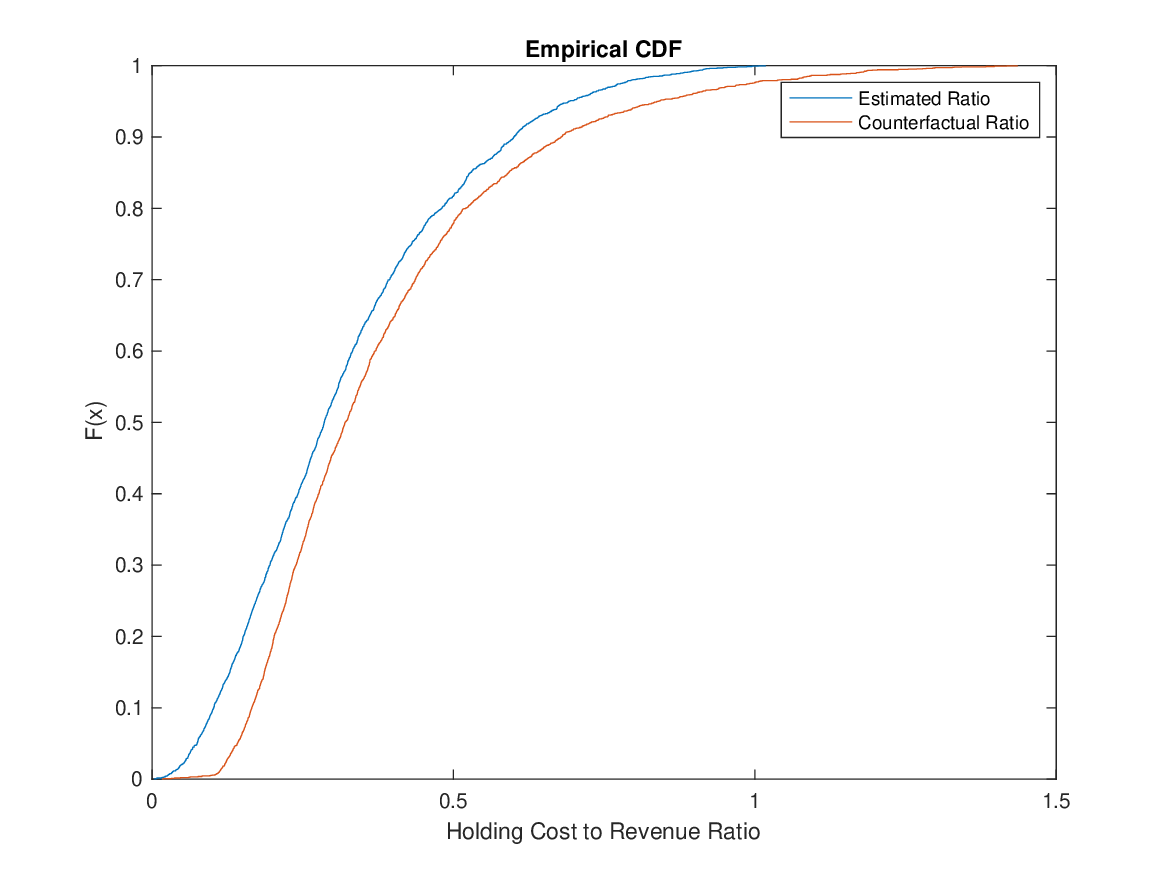}
\end{subfigure}
\begin{subfigure}{.4\textwidth}
    \caption{Stockout Cost to Revenue}
    \centering
    \includegraphics[width=6cm]{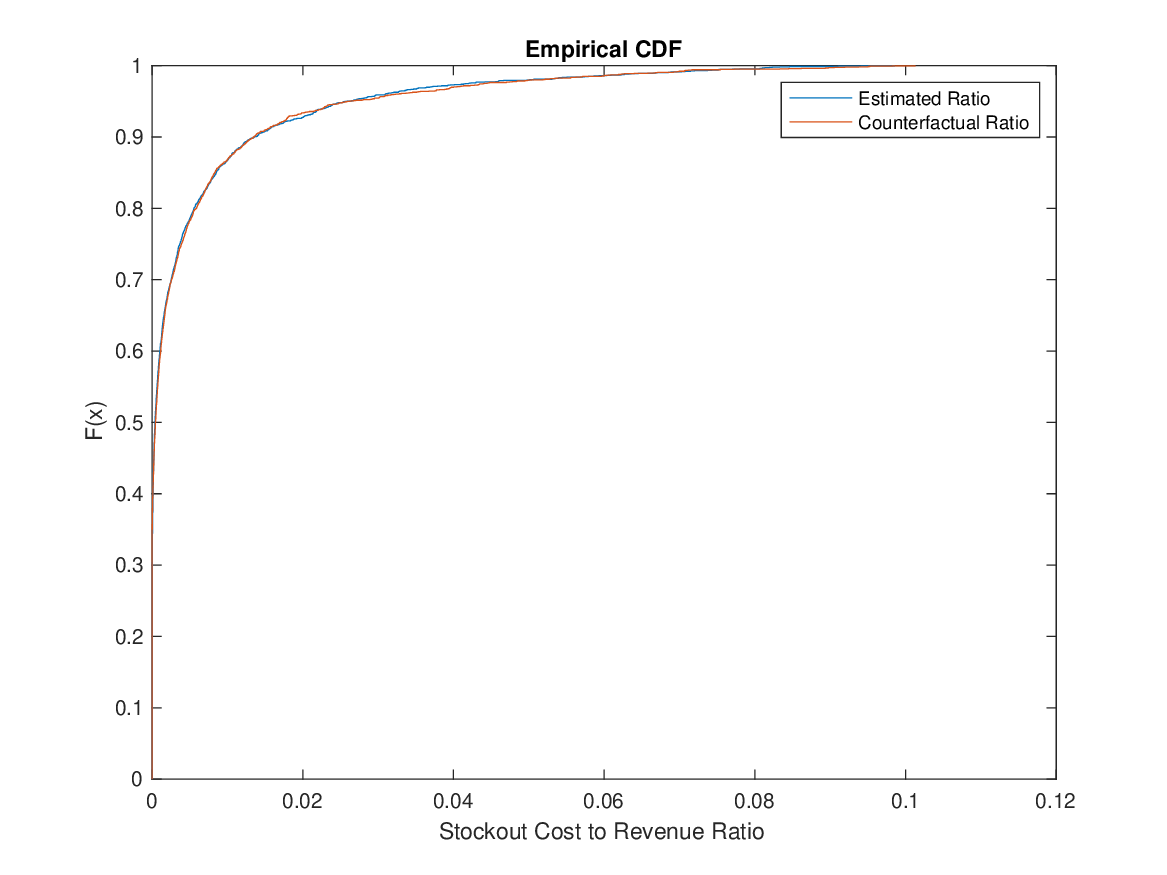}
    \vspace*{2mm}
\end{subfigure}
\begin{subfigure}{.4\textwidth}
    \caption{Fixed Ordering Cost to Revenue}
    \centering
    \includegraphics[width=6cm]{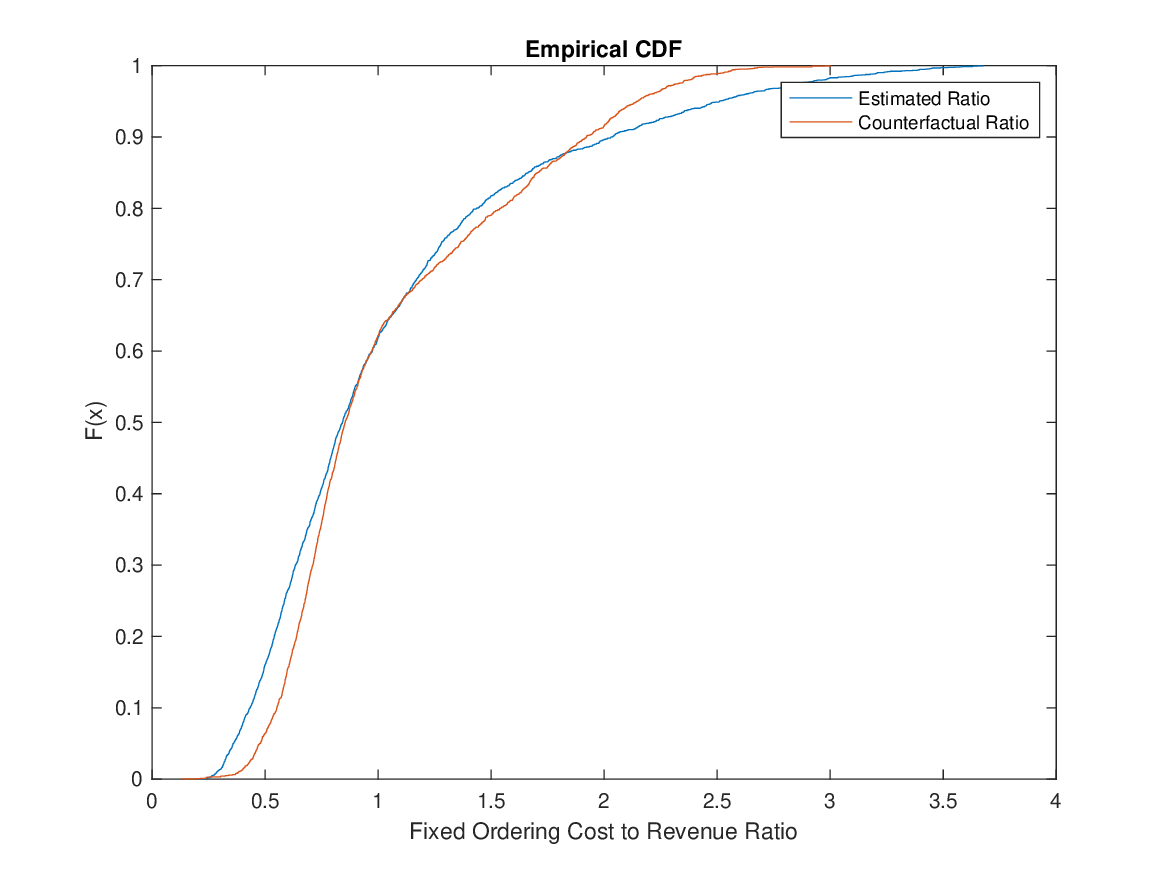}
\end{subfigure}
\begin{subfigure}{.4\textwidth}
    \caption{Variable Ordering Cost to Revenue}
    \centering
    \includegraphics[width=6cm]{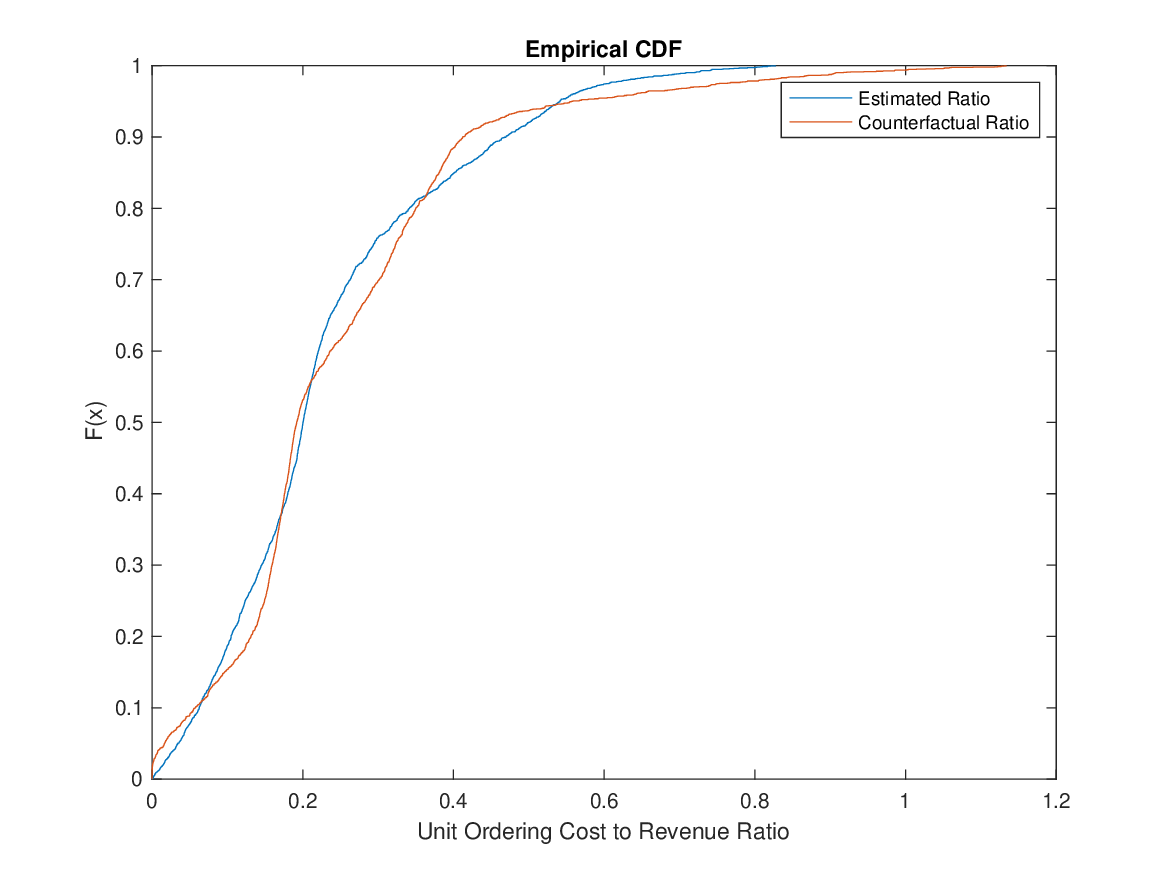}
\end{subfigure}
\end{center}
\end{figure}

In Figure \ref{fig: inv_ratios}, the blue curves represent the CDFs across stores and products of each of the four cost-to-revenue ratios. These distributions show the following ranges between percentiles $5\%$ and $95\%$: $[0.1\%, 0.7\%]$ for the inventory holding cost; $[0.0\%, 0.03\%]$ for the stockout cost; $[0.4\%, 2.75\%]$ for the fixed ordering cost; and $[0.05\%, 0.6\%]$ for the variable ordering cost. We can see that the realized fixed ordering costs are not only the costs with larger contribution to the firms' profit, but also with larger heterogeneity across stores.

The dispersion across stores in these cost-to-revenue ratios is the combination of dispersion in structural parameters and dispersion in decision and state variables affecting these costs. In particular, managers' optimal inventory decisions can partly compensate for the heterogeneity in the structural parameters. For instance, the inventory holding cost to revenue ratio for store $i$ and product $j$ is $\gamma^{h}_{i,j}$ $\Bar{k}_{i,j} / \Bar{r}_{i,j}$. A store-product with large per-unit inventory holding cost, $\gamma^{h}_{i,j}$, will tend to keep smaller levels of inventory than a store-product with a small value of this parameter such that the difference between these stores in the ratio $\gamma^{h}_{i,j}$ $\Bar{k}_{i,j} / \Bar{r}_{i,j}$  will be smaller than the difference between their per unit inventory holding cost. To measure the magnitude of this behavioural response by store managers, the red curves in Figure \ref{fig: inv_ratios} present the CDFs of the cost ratios when we replace the store-product specific structural parameters by their means across products, but we keep the values of decisions and state variables. That is, for the inventory holding cost ratio, the red curve is the CDF of variable $\Bar{\gamma}^{h}_{i}$ $\Bar{k}_{i,j} / \Bar{r}_{i,j}$. For each of the four inventory ratios, the counterfactual CDFs in the red curves are steeper than the factual CDFs in the blue curves. Store managers with a perception of higher inventory costs make decisions that entail lower costs of managing their inventories relative to revenue.

\subsection{Dispersion of cost parameters \label{appendix_cost_dispersion}}

In Table \ref{tab: dispersion_store}, we explore how the dispersion of the manager component of costs depends on store and location characteristics. Consistent with the interpretation of managers' biased perception of true costs, we find that managers in high-type stores have a smaller dispersion in this component of costs. 

\clearpage

\begin{table}[ht]
\begin{center}
\caption{Regression of Cost Dispersion on Store Characteristics \label{tab: dispersion_store}}
\resizebox{\textwidth}{!}
{\begin{tabular}{lcccc}
\hline \hline 
& ($|res(\gamma^{h})|$) & ($|res(\gamma^{z})|$)
& ($|res(\gamma^{f})|$) & ($|res(\gamma^{c})|$)
\\ \hline 
& \textit{Est.} & \textit{Est.} 
& \textit{Est.} & \textit{Est.} 
\\
& \textit{(s.e.)} & \textit{(s.e.)} 
& \textit{(s.e.)} & \textit{(s.e.)} 
\\ \hline
\addlinespace
Store Class & & & & \\
\multicolumn{1}{c}{\textit{AA}}       &    0.000822\sym{***}&      0.0210\sym{**} &      0.0149         &     0.00154         \\
                    &  (0.000212)         &   (0.00975)         &    (0.0589)         &   (0.00349)         \\
\addlinespace
\multicolumn{1}{c}{\textit{A}}       &    0.000793\sym{***}&      0.0343\sym{***}&     -0.0271         &     0.00295         \\
                    &  (0.000167)         &   (0.00928)         &    (0.0583)         &   (0.00341)         \\
\addlinespace
\multicolumn{1}{c}{\textit{B}}       &    0.000690\sym{***}&      0.0359\sym{***}&    -0.00822         &     0.00179         \\
                    &  (0.000193)         &    (0.0112)         &    (0.0656)         &   (0.00354)         \\
\addlinespace
\multicolumn{1}{c}{\textit{C}}       &    0.000529\sym{**} &      0.0270\sym{*}  &      0.0814         &     0.00149         \\
                    &  (0.000241)         &    (0.0139)         &    (0.0831)         &   (0.00393)         \\
\addlinespace
\multicolumn{1}{c}{\textit{D}}       &   0.0000842         &      0.0263\sym{*}  &       0.120         &    0.000249         \\
                    &  (0.000265)         &    (0.0153)         &    (0.0989)         &   (0.00411)         \\
\addlinespace
ln(Product Assortment Size)       &    0.000138         &     0.00490         &     -0.0536         &     0.00204\sym{*}  \\
                    &  (0.000109)         &   (0.00616)         &    (0.0367)         &   (0.00121)         \\
\addlinespace
ln(Population in City)       &   0.0000201         &    0.000206         &      0.0150         &    0.000304         \\
                    & (0.0000198)         &   (0.00118)         &   (0.00925)         &  (0.000301)         \\
\addlinespace
ln(Median Income in City)    &   -0.000583\sym{**} &   -0.000222         &       0.176\sym{*}  &     0.00133         \\
                    &  (0.000287)         &    (0.0157)         &     (0.106)         &   (0.00410)         \\
\addlinespace
\hline
Location dummies (25 regions, 4 districts) &  YES  & YES   & YES   & YES \\
Product dummies (5 products)  &  YES  & YES   & YES   & YES \\
\addlinespace
\hline
R-squared & 0.1038 & 0.0403 & 0.0712 & 0.0215 \\
Observations        &        2,589         &        2,589         &        2,589         &        2,589         \\
\hline \hline
\multicolumn{5}{l}{\footnotesize{(1) Location dummies based on LCBO's own division of Ontario into 25 regional markets and 4 districts.}}\\
\multicolumn{5}{l}{\footnotesize{(2) Robust standard errors clustered at the store level in parentheses}}\\
\multicolumn{5}{l}{\footnotesize{(3) * means p-value<0.10, ** means p-value<0.05, *** means p-value<0.01}}
\end{tabular}}
\end{center}
\end{table}

\subsection{Manager bias: granular examples \label{tab: manager_bias_examples}}

The first pair of LCBO stores we explore in Table \ref{tab: granular_ex_1} is store \#452 (1138 Avenue Road) and store \#572 (1245 Dupont Street), both located in the Toronto-North area. Although both stores are of type "\textit{A}" and have similar average weekly sales, their managers have very different years of experience at LCBO and significantly different estimates of cost parameters. Specifically, the manager of store \#572 has an additional 16 years of experience at LCBO, a 29\% higher average holding cost, a 65\% lower average stockout cost, a 13\% higher average fixed ordering cost, and a 19\% lower average unit ordering cost. The second pair of stores we examine is store \#538 (122 Rideau Street) and store \#547 (111 Albert Street), both located in the Ottawa-Central area. Again, although the two stores are of type "\textit{B}" and have similar average weekly sales, the managers have very different experience levels and significantly different cost estimates. Store \#547 has 34 fewer years of experience, a 34\% lower average holding cost, a 27\% higher average stockout cost, a 39\% higher fixed ordering cost, and a (mere) 1\% lower average unit ordering cost.

\medskip

\begin{table}[ht]
\begin{center}
\caption{Manager Biases in Cost Parameters \label{tab: granular_ex_1}}
\resizebox{\textwidth}{!}
{\begin{tabular}{r|ccccccc}
\hline \hline
&
\multicolumn{1}{c}{Store Type}
&
\multicolumn{1}{c}{Average Weekly Sales}
&
\multicolumn{1}{c}{Manager Years at LCBO}
&
\multicolumn{1}{c}{$\hat{\gamma}^{h}$}
&
\multicolumn{1}{c}{$\hat{\gamma}^{z}$}
&
\multicolumn{1}{c}{$\hat{\gamma}^{f}$}
&
\multicolumn{1}{c}{$\hat{\gamma}^{c}$}
\\ \hline
& 
\\
Store \#452
& $A$  & $92,422$ & $13$  &  $0.003347$ &  $0.097933$ &  $2.17968$ &  $0.061739$ \\
Store \#572
& $A$  & $80,376$ & $29$  &  $0.004344$ &  $0.033821$ &  $2.45856$ &  $0.050037$ \\
\\
\hline \\
Store \# 538
& $B$  & $44,254$ & $35$  &  $0.006972$ &  $0.099627$ &  $2.05306$ &  $0.047136$ \\
Store \# 547
& $B$  & $37,888$ & $1$  &  $0.004608$ &  $0.126806$ &  $2.86206$ &  $0.046638$ \\
& 
\\
\hline \hline
\end{tabular}}
\end{center}
\end{table}

\end{onehalfspacing}

\end{document}